\pdfoutput=1
\documentclass{JFM-FLM_Au}
\usepackage{graphicx}
\usepackage{amsmath}
\usepackage{subfigure}
\usepackage{soul, color, xcolor}

\lefttitle{Tao Li, Chunze Zhang}
\righttitle{Journal of Fluid Mechanics}

\title{How do trout regulate patterns of muscle contraction to optimize propulsive efficiency during steady swimming?}

\author{Tao Li\aff{1}, Chunze Zhang\aff{1}, Weiwei Yao\aff{2}, Junzhao He\aff{1}, Ji Hou\aff{1}, Qin Zhou\aff{1}, Xujin Zhang\aff{1}, Lu Zhang\aff{3}}

\affiliation{\aff{1}The College of River and Ocean Engineering, Chongqing Jiaotong University,Chongqing, 400016, China
	\aff{2}State Key Laboratory of Hydraulics and Mountain River Engineering, Sichuan University, Chengdu, 610065, China
	\aff{3}Chongqing Deepflow Intelligent Computing Technology Co., Ltd, Chongqing, 401122, China}

\corresau{Chunze Zhang, \email{zhangchunze@whu.edu.cn}}

\begin{document}
\maketitle

\begin{abstract}
	Understanding the mechanisms behind efficient fish locomotion has broad implications across various fields, including biomechanics, fluid dynamics, evolutionary biology, eco-hydraulics, and ocean engineering. Traditional researchs, however, often overlook the critical connection between neuromuscular control at the muscle–skeletal level and whole-body locomotor patterns. To study energy transmission and integration across scales in carangiform swimming and uncover high locomotor efficiency principles from an embodied intelligence perspective, we developed a bio-inspired digital trout model. The model integrated multibody dynamics, Hill-type muscle modeling, and a high-fidelity FSI algorithm, and closely replicated the morphology and mechanical properties of live trout. By leveraging deep reinforcement learning, the digital trout's neural system achieved hierarchical spatiotemporal control of muscle activation patterns. This allowed for the systematic investigation of how different muscle activation strategies influence propulsion speed and energy consumption. The results show that, under steady swimming conditions, the axial serial coupling of myomeres—with activation spanning more than 0.5 body lengths—is essential for the stable propagation of body curvature waves. Moreover, moderate muscle contraction duration  ($[0.1,0.3]$ of a tail-beat cycle) allows the fish body and surrounding fluid to function together as a passive damping system, reducing deformation velocity without active input and significantly lowering energy consumption. Finally, the activation phase lag of axial myomeres plays a crucial role in shaping the body wave. If the phase lag becomes too large, it causes antagonistic contractions on opposite sides of the body, which interfere with thrust generation. These findings enhance our understanding of bio-inspired locomotion and offer valuable insights for designing energy-efficient underwater systems.
\end{abstract}



\section{Introduction}
\label{sec:intro}

Over hundreds of millions of years, fish have evolved highly optimized propulsion systems uniquely suited to aquatic environments. These systems rely on the seamless integration of muscular, skeletal, and neural mechanisms, enabling precise and energy-efficient control of body undulations. This form of biological propulsion achieves a level of energy conversion efficiency, responsiveness, and maneuverability that still far surpasses what current bio-inspired underwater robotic technologies can replicate (\cite{reddy2015caudal,gu2024deformation}). Therefore, the study of fish locomotion holds significant importance and broad applications across various fields, including biomechanics, fluid dynamics, evolutionary biology, eco-hydraulics, and ocean engineering (\cite{molloy2009effects, liao2003fish, zhang2014prediction, shi2022experimental, bodaghi2023effects}).

Previous studies have used experimental techniques such as high-speed videography and particle image velocimetry (PIV) to systematically quantify the macroscopic kinematics of fish locomotion. These efforts have led to the development of classification models based on movement patterns, such as carangiform and anguilliform swimming modes (\cite{liao2003fish, liao2003karman, di2021convergence, yan2013interspecific}). However, these kinematics-based studies—rooted in a black-box observational approach focused on external motion—provide limited insight into the internal dynamics of fish locomotion. Specifically, they overlook the temporal patterns of muscle activation, neural encoding strategies, and the mechanisms that govern energy transfer across scales. In contrast, other researchers have attempted to replicate fish-like body wave propagation using multi-link rigid-body mechanisms driven by pre-programmed actuators, such as servo motors, to produce biomimetic undulatory motion (\cite{salazar2018optimal, katzschmann2018exploration, yan2020efficient}). While these engineering approximations can replicate the macroscopic movement patterns of fish at the kinematic level, they fundamentally mimic external motion rather than capture the internal mechanisms that generate it. At its core, fish locomotion arises from a complex interplay among actively controlled muscle contractions and relaxations, passive body deformations, and dynamic interactions with the surrounding fluid environment. Limited by its inability to achieve real-time coupling of the musculoskeletal system, body fluctuations, and surrounding flow fields, the traditional kinematic paradigm struggles to reveal the dynamic causal chains and cross-scale energy transfer mechanisms underlying neuromuscular coordination. This includes phenomena such as the neural regulation of muscle activation patterns or strain energy transfer at the tissue level.

To study fish locomotion from an integrated locomotor chain perspective, some researchers have modeled the fish body as an elastic beam using strain-based mechanics. By incorporating either simplified representations of local hydrodynamic forces (\cite{mcmillen2008nonlinear, zhang2014prediction}) or high-fidelity computational fluid dynamics (CFD) simulations (\cite{curatolo2016modeling, li2023roles}), these studies have explored how body deformation interacts with the surrounding flow field. Other researchers have used electromyography (EMG) combined with motion capture techniques to empirically link muscle activation patterns with swimming kinematics, aiming to reveal how the muscle–skeletal dynamic chain influences overall fish propulsion (\cite{williams1989locomotion,van1990function,rome1993fish,wardle1993timing,gillis1998neuromuscular}). However, these approaches still face several limitations. First, fish musculature is composed of multiple discrete myomeres, and strain-based mechanical models cannot fully capture the nonlinear variations in body stiffness that arise from the independent contractions of these muscle units. Additionally, passive stiffness along the axial length of the body is non-uniform, and the spatial neural rhythm of muscle activation and deactivation across segments can significantly influence the distribution of body stiffness. Moreover, experimental methods have difficulty quantifying how flow-induced reaction forces feed back into and influence muscle activation patterns. In summary, at least two critical questions remain unanswered: How do different muscle activation patterns dynamically regulate body stiffness? And how does this regulation at the organ scale affect swimming speed and propulsive efficiency at the macroscopic level? Addressing these questions will likely require a cross-scale investigative framework that integrates musculoskeletal mechanics, body–fluid interactions, and adaptive control mechanisms.

The high-performance locomotion of fish is not the result of pre-programmed motion patterns, but instead emerges from embodied intelligence—generated through real-time interactions between the neuromuscular-skeletal system and the surrounding fluid environment. Fish continuously adapt their swimming strategies to changing flow conditions, physiological states, and behavioural goals through a closed-loop system that integrates distributed sensing (e.g., the lateral line and vision), proprioceptive feedback (from muscles and the skeleton), and goal-directed decision-making (from the fish brain). In recent years, advances in deep reinforcement learning have opened up promising new avenues for studying fish locomotion through the lens of embodied intelligence. These intelligent algorithms enable researchers to emulate biologically inspired decision-making processes, allowing for the modeling of autonomous behaviours and adaptive strategies. Such insights are increasingly informing the design of autonomous underwater vehicles (\cite{Li2021a, zheng2021learning, TianhaoZhang2022}). This approach has also been successfully applied to a variety of fish locomotion scenarios, including multimodal swimming (\cite{gazzola2014reinforcement,Yan2020a, Zhu2021,  cui2024enhancing, zhu2025intermittent}), collective behaviour and schooling (\cite{Verma2018}), adaptive swimming in complex flow fields (\cite{PeterGunnarson2021, hou2024learning, zhang2024numerical, li2025simulating}), ecologically adaptive swimming (\cite{li2024numerical}), and fishway passage simulations (\cite{li2025conducting}). Although current studies still rely on kinematics-based body wave fitting without incorporating organ-level muscular coordination, this emerging paradigm—grounded in perception–feedback–decision–execution loops—offers a promising conceptual framework for future research on bio-inspired locomotion from the perspective of embodied intelligence.

This study aims to investigate the integrative locomotor chain of fish swimming from multiple perspectives—including internal musculoskeletal dynamics, external fluid dynamics, and internal-external fluid–structure interaction. It also seeks to elucidate the multiscale mechanisms linking muscle activity level, body undulation, and swimming behaviour. Using rainbow trout (Oncorhynchus mykiss) as the model organism, we developed a high-fidelity digital trout in a virtual environment, faithfully replicating the morphological and biomechanical characteristics of a live trout. The digital trout integrates multi-scale modeling of the locomotor chain by combining the Hill-type muscle model, multibody dynamics, and fluid–structure interaction (FSI) algorithms. Furthermore, by incorporating deep reinforcement learning, we enabled the digital trout’s neural system to achieve hierarchical spatiotemporal control of muscle activation patterns. The remainder of this paper is organized as follows: Section \ref{sec:method} presents the numerical methods. Section \ref{sec:trout} details the construction of the digital trout. Section \ref{sec:result} discusses the results and provides analysis. Finally, Section \ref{sec:conclusion} offers the conclusions.

\section{Methods}\label{sec:method}

\subsection{Fluid-structure interaction}

Developing a dedicated fluid–structure interaction (FSI) solver with bidirectional coupling capabilities is essential for investigating the integrated locomotor chain of fish within the embodied intelligence framework. To address this need, we developed a high-fidelity FSI module based on the Immersed Boundary–Lattice Boltzmann Method (IB-LBM). The Immersed Boundary Method (IBM), originally proposed by Peskin (1972), embeds the dynamic influence of Lagrangian flexible structures into Eulerian fluid grids via force terms, thus eliminating mesh distortion—a common issue in traditional moving mesh methods when handling large structural deformations (\cite{salehi2023semi}). The IBM has been successfully applied to various complex boundary problems in fluid dynamics, such as particulate suspensions (\cite{uhlmann2005immersed}), fluid–structure interactions in wind turbines (\cite{stival2022wake}), heart valve dynamics (\cite{borazjani2013fluid}), and fish swimming (\cite{cui2020sharp}). The Lattice Boltzmann Method (LBM), grounded in mesoscopic particle kinetics, solves the Navier–Stokes(N-S) equations through the evolution of discrete velocity distribution functions. Its explicit time-stepping scheme and localized data dependencies make it highly suitable for large-scale parallel computation on GPU clusters (\cite{guo2002discrete, obrecht2013multi}). The IB-LB coupling algorithm integrates the flexible boundary-handling of IBM with the computational efficiency of LBM (\cite{feng2004immersed, wu2009implicit}). Specifically, LBM reconstructs flow field evolution via discrete velocity models, while IBM captures structural deformation through Lagrangian markers—enabling bidirectional momentum exchange at the fluid-solid interface. The primary governing equations are as follows:

\begin{equation}\label{equ_1}
	\begin{array}{l}f_{\alpha}^d(\mathbf{x}+\mathbf{e_{\alpha}}\Delta t,t+\Delta t)-f_{\alpha}^d(\mathbf{x},t)\\ =-M^{-1}\hat{S}[m_{\alpha}(\mathbf{x},t)-m_{\alpha}^{eq}(\mathbf{x},t)]+\Delta t F_{\alpha}(\mathbf{x},t)\end{array}
\end{equation}
In equation \ref{equ_1}, the term with superscript $d$ represents the ideal reference value and strictly satisfies the non-slip boundary condition, \emph{t} represents the current moment of the transient model, $\Delta t$ is the calculation time step, ${f_{\alpha}}^d(\mathbf{x},t):\alpha = 0,1,...,8$ denotes the particle equilibrium distribution function that strictly satisfies the non-slip boundary condition at the position \textbf{x} and moment \emph{t}, \emph{M} is the collision matrix, $\hat{S}$ is the diagonal matrix corresponding to the relaxation factor in the MRT model, and ${m_{\alpha}}(\mathbf{x},t):\alpha = 0,1,...,8$ is the equilibrium distribution function of the flow field lattice point at the position \textbf{x} of moment \emph{t}. The relationship between ${m_{\alpha}}(\mathbf{x},t)$ and ${f_{\alpha}}^d(\mathbf{x},t)$ can be expressed as ${m_{\alpha}}(\mathbf{x},t) = M{f_{\alpha}}^d(\mathbf{x},t)$. ${m_{\alpha}}^{eq}(\mathbf{x},t)$ is the ${m_{\alpha}}(\mathbf{x},t)$ equilibrium state in the moment space and $F_{\alpha}(\mathbf{x},t):\alpha = 0,1,...,8$ is the external force on the fluid point at position \textbf{x} and moment \emph{t}.

For the fluid-interface interaction process in the immersed boundary method:

\begin{equation}\label{equ_2}
	{\bf{U}}(s,t) = \int_{{\Omega _{\rm{f}}}} {\bf{u}} (\mathbf{x},t)\delta (\mathbf{x} - {\bf{X}}(s,t)){\rm{d}}{\bf{x}}
\end{equation}	

\begin{equation}\label{equ_3}
	{\bf{f}}(\mathbf{x},t) = \int_{{\Gamma _{\rm{b}}}} {\bf{F}} (s,t)\delta (\mathbf{x} - {\bf{X}}(s,t)){\rm{d}}s
\end{equation}
where lowercase letters denote the Eulerian variables defined on fluid region $\Omega_{\rm{f}}$, uppercase letters denote the Lagrangian variables defined on immersed boundary $\Gamma_{b}$, and ${\bf{x}}$, ${\bf{f}}$, and ${\bf{u}}$ are the position vector in Cartesian coordinates, external force term in the N-S equation, and velocity of the Eulerian variables, respectively. ${\bf{X}}$, ${\bf{F}}$, and ${\bf{U}}$ are the position of the boundary in Lagrangian coordinates, force of the immersed boundary on the fluid, and velocity of the interface, respectively. $\delta$ is the Dirac delta function, which is the key function for the implementation of fluid-structure interaction. It can interpolate the fluid velocity ${\bf{u}}$ in equation \ref{equ_2} to obtain the velocity of the immersed boundary ${\bf{U}}$ or diffuse the force ${\bf{F}}$ of the immersed boundary to the surrounding fluid nodes in equation \ref{equ_3} to obtain ${\bf{f}}$. To enable GPU parallel computing for LBM, we designed a parallel program based on the NVIDIA CUDA framework. Here, each thread handles calculations at a single grid point, using a one-dimensional block structure with a two-dimensional thread grid. To enhance performance and reduce latency during kernel execution, data was efficiently exchanged between shared memory and global memory.

\subsection{Multibody and muscle dynamics}

We formulate the multibody dynamics equations using the Newton–Euler method (\cite{todorov2012mujoco, da2024multibody}):

\begin{equation}\label{equ_4}
	{\bf{M}}\left( {\bf{q}} \right)\mathop {\bf{q}}\limits^{..}  = \left( {{\bf{b}}\left( {{\bf{q}},{\bf{v}}} \right) + \boldsymbol{\tau}} \right){\rm{d}}t + {J}_E^ \top \left( {\bf{q}} \right){{\bf{f}}_E}\left( {{\bf{q}},{\bf{v}},{\boldsymbol{\tau}}} \right)
\end{equation}
where, $\mathbf{q}$ denotes the generalized position coordinates, representing the positions and orientations of all rigid bodies. $\mathbf{v}$ refers to the generalized velocity, with the relationship $\mathbf{v} = \dot{\mathbf{q}}$. The term $\mathbf{b}(\mathbf{q}, \mathbf{v})$ represents the internal force components, which include Coriolis forces, centrifugal forces, elastic forces, and other inertial effects. $\boldsymbol{\tau}$ denotes the external force components, primarily comprising hydrodynamic forces and active muscle-generated forces, $J_E$ is the Jacobian matrix of equality constraints, used to project joint velocities into the constraint coordinate space. $\mathbf{f}_E$ represents the generalized forces (or impulses) arising from the equality constraints, which are introduced to rigorously enforce specific kinematic conditions.

A.V. Hill proposed the Hill-type muscle model in 1938 to describe the mechanical properties of skeletal muscle (\cite{hill1938heat}). Its core concept is to represent muscle force generation as the combined effect of a contractile element (active component) and an elastic element (passive component), making it widely adopted in biomechanical simulations, biomimetic muscle modeling, and actuator development to replicate the dynamic behaviour of muscle–tendon units (\cite{delp2007opensim, millard2013flexing, zhang2017modeling}).

The Hill-type model consists of two primary components:

(1) Contractile Element (CE): Represents the force–length–velocity relationship of active muscle contraction. It captures the dynamic behaviour driven by muscle activation and corresponds to the relative sliding of myosin and actin filaments. The generated tension is proportional to the number of cross-bridges formed between these filaments. (2) Parallel Elastic Element (PEE): Describes the nonlinear passive elastic force generated during muscle stretching, capturing the elasticity of muscle tissue that acts in parallel with the contractile component. In some studies, an additional component called the Series Elastic Element (SEE) is introduced to model muscle elasticity in series with the CE. However, since the energy stored within cross-bridges is significantly smaller than the total energy stored in tendinous structures both inside and outside the muscle (\cite{zajac1989muscle}), the SEE was omitted in our work. Thus, the total muscle output force is computed as:

\begin{equation}\label{equ_5}
	{{\rm{f}}_{{\rm{muscle}}}} = {f_0} \cdot \left[ {{{\bf{f}}_L}({L_{muscle}}) \cdot {{\bf{f}}_V}({V_{muscle}}) \cdot \alpha  + {{\bf{f}}_P}({L_{muscle}})} \right]
\end{equation}
Here, ${f_0}$ represents the peak active force generated by the muscle at its optimal resting length $L_{0}$ , which allows for explicit specification of muscles with varying strength capacities. ${{{\bf{f}}_L}({L_{muscle}})}$ denotes the active force–length relationship, $ {{{\bf{f}}_V}({V_{muscle}})}$ represents the force–velocity relationship, and ${{{\bf{f}}_P}({L_{muscle}})}$ corresponds to the passive force–length relationship of the muscle. The term $\alpha$ indicates the activation level, which is generated from the control input using a first-order nonlinear filter. This parameter captures the dynamics of muscle activation (\cite{millard2013flexing}). See equation \ref{equ_6}:

\begin{equation}\label{equ_6}
	\frac{d}{{dt}}\alpha  = \frac{{{\rm{u_{n}}} - \alpha }}{{\sigma (\alpha )}}
\end{equation}
Here, $\rm{u_{n}}$ is neural excitation; the time constant $\sigma$ is divided into two phases: activation and deactivation. The computation of  $\sigma$ is given in equation \ref{equ_7}:

\begin{equation}\label{equ_7}
	\sigma(\alpha) =
	\begin{cases}
		\sigma_{\text{act}} \cdot (0.5 + 1.5 \cdot \alpha) & \text{if } \rm{u_{n}} - \alpha > 0 \\
		\sigma_{\text{deact}} / (0.5 + 1.5 \cdot \alpha) & \text{if } \rm{u_{n}} - \alpha \leq 0
	\end{cases}
\end{equation}
where $\sigma_{\text{act}}$ and $\sigma_{\text{deact}}$ are, respectively, the activation and deactivation time constants, which were set to 10 $ms$ and 40 $ms$ in this study.

\subsection{Deep reinforcement learning}

Deep  Reinforcement  Learning  (DRL)  is central to the embodiment-based paradigm of this study. It provides a powerful computational framework to address a core challenge of embodied intelligence: how an agent, grounded  in physical  reality (or high-fidelity simulations), learns to perceive, act, and interact with its environment to achieve goal-directed tasks. By integrating deep learning with reinforcement learning, DRL has emerged as a dominant paradigm for solving complex decision-making problems. DRL operates within the theoretical framework of a Markov Decision Process (MDP), defined by the tuple $(S, A, P, R, \gamma)$, where $S$ is the state space, $A$ is the action space, $P$ represents the state transition probability, $R$ is the reward function, and $\gamma \in [0,1]$ is the discount factor that balances immediate and future rewards. In this study, we adopt the Soft Actor-Critic (SAC) algorithm—a stochastic policy-based DRL method—to achieve spatiotemporal muscle activation pattern control at the neural system level (\cite{haarnoja2018soft}). SAC is an off-policy algorithm that incorporates the maximum entropy principle to balance exploration and exploitation by encouraging stochastic behaviour. Its core components include: a stochastic policy network (actor), dual Q-value networks, dual target Q-value networks, a state value network, and an automatic entropy tuning mechanism. The core optimization objective augments the standard reward signal with a policy entropy term $\zeta \mathcal{H}(\pi(\cdot | s_t))$, encouraging exploration and enhancing the robustness of the learned control policies:

\begin{equation}\label{equ_8}
	J(\pi) = \mathbb{E} \left[ \sum_{t=0}^{\infty} \gamma^t \left( r(s_t, a_t) + \zeta \mathcal{H}(\pi(\cdot | s_t)) \right) \right]
\end{equation}
Here, $\gamma$ is the discount factor, $\mathcal{H}(\pi(\cdot|s)) = -\mathbb{E}_{a\sim\pi}[\log\pi(a|s)]$ denotes the policy entropy, and $\zeta$ is an adaptive temperature coefficient.

A double Q-network architecture is employed to mitigate overestimation bias in value function learning. The Q-loss function is defined as:

\begin{equation}\label{equ_9}
	\mathcal{L}_Q(\theta_i) = \mathbb{E}_{(s, a, r, s') \sim \mathcal{D}} \left[ \left( Q_{\theta_i}(s, a) - \hat{Q}(s, a) \right)^2 \right] \quad i = 1, 2
\end{equation}
Here, $\mathcal{D}$ denotes the experience replay buffer.

The target Q-value is computed as follows:

\begin{equation}\label{equ_10}
	\hat{Q}(s, a) = r(s, a) + \gamma \mathbb{E}_{s' \sim P} \left[ \min_{j=1,2} Q_{\theta_j}(s', \tilde{a}') - \zeta \log \pi_{\psi}(\tilde{a}' \mid s') \right]
\end{equation}
where $\tilde{a}' \sim \pi_\psi(\cdot \mid s')$ denotes an action sampled from the policy $\pi_\psi$ given the next state $s'$.

The target network parameters $\theta^-$ are updated via Polyak averaging (\cite{lillicrap2015continuous}), which helps maintain stable target estimates over time:

\begin{equation}\label{equ_11}
\theta _i^ -  \leftarrow \chi {\theta _i} + (1 - \chi )\theta _i^ - \quad (\chi  = 0.01)
\end{equation}

The policy network is trained by minimizing the  Kullback–Leibler (KL) divergence, encouraging the learned policy to align with the soft optimal distribution: 

\begin{equation}\label{equ_12}
	\mathcal{L}_{\pi}(\psi)=\mathbb{E}_{s\sim \mathcal{D}}\left[\mathbb{E}_{a\sim\pi_{\psi}}\left[\zeta\log\pi_{\psi}(a|s)-\min_{j=1,2}Q_{\theta_{j}}(s,a)\right]\right]
\end{equation}

Additionally, the entropy temperature coefficient is adaptively tuned by minimizing a loss function that encourages the policy entropy to approach a target value $\mathcal{H}_0$, typically set to the negative dimensionality of the action space $\mathcal{A}$, where $\mathcal{H}_0 = -\dim(\mathcal{A})$:

\begin{equation}\label{equ_13}
	\mathcal{L}(\zeta) = \mathbb{E}_{s \sim \mathcal{D}} \left[ -\zeta \log \pi_{\psi}(a \mid s) - \zeta \mathcal{H}_0 \right]
\end{equation}

\section{Constructing a multiscale digital trout model}\label{sec:trout}

\subsection{Fish body modeling}

\begin{figure}[htbp]
	\centering
	\includegraphics[width=4in]{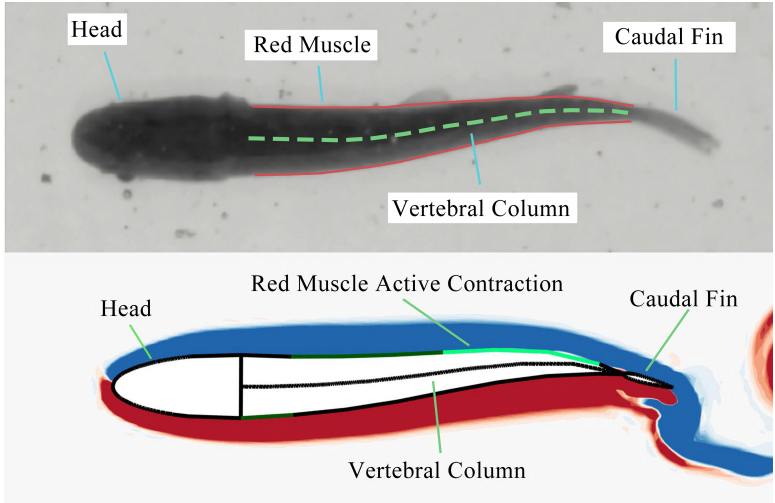}
	\caption{Schematic comparison between a live trout and the digital trout model.}
	\label{pic/f1}
\end{figure}

\begin{figure}[htbp]
	\centering
	\includegraphics[width=6in]{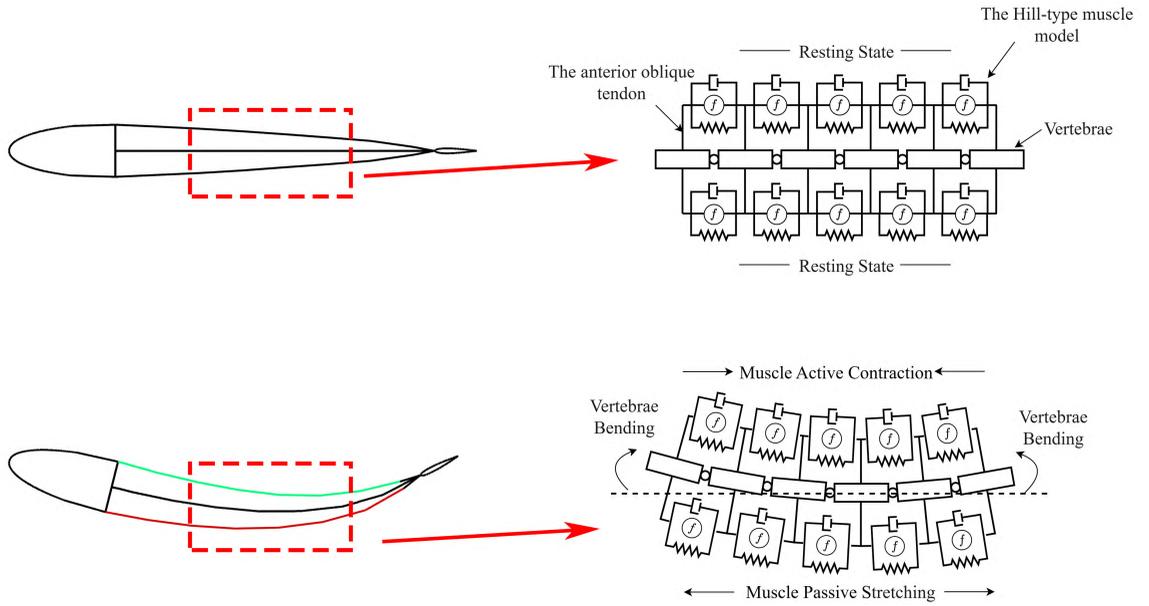}
	\caption{Anatomical schematic of the digital trout (the anterior 0.25 BL represents the rigid head region).}
	\label{pic/f2}
\end{figure}

Trout, members of the family Salmonidae, are closely related to salmon. Among them, the rainbow trout (Oncorhynchus mykiss) is widely regarded as a representative species for studying trout kinematics and biology, owing to its considerable economic and scientific importance (\cite{contreras1998effects, liao2003fish, drucker2005locomotor, akanyeti2013kinematic}). Rainbow trout typically range in length from 2.0 to 40.0 cm, with adult individuals commonly measuring between 20.0 and 40.0 cm. They have approximately 60 vertebrae (\cite{froese2010fishbase}) and primarily employ sub-carangiform swimming. In this mode of locomotion, energy is transmitted along the body through undulatory waves of muscle contraction, culminating in significant thrust generation via the tail fin. To construct a trout model in a digital environment, we first referenced the morphological data from \cite{di2021convergence} to build a 2D body shape. The vertebral column was modeled as 11 interconnected segments, with the anterior 25\% defined as a rigid head section. The remaining posterior portion was actively involved in the body–caudal fin (BCF) propulsion process. It is important to note that a live rainbow trout has approximately 60 vertebrae. However, in our study, we found that modeling the vertebral column with about 10 links is sufficient to efficiently and accurately capture the key physical dynamics of interest. Therefore, we simplified the number of vertebrae in the model. The resulting model had a total of 13 degrees of freedom and included 7 pairs of myomeres (14 actuators in total) symmetrically distributed along both sides of the body. Figure \ref{pic/f1} compares a live trout with the digital model, while figure \ref{pic/f2} shows the anatomical schematic of the digital trout.

\begin{figure}[htbp]
	\centering
	\includegraphics[width=5in]{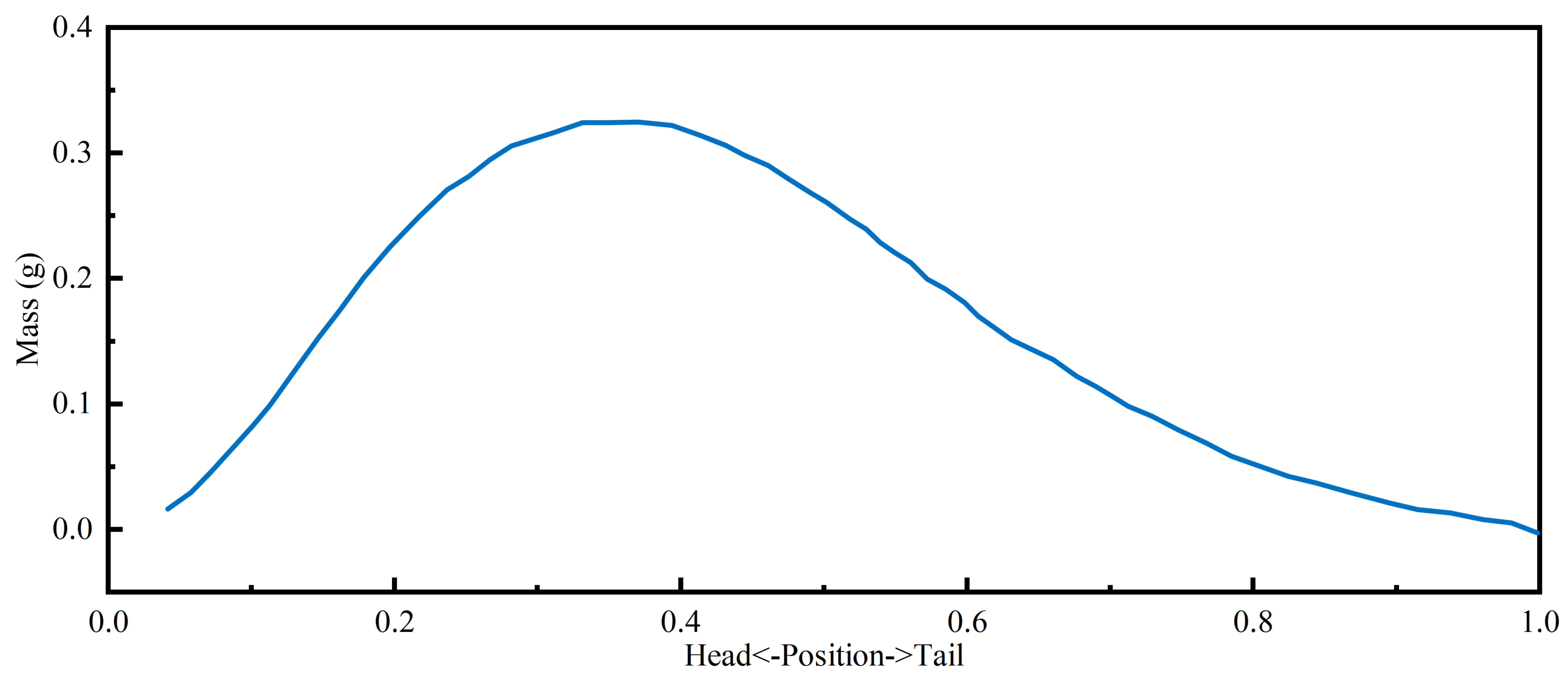}
	\caption{Axial mass distribution of the digital trout.}
	\label{pic/f3}
\end{figure}

Then, it is necessary to specify the physical property parameters of the digital trout body. Regarding the elastic modulus $E_{m}$ of the spinal and muscular regions, the elastic modulus of the muscle region $E_{muscle}$ was set to $10^5pa$(\cite{cheng1998continuous}), while the elastic modulus of the spinal region $E_{spinal}$ was set to $2{\rm{ \times }}{10^5}Pa$ (\cite{mchenry1995mechanical}). The equivalent passive tensile and compressive stiffness of the muscle region was calculated according to formulation ${k} = EA$, where $A$ denotes the average effective cross-sectional area of the corresponding segment. The equivalent bending stiffness at the vertebral (spinal) joints was calculated using formulation $\kappa  = EI$, where $I$ denotes the second moment of area (also known as the area moment of inertia) of the vertebral cross-section. For estimating the mass of the digital trout, the classical length–weight relationship for fish is used as follows (\cite{le1951length}):

\begin{equation}\label{equ_14}
	M = m{L^n}
\end{equation}
Here, $m$ denotes the condition factor, which reflects the fullness and nutritional status of the fish body. $n$ represents the allometric growth parameter, describing the scaling relationship between body weight and body length. For rainbow trout, based on data from FishBase—the world’s largest open-access database on fish species (\cite{froese2010fishbase})—the parameter $m = 0.0110, n = 2.997$ was adopted. This value originates from Abashiri River basin, eastern Hokkaido / 2007-2011, with a sample size of N=3410. Using this parameter, the estimated body weight of a rainbow trout with a length of 0.1 m was approximately 10 grams. For the distribution of body mass along the axial direction, a spindle-shaped mass distribution model is employed (\cite{guo2007fish}), as illustrated in figure \ref{pic/f3}.

\subsection{Computational parameters and simulation conditions}
\begin{figure}[htbp]
	\centering
	\includegraphics[width=4in]{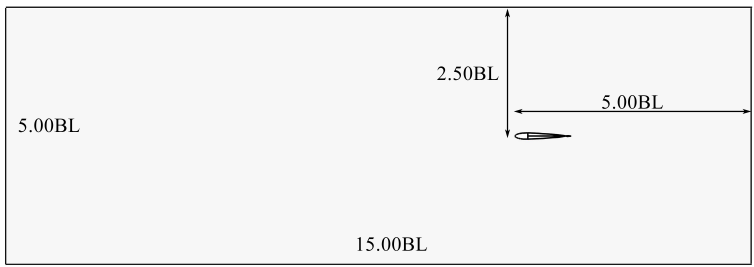}
	\caption{Schematic diagram of the computational domain.}
	\label{pic/f4}
\end{figure}

To standardize results, all length-related quantities were normalized by body length (BL). Table \ref{tab1} summarizes the physical dimensions and lattice resolution of the computational domain, with dimensional conversion for LBM following \cite{baakeem2021novel} (2021) (this scheme enhances parameter flexibility and solver stability). The computational domain was discretized using a $15.00 BL \times 5.00 BL$L Cartesian grid (750,000 mesh points; see figure \ref{pic/f4} for schematic). Boundary conditions were defined as follows: the top, bottom, and left boundaries of the flow field were set as solid walls, while the right boundary was implemented as a zero-gradient outflow boundary. When the high-speed jet induced by fish undulation reached the right boundary, the outflow condition allowed the jet to exit the domain effectively. All solid walls were subject to non-slip boundary conditions. Each simulation episode had a maximum time duration of 10 seconds. A total of 10,000 episodes were simulated. An episode ended when the digital trout either reached the left boundary or the 10-second time limit was reached.

\begin{table}
	\begin{center}
		\def~{\hphantom{0}}
		\begin{tabular}{ccc}
			  & real world   &  lattice unit  \\[3pt]
			fish body length   & 0.1 $m$ & 100 $lu$ \\
			characteristic time   & 1.0 $s$ & 1000$ts$ \\
			characteristic velocity  & 0.1 $m/s$ & 0.1 $lu/ts$ \\
			tail beat cycle   & 0.5 $s$ & 500 $ts$ \\
			characteristic mass & 0.01 $kg$ & 10000 $mu$ \\
		\end{tabular}
		\caption{Physical dimensions and lattice resolution of the computational domain.}
		\label{tab1}
	\end{center}
\end{table}

\subsection{DRL module configuration}

During fish locomotion, the neural activation sequence and activation duration of myomeres have a significant influence on macroscopic swimming performance, including propulsion speed and energy consumption. In this study, we enable the digital trout to autonomously control the independent contraction and relaxation of each myomere using a deep reinforcement learning algorithm. This allows the agent to achieve swimming objectives with varying demands on energy efficiency and propulsion speed.

\subsubsection{Hyperparameters setting}

For the hyperparameters of the reinforcement learning module, the temperature coefficient $\zeta$ controlling the stochasticity of the SAC policy was computed adaptively (\cite{haarnoja2018soft}). The learning rate $\eta$ for the neural networks was set to 0.0005, the discount factor $\gamma$ was set to 0.99, and each training iteration used a batch size of 512. The target networks were updated softly with a coefficient $\chi=0.01$. The simulation time for each episode was set to 10.0 seconds (approximately 10,000 lattice timesteps), with a total of 10,000 episodes conducted during training.

\subsubsection{State space}

The state space is defined as the set of all possible environmental states and serves as the perceptual input for the agent, providing a structured representation of the environment. In this study, the DRL state space was designed with a total of 36 dimensions. To account for the temporal dependencies inherent in decision-making, the states from two consecutive time steps were stacked together, as shown in equation \ref{equ_15}. Specifically:

\begin{equation}\label{equ_15}
	\resizebox{\textwidth}{!}{$
	\left[ {\begin{array}{*{20}{c}}
			{{{[\begin{array}{*{20}{c}}
							{{\phi _1}}&{{\phi _2}}&{{\phi _3}}&{{\phi _4}}&{{\phi _5}}&{{\phi _6}}&{{\phi _7}}&{{d_1}}&{{d_2}}&{{d_3}}&{{d_4}}&{{d_5}}&{{d_6}}&{{d_7}}&{{W_{active}}}&{{U_{mean}}}&{{\alpha _t}}&{{\alpha _{t - 1}}}
						\end{array}]}_t}}\\
			{{{[\begin{array}{*{20}{c}}
							{{\phi _1}}&{{\phi _2}}&{{\phi _3}}&{{\phi _4}}&{{\phi _5}}&{{\phi _6}}&{{\phi _7}}&{{d_1}}&{{d_2}}&{{d_3}}&{{d_4}}&{{d_5}}&{{d_6}}&{{d_7}}&{{W_{active}}}&{{U_{mean}}}&{{\alpha _{t - 1}}}&{{\alpha _{t - 2}}}
						\end{array}]}_{t - 1}}}
	\end{array}} \right]
$}
\end{equation}
where ${\phi _i}$ represents the activation phase of the i-th myomere (i.e., the onset timing of muscle activation, expressed as a percentage of half the tail beat cycle), ${d_i}$ denotes the activation duration of the $i$-th myomere, ${W_{active}}$ is the mechanical work performed by active muscle contraction during the current episode, ${U_{mean}}$ refers to the average propulsion speed achieved during the episode, where ${U_{{\rm{mea}}n}} = D/{T_{episode}}$, $D$ is the total forward distance traveled in the current episode, ${\alpha _{t}}$ denotes the action taken in the current time step, ${\alpha _{t-1}}$ represents the action taken in the previous time step.

\subsubsection{Action space}

During steady swimming, the muscles on either side of the body contract and relax in an alternating pattern, generating coordinated wave-like motions that propagate toward the tail. This pattern is somewhat analogous to artificially programmed "traveling wave" control. In the field of fish locomotor biology, the sequential activation and relaxation of myomeres are commonly described using the concepts of the activation wave and deactivation wave. The activation wave refers to the initiation of sequential muscle contractions, triggered by neural signals that induce calcium ion release, leading to muscle fiber contraction. Under steady swimming conditions, the activation wave travels from the anterior part of the body (near the head) toward the posterior (toward the tail), generating a traveling curvature wave that drives propulsion. The propagation speed of the activation wave along the body axis is denoted as $V_{act}$. In contrast, the deactivation wave represents the sequential return of muscles from a contracted to a relaxed state. This process is triggered by inhibitory neural signals or calcium ion reuptake, leading to muscle relaxation. Its axial propagation speed is denoted as $V_{deact}$. The time required for the activation and deactivation waves to travel from the rostral myomere (first) to the caudal myomere (last) is represented by ${t_{act}}$ and ${t_{deact}}$, respectively. Because muscle contraction requires time to accumulate intracellular calcium concentrations ($Ca^{2+}$ transients), whereas muscle relaxation is primarily driven by calcium reuptake and the passive elastic recoil of structural proteins such as titin, the passive elements dominate the rapid deactivation process. Consequently, in carangiform and subcarangiform swimmers, the propagation speed of the activation wave is generally slower than that of the deactivation wave (\cite{wardle1995tuning, gillis1998neuromuscular}).

Therefore, a constraint must be imposed in the design of the DRL action space:
(1) Boundary condition: $d_i \le 0.5TBC$, $0 < {t_{act}} \le 0.5TBC$, $0 < {t_{deact}} \le \min (0.5TBC,{t_{act}})$, where TBC is the fish’s tail beat cycle;
(2) ${V_{act}} < {V_{deact}}$.

\begin{figure}[htbp]
	\centering
	\includegraphics[width=5in]{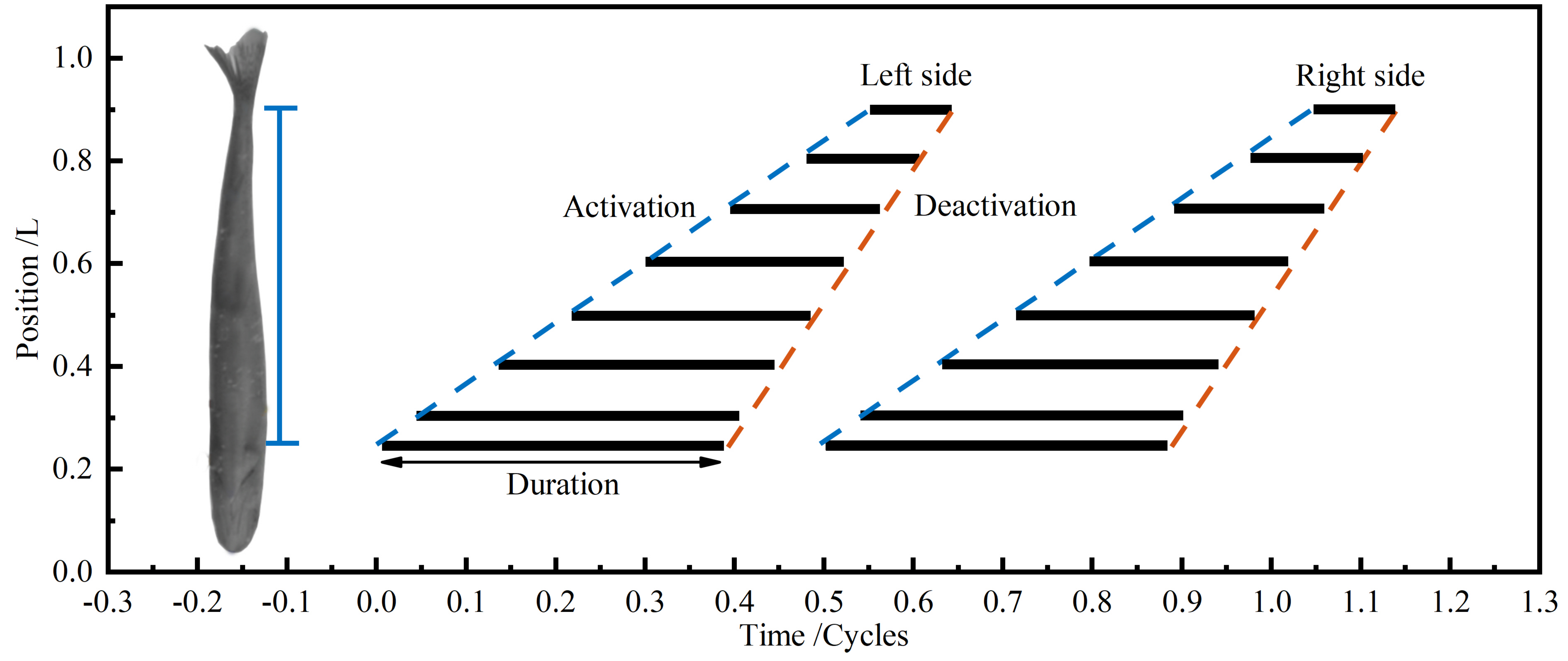}
	\caption{Neural activation patterns of bilateral muscles during steady swimming in trout. The start of each black bar indicates the muscle activation phase, and its length represents the contraction duration time (CDT). EMG data are adapted from (\cite{williams1989locomotion}); the trout image on the left is adapted from (\cite{gibbs2024kinematics}).}
	\label{pic/f5}
\end{figure}

Figure \ref{pic/f5} illustrates the neural rhythm map of bilateral muscle activity in trout during steady swimming. The underlying EMG data is sourced from Williams et al.(\cite{williams1989locomotion}). It is important to note that we made certain modifications to the original neural activation diagram presented by Williams et al.(\cite{williams1989locomotion}). In our model, we did not account for the nonlinear propagation effects of the activation wave. Instead, we assumed that both the activation and deactivation neural signals propagate along the body at constant speeds, resulting in uniform contraction and relaxation of muscles along the same side of the body. As shown in the figure \ref{pic/f5}, the activation wave speed is clearly slower than the deactivation wave speed. This is reflected in the slopes of the dashed lines: the blue dashed line (activation) has a shallower slope than the orange dashed line (deactivation). Therefore, by adjusting just three parameters ${t_{act}}$, ${t_{deact}}$ and the contraction duration time (CDT) at the initial muscle segment—we can fully define the spatiotemporal neural rhythm map for unilateral muscle activation. This neural pattern drives vertebral motion, ultimately generating body undulations at the macroscopic scale. The subsequent design of the action space in our DRL framework will be guided by these principles.

The action space was defined as a discrete action space consisting of 27 dimensions. The agent was allowed to simultaneously adjust three key parameters: activation wave speed $V_{act}$, deactivation wave speed $V_{deact}$, and CDT. The activation and deactivation wave speeds were controlled by varying parameters ${t_{act}}$ and ${t_{deact}}$, respectively. Smaller values of ${t_{act}}$ and ${t_{deact}}$ resulted in faster propagation of the activation and deactivation waves, whereas larger values led to slower propagation. The structure of the action space is illustrated below, where each action is assigned an action ID, denoted as $a \in \{ 0,1, \ldots ,26\} $, all temporal units within the action space are represented using discrete lattice time units (ts):

\begin{equation}\label{equ_16}
		\Delta t(a) = \left\{ 
		\begin{array}{l}
			\Delta t_{\text{act}}(a) = 10 \cdot \sigma_{\text{act}} \left\lceil \dfrac{a}{9} \right\rceil \\[6pt]
			\Delta t_{\text{deact}}(a) = 10 \cdot \sigma_{\text{offset}} \left\lceil \dfrac{a \bmod 9}{3} \right\rceil \\[6pt]
			\Delta CDT(a) = 10 \cdot \sigma_{\text{offset}} (a \bmod 3)
		\end{array} 
		\right.	
\end{equation}

${\sigma _{{\rm{offset}}}}$ is the direction selection function, computed as follows:

\begin{equation}\label{equ_17}
{\sigma _{{\rm{ac}}t}}(x) = \left\{ {\begin{array}{*{20}{l}}
		1&{{\rm{if }}x = 0}\\
		{ - 1}&{{\rm{if }}x = 1}\\
		0&{{\rm{if }}x = 2}
\end{array}} \right.
\end{equation}

${\sigma _{{\rm{offset}}}}$ is the offset generation function, defined as:

\begin{equation}\label{equ_18}
	{\sigma _{{\rm{offset}}}}(y) = \left\{ {\begin{array}{*{20}{l}}
			0&{{\rm{if }}y = 0}\\
			1&{{\rm{if }}y = 1}\\
			{ - 1}&{{\rm{if }}y = 2}
	\end{array}} \right.
\end{equation}

Using the above equations, the agent could determine the corresponding changes in the three control dimensions based on the selected action ID. These computed values were then used to modify the activation phase and contraction duration of each myomere on the same side of the body.

\subsubsection{Reward function}

In studies of steady swimming, assuming a fixed tail-beat cycle, the primary behavioural differences are governed by two key neuromechanical parameters: the muscle activation phase—defined as the ratio of activation onset timing to half the tail-beat period—and the activation duration time, which refers to the time interval between activation onset and offset. Different combinations of these parameters significantly influence macroscopic swimming performance, specifically: (1) Swimming energy consumption, defined as the total mechanical work performed by active muscle contractions during a single swimming episode, ${W_{active}} = \int\limits_0^t {\left| p \right|dt} $, where $p$ denotes the instantaneous power at time $t$, $p = {\rm{f}_{active}} \cdot {V_{muscle}}$; (2) Average swimming speed $U_{mean}$, representing the net forward locomotion achieved per unit time. These performance metrics reflect a classic Pareto trade-off: achieving higher swimming speeds typically requires increased energy consumption, whereas minimizing energy use too aggressively can lead to a significant reduction in speed (\cite{weihs1973optimal}). The latter is particularly disadvantageous in ecological contexts where fish must evade predators or perform other energetically demanding activities. The reward function serves as a numerical and functional expression of the digital trout's behavioural goals, acting as a crucial bridge in the evolutionary process of optimizing neuromuscular control. Based on the dual objectives of minimizing energy cost and maximizing forward speed, the DRL reward function is formulated as follows:

\begin{equation}\label{equ_19}
	{r_{reward}} = \left\{ {\begin{array}{*{20}{c}}
			{ - {w_1} \cdot ({U_{mean}} - {W_{active}})}\\
			{w_2} \cdot {{r_{now}} - {r_{pre}}}\\
			{ - {w_3} \cdot 1[{U_{mean}} < 0.3BL/s]{\kern 1pt} {\kern 1pt} {\kern 1pt} {\kern 1pt} {\kern 1pt} {\kern 1pt} {\kern 1pt} }
	\end{array}} \right.
\end{equation}

The three components of the reward function are interpreted as follows: The first term is a sub-reward that balances swimming speed and energy cost. All variables are normalized to prevent differences in scale from distorting the reward landscape; The second term represents a reward gradient, which encourages the agent to explore directions that lead to improved performance while helping to avoid convergence to local optima; The third term is a penalty applied when swimming speed falls below a critical threshold. At such low speeds, the fish would be unable to perform essential behaviours like evading predators, posing a serious threat to survival. This penalty is triggered only under these low-speed conditions. The parameters ${w_1}$, ${w_2}$, and ${w_3}$ are dynamic weighting coefficients that balance the contributions of each reward component. By systematically integrating the defined state space, action space, and reward function, and exploring the solution space using the SAC algorithm, the model aims to progressively approach the Pareto frontier of optimal swimming behaviour.

\subsection{Hardware configuration}\label{subsec:harware}

The simulations in this study were conducted on a high-performance workstation equipped with an NVIDIA GeForce RTX 4070Ti GPU, utilizing the CUDA parallel computing architecture to enable GPU-accelerated computation. The system is powered by an Intel Core i7-13700F CPU, and the RTX 4070 Ti features 12 GB of GDDR6X memory (192-bit interface, 504 GB/s memory bandwidth), 7680 CUDA cores, and 240 fourth-generation Tensor Cores. In addition, the workstation is configured with 32 GB of Kingston DDR4-3600 memory in a dual-channel setup.

\section{Results and discussion}\label{sec:result}

\subsection{Training results}\label{secsec:traing_result}

In this study, the DRL module was trained for 10,000 episodes, with each episode averaging 180 seconds. Computational costs primarily stemmed from the high-fidelity FSI solver (dominating the workload), alongside simulation stepping, policy inference, and network updates. Total training time was $\sim$500 hours (see section\ref{subsec:harware} for hardware details).

\begin{figure}[htbp]
	\centering
	\includegraphics[width=5in]{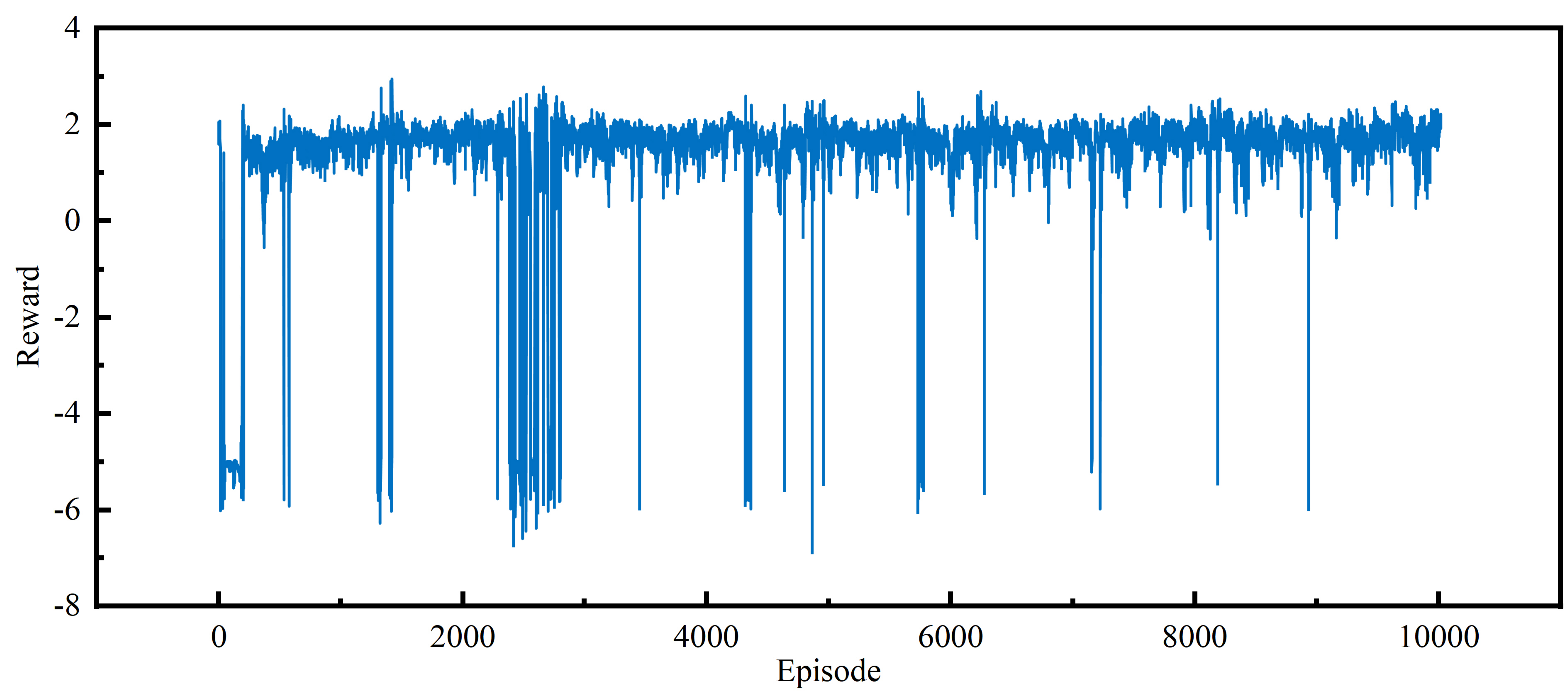}
	\caption{Reward evolution during the training process.}
	\label{pic/f6}
\end{figure}

\begin{figure}[htbp]
	\centering
	\includegraphics[width=5in]{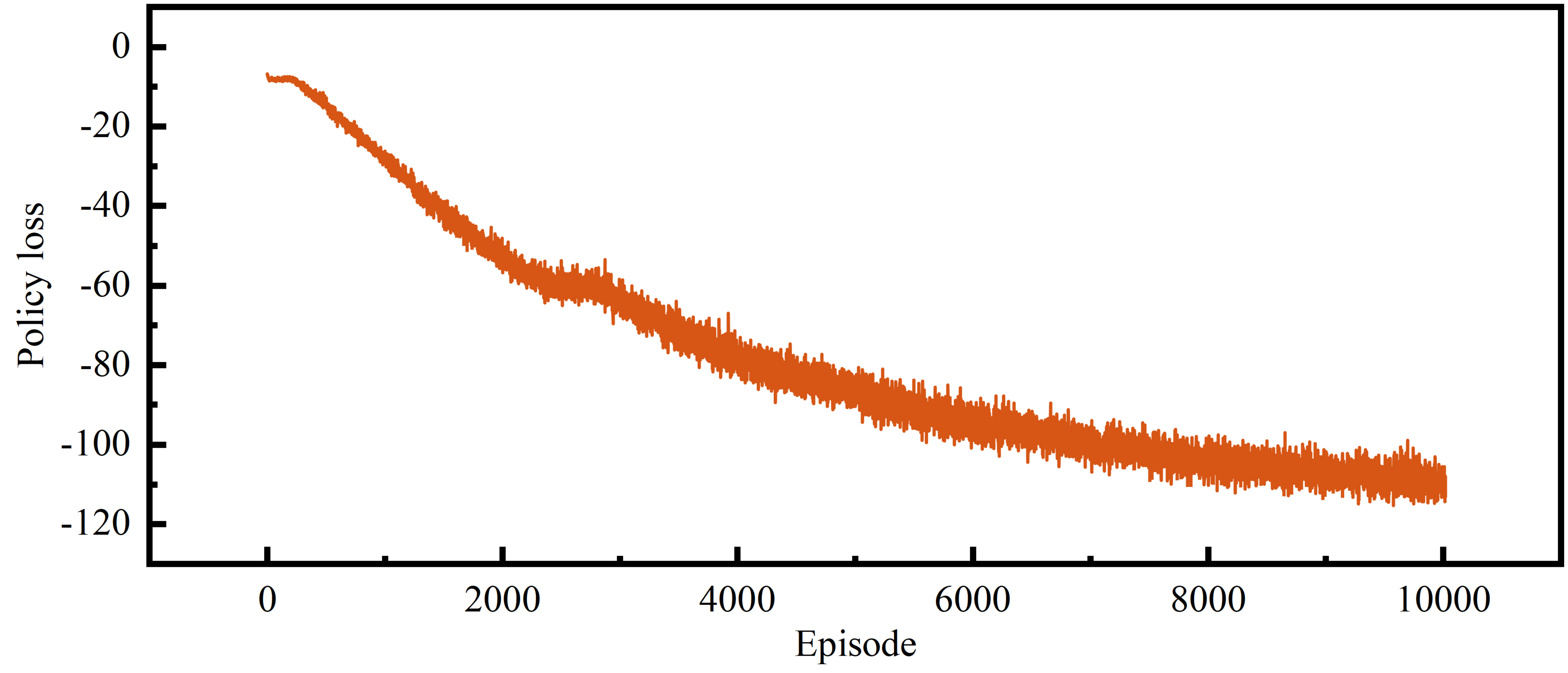}
	\caption{Policy loss evolution during the training process.}
	\label{pic/f7}
\end{figure}

Figures \ref{pic/f6} and \ref{pic/f7} respectively illustrate the evolution of the reward and policy loss during the optimization of the digital trout's swimming behaviour using DRL. The reward curve (figure \ref{pic/f6}) shows non-monotonic fluctuations during multi-objective optimization (maximizing speed while minimizing energy), reflecting the agent’s exploration and exploitation of the solution space. The agent continuously tested spatiotemporal muscle activation patterns to find Pareto-optimal strategies, with occasional negative reward peaks indicating ineffective activation combinations (insufficient thrust). When the agent encountered these adverse conditions, the SAC algorithm—through its stochastic policy formulation and the optimization objective of maximizing expected return and policy entropy—encouraged efficient exploration of alternative regions within the policy space. This mechanism allowed the agent to escape local optima and transition rapidly to higher-reward strategies, thereby preventing training stagnation. The policy loss curve in figure \ref{pic/f7} shows a steadily decreasing and converging trend, suggesting that the agent progressively learns more effective muscle control strategies through policy gradient updates. Overall, for the majority of the training process, the agent successfully balanced the competing objectives of swimming speed and muscular energy consumption. This enables the digital trout to achieve stable and efficient self-propelled locomotion.

\subsection{Modes classification}

\begin{figure}[htbp]
	\centering
	\includegraphics[width=4in]{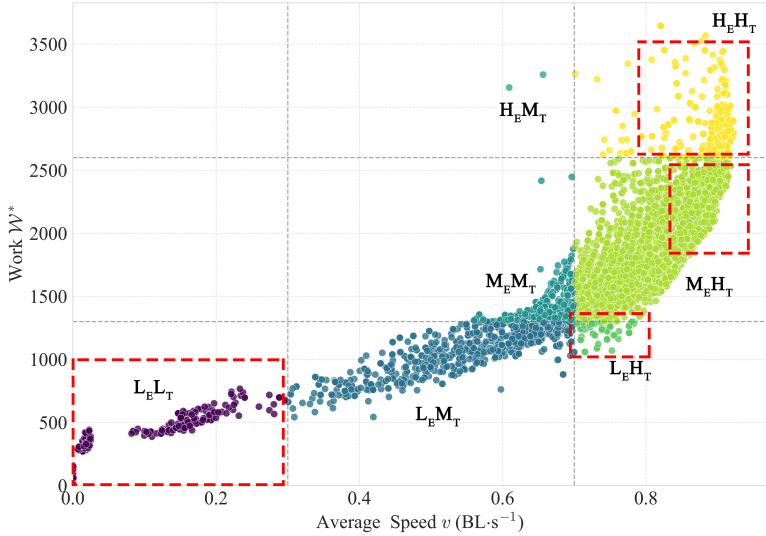}
	\caption{Modes classification of all results ($L$, $M$, and $H$ denote low, medium, and high parameter ranges, respectively; subscripts $E$ and $T$ represent energy consumption and thrust, respectively).}
	\label{pic/f8}
\end{figure}

Figure \ref{pic/f8} shows a scatter plot of results, categorized by muscular work (energy consumption, $E$) and average propulsion speed (thrust performance, $T$). Based on these two metrics, the dataset can be divided into several distinct areas. A large portion of the data falls within the regions labeled $L_{E}L_{T}$, $L_{E}M_{T}$, $M_{E}H_{T}$, and $H_{E}H_{T}$, while fewer samples are located in the $L_{E}H_{T}$, $M_{E}M_{T}$, and $H_{E}M_{T}$ regions. Here, $L$, $M$, and $H$ indicate low, medium, and high ranges of the parameter values, respectively. From these clusters, we selected representative modes from several interesting regions to further investigate their underlying dynamics and swimming characteristics.

\subsection{Mode analysis}

\subsubsection{Muscle behaviour and body wave in representative modes}

During low-speed steady swimming, fish generate a traveling curvature wave along the longitudinal axis, driven by the sequential contraction and relaxation of red muscle from head to tail (\cite{johnston1980endurance}). This wave converts lateral undulations into backward thrust via a high-speed tail jet, generating propulsive force, while minimizing drag and maximizing energy efficiency.

\begin{figure}[htbp]
	\centering
	\includegraphics[width=5in]{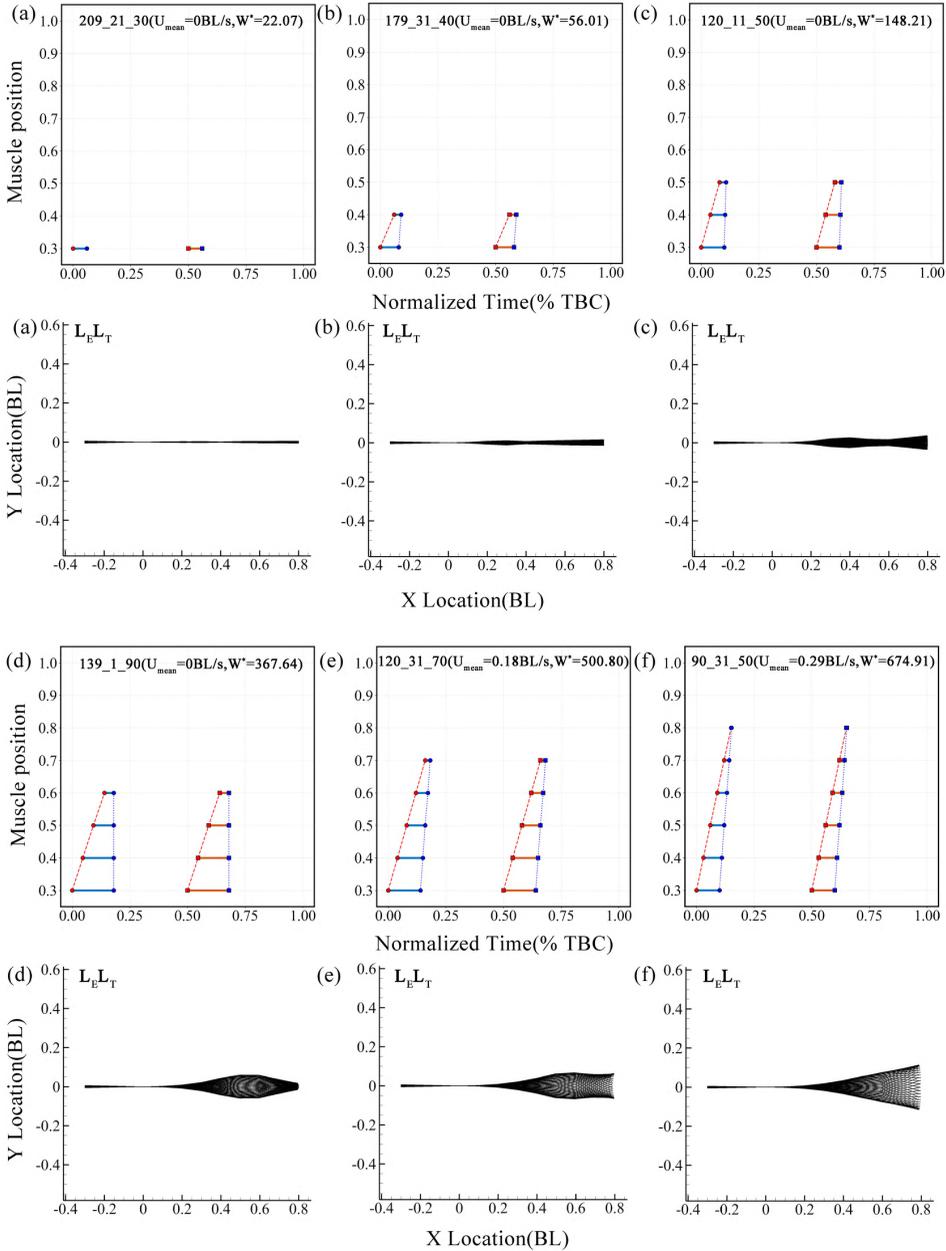}
	\caption{Comparison of typical muscle activation patterns and the envelope of body waves in the $L_{E}L_{T}$ region.}
	\label{pic/f9}
\end{figure}

During training, we identified a set of low-speed swimming modes characterized by both low energy consumption and low swimming speed (located in the $L_{E}L_{T}$ region of figure \ref{pic/f8}). Figures \ref{pic/f9} a-d illustrate the spatiotemporal muscle activation patterns and body wave dynamics under different axial activation intervals (expressed as a proportion of body length, BL). In each subfigure, the upper panel shows the spatiotemporal muscle activation pattern, while the lower panel presents the corresponding body wave pattern; the naming convention for different modes is described as follows: ${t_{{act}}}{\rm{\_}}{t_{{deact}}}{\_CDT}$, all of them are expressed in lattice time units ($t_s$). All modes analyzed in this group failed to generate effective propulsion. When the rostrocaudal muscle activation range was too narrow (e.g., only 0.3 BL in figure \ref{pic/f9}-a), the stimulation was confined to a small region of the body, producing a fragmented curvature wave. As a result, the fish displayed inefficient undulations, resembling the passive elastic oscillations of a damped beam in a viscous fluid (\cite{ramananarivo2013passive}). When the activation range was extended to 0.3–0.4 BL (figure \ref{pic/f9}-b), axial wave transmission began to emerge; however, the resulting waveform appeared irregular and "pinched," resembling a slender hourglass. Due to inefficient energy transfer, the fish was still unable to generate effective thrust. Further extending the activation range to 0.3–0.5 BL (figure \ref{pic/f9}-c) led to increased oscillation amplitude, but the curvature wave still failed to reach the tail. Even at 0.3–0.6 BL (figure \ref{pic/f9}-d), where the anterior wave amplitude nearly reached normal levels, a sharp drop in amplitude around 0.7 BL hindered the formation of a high-speed jet at the tail. These results confirm that an insufficient axial activation range limits the propagation of the curvature wave and disrupts the coherence of the propulsive wave pattern. When the muscle activation range was further extended to 0.7 BL or even 0.8 BL (figures \ref{pic/f9} e–f), the body wave began to generate effective propulsion; however, its intensity was still insufficient to propel the fish beyond the $L_{E}L_{T}$ region. In figure \ref{pic/f9}-e (activation range 0.3–0.7 BL), the traveling wave attenuates after 0.8 BL, leading to a modest increase in average swimming speed to 0.18 BL/s but without noticeable tail acceleration. In contrast, figure \ref{pic/f9}-f (activation range 0.3–0.8 BL) demonstrates full-sequence muscle contraction and a cumulative curvature wave, resulting in effective propulsion and a higher average speed of 0.29 BL/s.

\begin{figure}[htbp]
	\centering
	\includegraphics[width=5in]{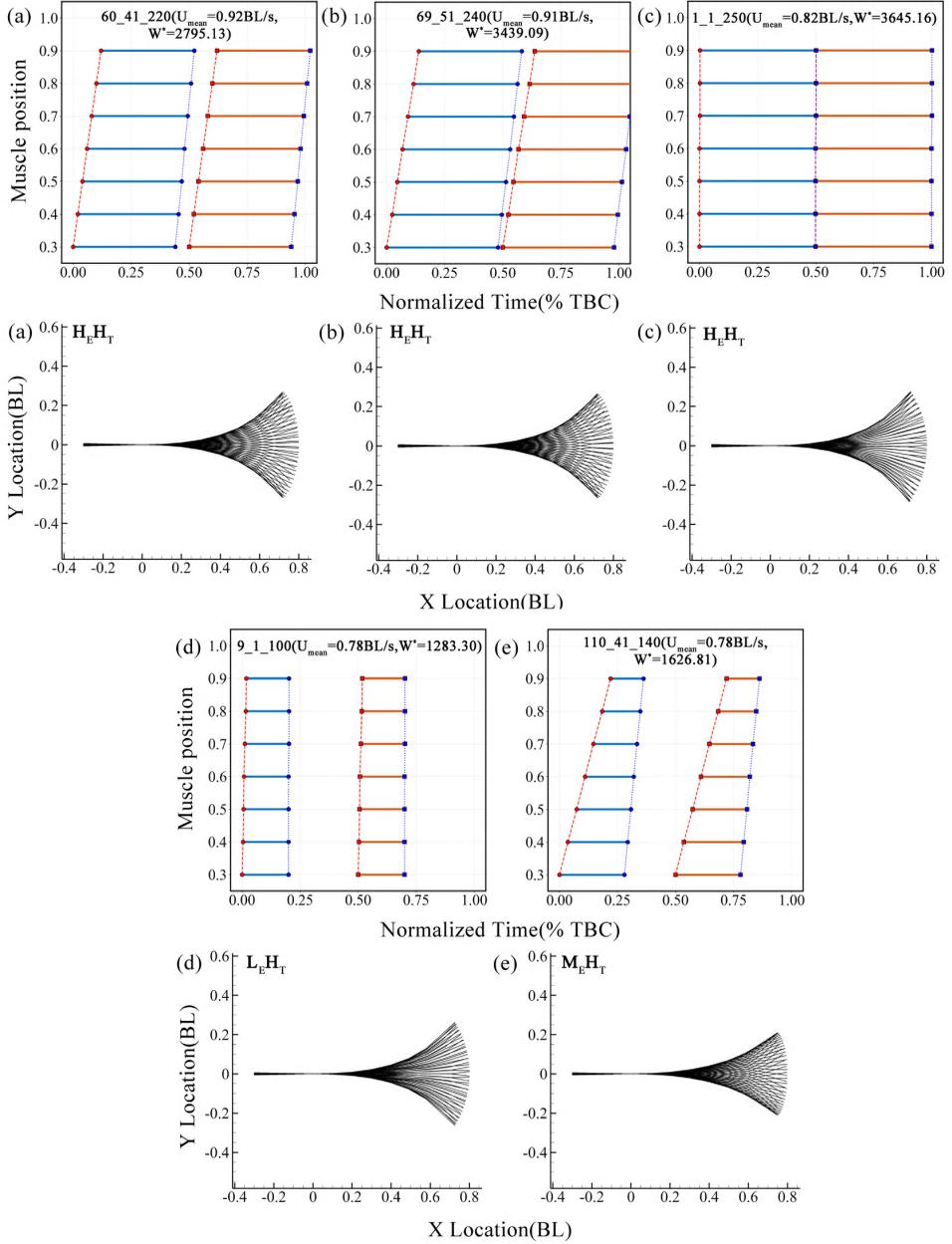}
	\caption{Comparison of typical muscle activation patterns and the envelope of body waves in the $H_{T}$ region.}
	\label{pic/f10}
\end{figure}

On the other hand, we also identified a number of high-speed swimming modes, each exhibiting distinct energy consumption characteristics, as illustrated in the subplots of figure \ref{pic/f10}. The high-energy consumption high-thrust modes ($H_{E}H_{T}$) represent strategies aimed at achieving maximum swimming speed. In these modes, the digital trout significantly extended the muscle activation duration on both sides of the body to enhance energy transmission and generate powerful body waves. As a result, these modes delivered the highest thrust per unit of swimming time. Figures \ref{pic/f10} a-c show the spatiotemporal patterns of muscle activation and the corresponding body wave behaviour for representative $H_{E}H_{T}$ modes. In these cases, the trout achieved an average propulsion speed exceeding 0.8 BL/s, and after the initial acceleration phase, it sustained a steady cruising speed of up to 1.0 BL/s—comparable to that of live trout swimming at 2 Hz (\cite{di2021convergence}). The axial activation range of muscles spanned from 0.3L to 0.9L, and the body wave exhibited a quadratic envelope shape with increasing amplitude from head to tail. Additionally, within the $H_{E}H_{T}$ region, we observed some high-energy consumption but lower-speed modes (figure \ref{pic/f10}-c), in which the CDT for each myomere approachesd 0.5 TBC. Despite substantial energy consumption, the swimming speed achieved was lower than that of typical $H_{E}H_{T}$ cases (figure \ref{pic/f10} a-b). Figure \ref{pic/f10}-d presents a low-energy consumption high-thrust mode ($L_{E}H_{T}$). This mode achieved considerable swimming speed while requiring relatively low muscular work, making it a highly efficient locomotor strategy. It was characterized by a short CDT, avoiding prolonged activation of muscles on the same side. After a brief activation phase, the fish body used the kinetic energy generated during the initial contraction to complete the remaining body oscillations, while leveraging fluid–structure interactions to decelerate efficiently. A more detailed analysis of this swimming mechanism is provided later. Figure \ref{pic/f10}-e illustrates a medium-energy consumption high-thrust mode ($M_{E}H_{T}$), also showing both the muscle activation pattern and corresponding body wave. This mode maintained a satisfactory balance between energy consumption and thrust generation. Compared to $H_{E}H_{T}$ modes, $M_{E}H_{T}$ achieved relatively high propulsion performance while keeping energy usage at a moderate level. Similar to $L_{E}H_{T}$ modes, it also featured a short CDT and belonged to the class of energy-efficient swimming modes.

\subsubsection{Near-body flow field analysis of representative modes}

\begin{figure}[htbp]
	\centering 
	\subfigbottomskip=2pt 
	\subfigcapskip=-5pt 
	\subfigure[$L_{E}L_{T}(139-1-90)$]{
		\includegraphics[width=0.48\linewidth]{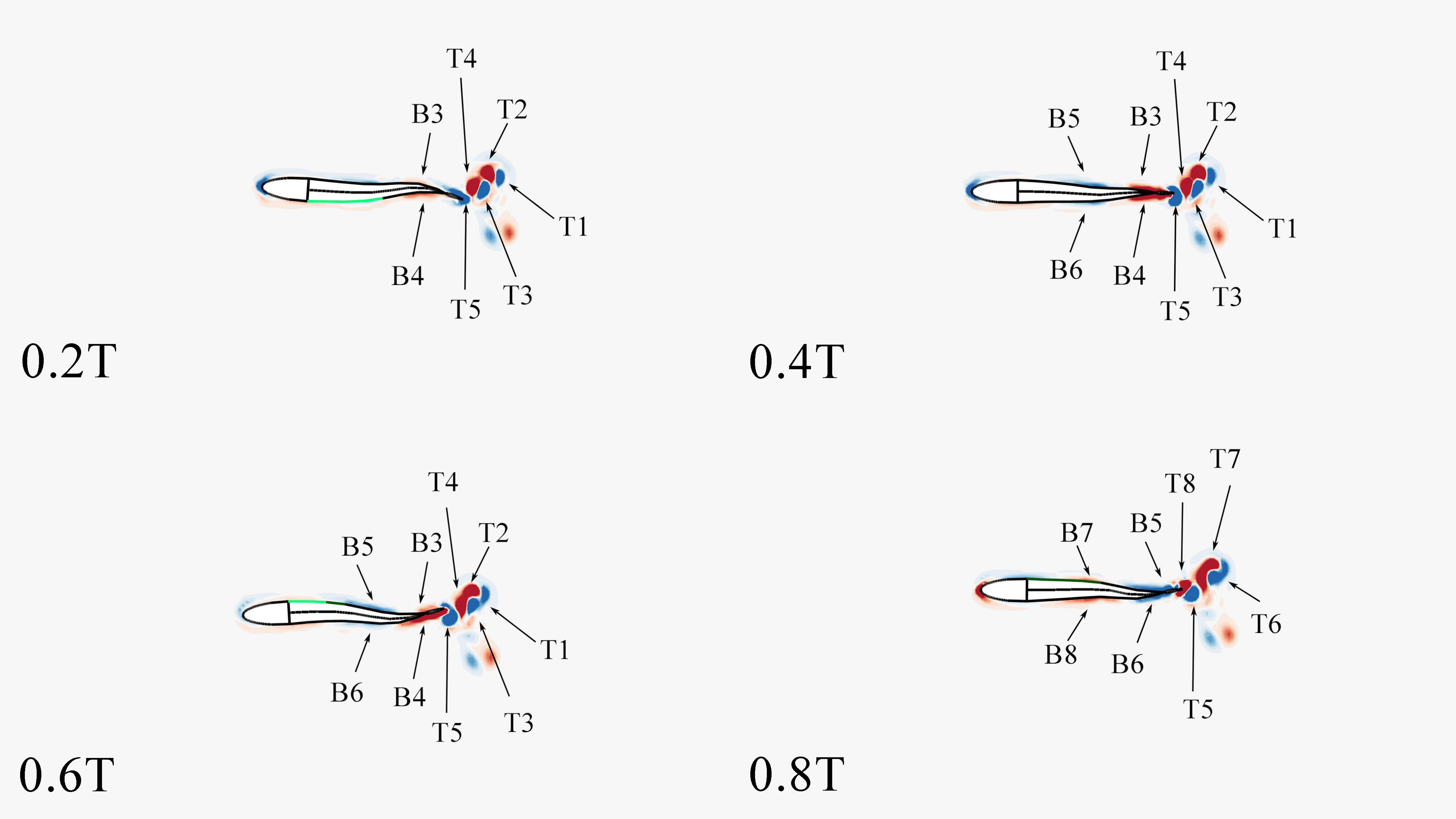}}
	\subfigure[$H_{E}H_{T}(60-41-220)$]{
		\includegraphics[width=0.48\linewidth]{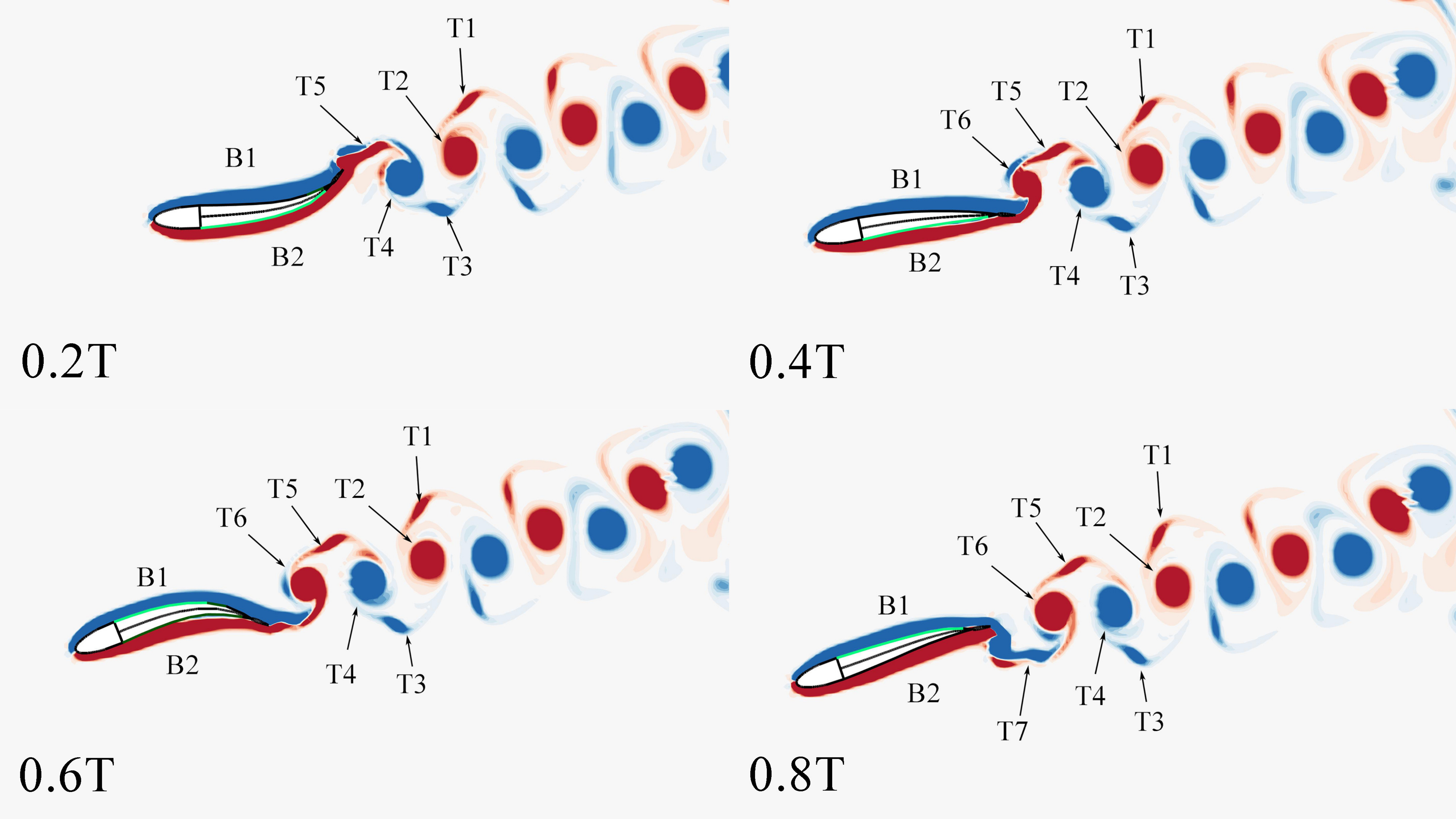}}
	\subfigure[$H_{E}H_{T}(1-1-250)$]{
		\includegraphics[width=0.48\linewidth]{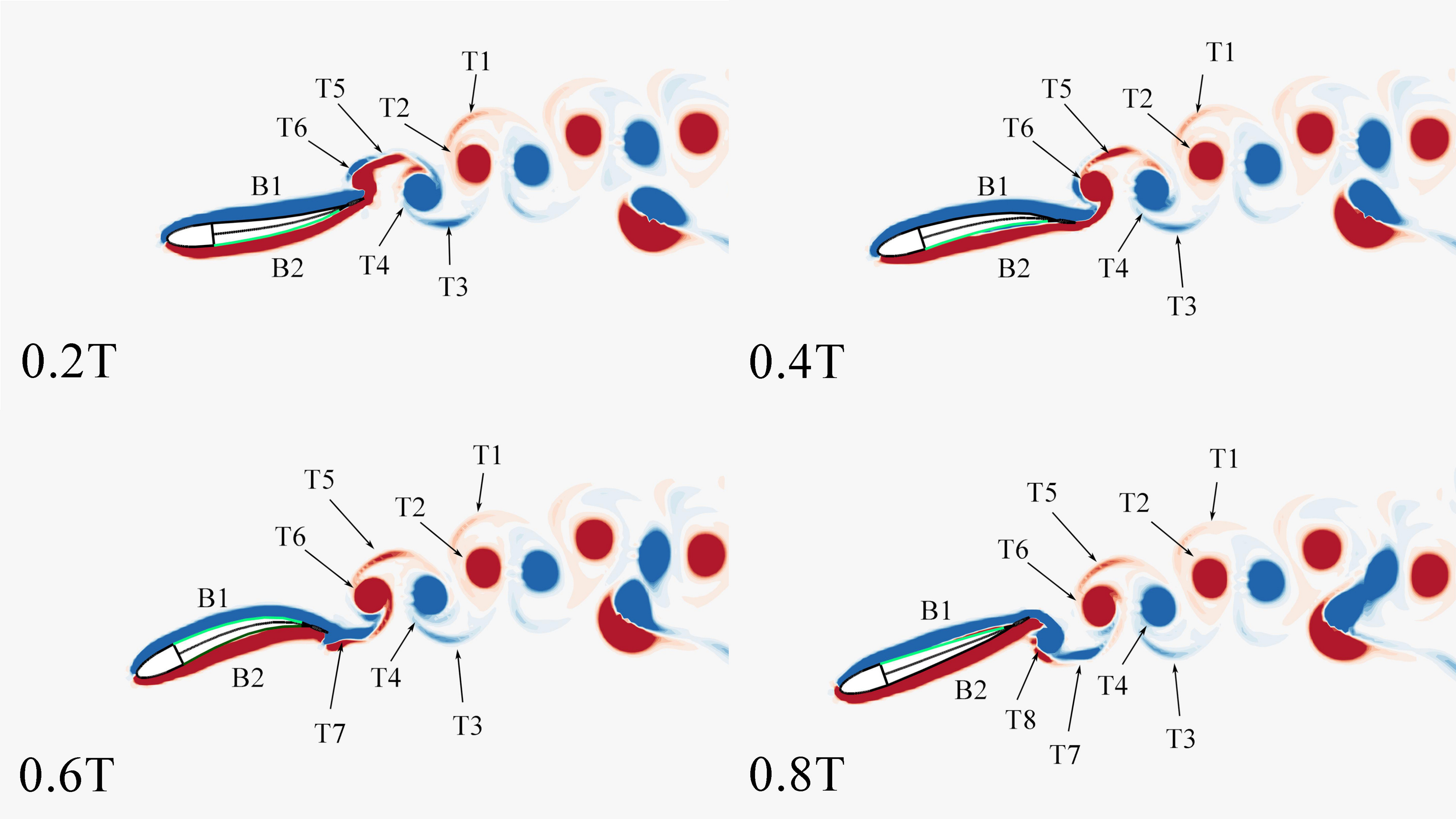}}
	\subfigure[$L_{E}H_{T}(9-1-100)$]{
		\includegraphics[width=0.48\linewidth]{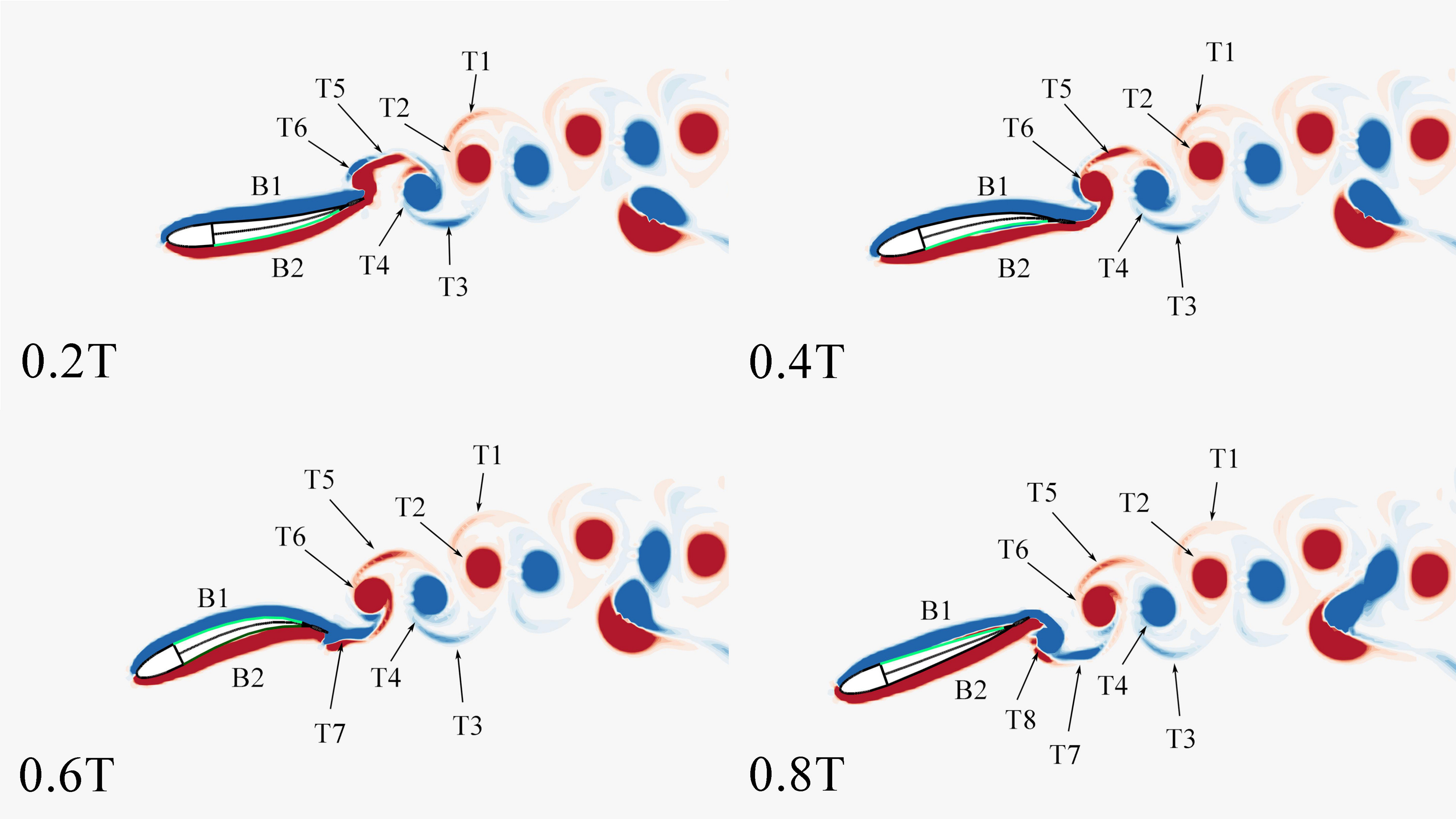}}
	\subfigure[]{
		\includegraphics[width=5in]{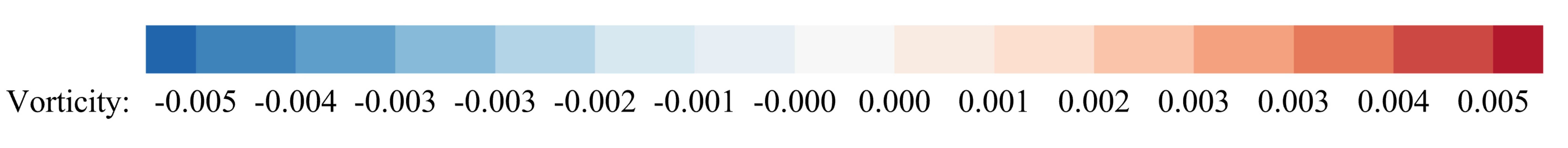}}
	\caption{Comparison of instantaneous flow vorticity fields (black indicates inactive muscles; bright green indicates strong activation; T denotes tail beat cycle).}
	\label{pic/f11}
\end{figure}

\begin{figure}[htbp]
	\centering
	\includegraphics[width=5in]{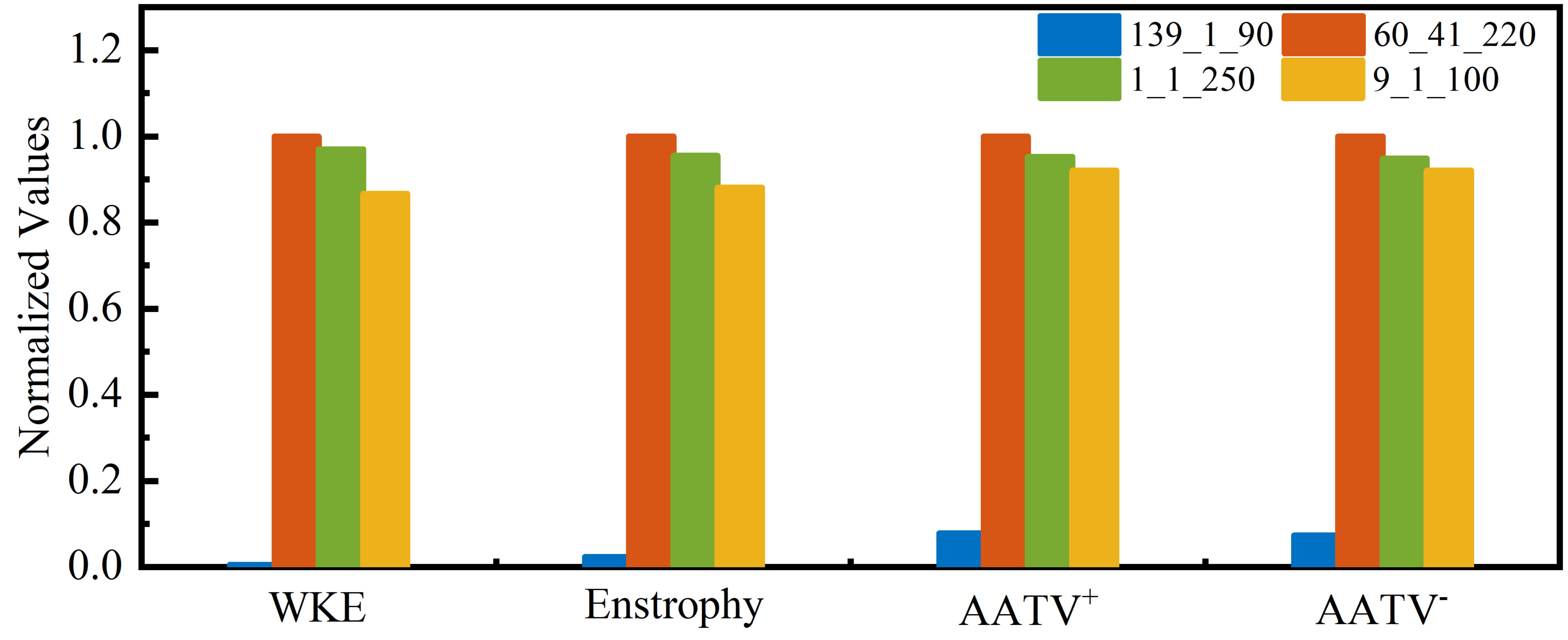}
	\caption{Quantitative comparison of WKE, Enstrophy, and AATV across four representative cases (all values are normalized).}
	\label{pic/f12}
\end{figure}

Figures \ref{pic/f11} a-d present instantaneous vorticity contours for four representative cases, revealing distinct vortex structures, energy distribution, and wake morphology. To quantitatively evaluate the wake dynamics across different modes, we defined several key metrics based on the framework of \cite{kang2023propulsive}:
(1) Wake Kinetic Energy (WKE), defined as the area-averaged integral of the perturbation velocity field within the wake region $S_W$ ($x \in [-L, 5L], y \in [-2L, 2L]$): $\text{WKE} = \frac{1}{S_W} \int_{S_W} \frac{|\mathbf{u}|^2}{2} \, dA$, where $\mathbf{u}$ denotes the perturbation velocity vector, and $S_W$ is the area of the wake region; (2) Enstrophy, defined as the area-averaged integral of vorticity magnitude (i.e., the squared norm of the velocity field's curl) within the same region: $\boldsymbol{\omega} = \frac{1}{S_W} \int_{S_W} \frac{|\boldsymbol{\omega}|^2}{2} \, dA$. Enstrophy quantifies the local rotational intensity of the fluid and serves as an indicator of energy fluctuations caused by disturbances in the wake. Higher enstrophy values signify stronger vortex shedding and greater wake instability; (3) Area-Averaged Total Vorticity (AATV), the positive total vorticity is computed by averaging all positive vorticity values in the integration domain: $\text{AATV}^{+}=\frac{1}{S_W} \int_{S_W} \frac{(|\omega| + \omega)}{2} \, dA$. Similarly, the absolute value of negative vorticity is given by: $\text{AATV}^{-}=\frac{1}{S_W} \int_{S_W} \frac{(|\omega| - \omega)}{2} \, dA$. These quantities serve to quantify the strength and asymmetry of vortex structures in the wake. A comparative analysis of these parameters across the four representative cases is provided in figure \ref{pic/f12}. For clarity and consistency, all computed values have been normalized to facilitate visual comparison.

In mode $L_{E}L_{T}(139-1-90)$ (figure \ref{pic/f10}-a), the thrust generated by the fish body was insufficient to produce effective propulsion. Specifically, the WKE was only 0.0048, enstrophy was 0.0232, the ${AATV}^{+}$ was 0.073, and the ${AATV}^{-}$ was 0.078. These values indicated that the trailing edge vortex (TEV) had weak intensity, with disorganized vortex structures and low energy density, failing to produce a clearly defined periodic jet in the wake. The bound vortex (BV), generated near the rostral region, nearly dissipated entirely during its transmission toward the caudal region. In this mode, red muscle contractions were primarily localized in the anterior portion of the body and occur as low-amplitude oscillations driven by alternating periodic activations. The resulting waveform resembled a standing wave, lacking significant phase velocity for axial propagation along the body. As a result, no effective propulsive traveling wave was formed, and the fluid did not separate properly at the trailing edge. The corresponding swimming Reynolds number for this mode was only 330.

In contrast, the high-energy consumption, high-speed mode $H_{E}H_{T}(60-41-220)$ (figure \ref{pic/f10}-b) exhibited entirely different hydrodynamic characteristics. All four wake metrics—the WKE, enstrophy, the ${AATV}^{+}$, and the ${AATV}^{-}$—reached a normalized value of 1.0, indicating the presence of densely packed and energetically strong vortex structures in the wake. Within a complete tail-beat cycle, a pair of strong, coherent, and counter-rotating TEVs can be clearly observed detaching in an orderly fashion from the trailing edge. Notably, a secondary TEV detached from the tail region prior to the formation of the primary TEV. This vortex-shedding pattern is typically classified as a canonical “2S” wake structure, characterized by the shedding of two distinct and symmetric vortices per tail-beat cycle and this wake mode is associated with high energy output and efficient thrust generation during self-propelled swimming (\cite{bodaghi2023effects}). In this mode, muscle activation was highly coordinated: both activation and deactivation waves propagated sequentially from the head (rostral end) to the tail (caudal end), resulting in orderly, sequential contractions of the myomeres and the formation of a continuous propulsive wave. The swimming Reynolds number in this case reached 11,770. Figures \ref{pic/f11} c-d display the instantaneous vorticity fields for two additional high-speed modes: high-energy consumption mode $H_{E}H_{T}(1-1-250)$ and low-energy consumption mode $L_{E}H_{T}(9-1-100)$. For mode $H_{E}H_{T}(1-1-250)$, the WKE was 0.970, enstrophy was 0.955, the ${AATV}^{+}$ was 0.952, and the ${AATV}^{-}$ was 0.949. For mode $L_{E}H_{T}(9-1-100)$, the WKE was 0.866, enstrophy was 0.881, the ${AATV}^{+}$ was 0.920, and the ${AATV}^{-}$ was 0.921. Although the wake energy and vortex density of both modes were slightly lower than those of mode $H_{E}H_{T}(60-41-220)$, they still exhibited high propulsive energy and reached significantly higher average swimming speeds than mode $L_{E}L_{T}(139-1-90)$. In both cases, clear formation of dominant BV and periodic shedding of TEV were observed throughout the swimming cycle, indicating that both modes share similar thrust generation mechanisms. The swimming Reynolds number for mode $H_{E}H_{T}(1-1-250)$ was 10,700, while for mode $L_{E}H_{T}(9-1-100)$, it was slightly lower at 10,141.

\subsubsection{Hydrodynamic force distribution}

\begin{figure}[htbp]
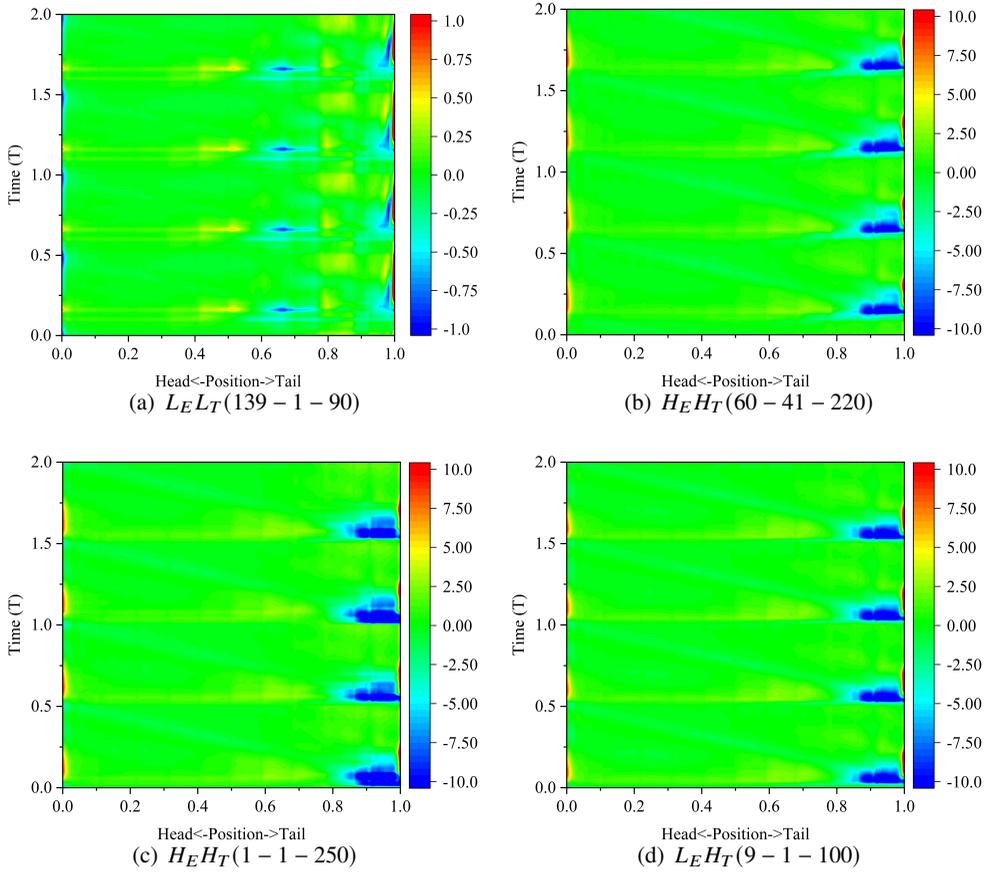

	\centering 
	\subfigbottomskip=2pt 
	\subfigcapskip=-5pt 
	\subfigure[$L_{E}L_{T}(139-1-90)$]{
		\includegraphics[width=0.48\linewidth]{f13a-eps-converted-to.pdf}}
	\subfigure[$H_{E}H_{T}(60-41-220)$]{
		\includegraphics[width=0.48\linewidth]{f13b-eps-converted-to.pdf}}
	\subfigure[$H_{E}H_{T}(1-1-250)$]{
		\includegraphics[width=0.48\linewidth]{f13c-eps-converted-to.pdf}}
	\subfigure[$L_{E}H_{T}(9-1-100)$]{
		\includegraphics[width=0.48\linewidth]{f13d-eps-converted-to.pdf}}
	\caption{Spatiotemporal distribution of hydrodynamic forces in the x-direction over two tail beat cycles (T denotes tail beat cycle).}
	\label{pic/f13}
\end{figure}

\begin{figure}[htbp]
	\centering
	\includegraphics[width=5in]{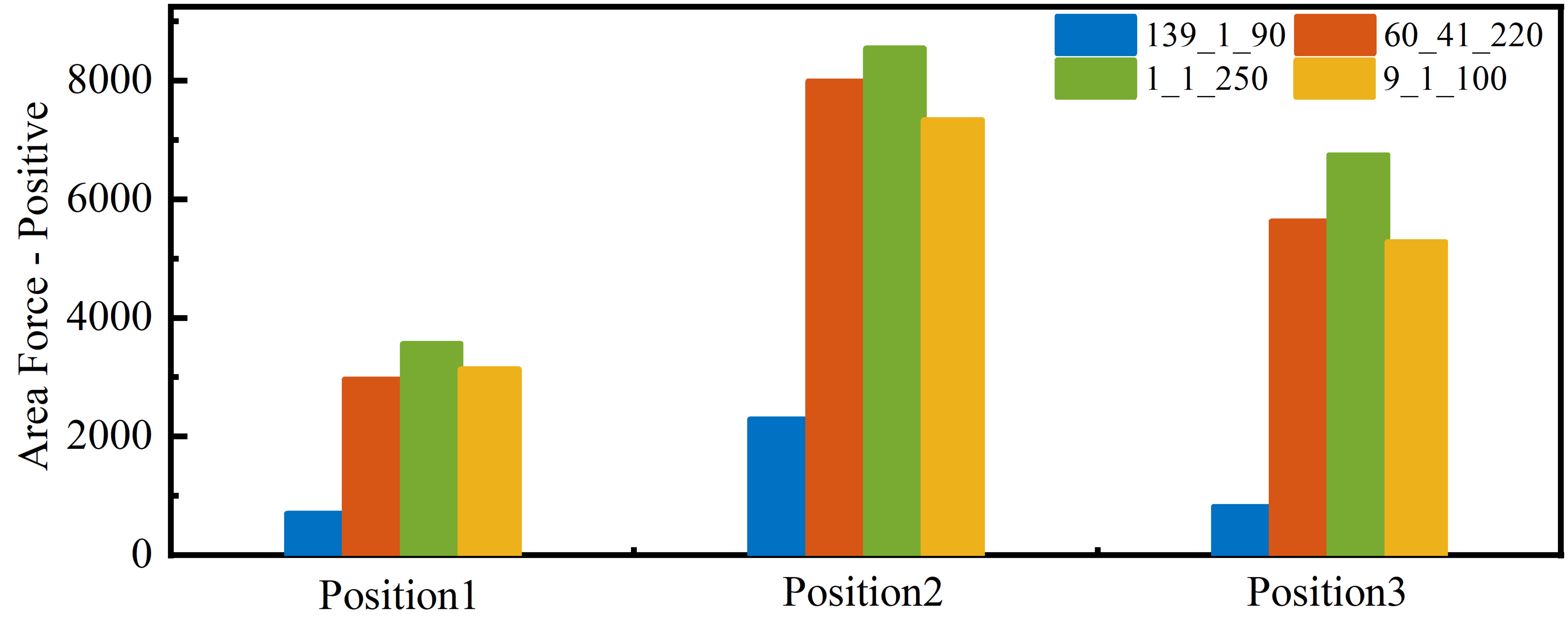}
	\caption{Comparison of spatiotemporal integration of lateral forces at different body positions across four representative cases.}
	\label{pic/f15}
\end{figure}

Figures \ref{pic/f13} and \ref{pic/f14} illustrate the spatiotemporal distribution of hydrodynamic forces in the x- and y-directions, respectively, across two tail-beat cycles for several characteristic swimming modes of the digital trout. Unlike anguilliform swimmers such as eels, which generate propulsion through full-body undulations, carangiform species concentrate thrust generation predominantly in the posterior region, while experiencing significant drag in the anterior body. This observation is consistent with previous findings (\cite{ming20193d, li2023roles}). In the $L_{E}L_{T}(139-1-90)$ mode, the distribution of hydrodynamic force in the x-direction exhibited pronounced nonuniformity. Due to the absence of a coherent axial traveling curvature wave mechanism, forward thrust could not be effectively generated via flow separation at the tail (figure \ref{pic/f13}-a). Analysis of the hydrodynamic forces in the y-direction (figure \ref{pic/f14}-a) further revealed that, although the $L_{E}L_{T}(139-1-90)$ mode generated a periodic lateral curvature wave in the anterior region, the wave rapidly attenuated along the body axis due to insufficient axial muscle coordination. As a result, the lateral hydrodynamic forces at the tail were exceedingly weak. By contrast, the other three modes established a fully developed chain of body wave transmission, significantly enhancing forward propulsion. Figure \ref{pic/f15} presents a comparison of the time-integrated lateral forces across different body regions for the four cases (only positive values are shown). Spatial integration was conducted over three segmented body regions: Position 1: $[0, 0.3BL]$ (anterior); Position 2: $[0.3BL, 0.8BL]$ (midbody); Position 3: $[0.8BL, 1.0BL]$ (posterior). The temporal integration interval was defined as ${\rm{t}} \in [0,2.0]$T.

Among the four cases, the $H_{E}H_{T}(1-1-250)$ mode exhibited the highest level of lateral force, but its average propulsion speed (0.82 BL/s) was lower than that of $H_{E}H_{T}(60-41-220)$ (0.92 BL/s). This discrepancy arises because, in the $H_{E}H_{T}(1-1-250)$ mode, the muscles on the same side of the body were activated almost simultaneously, causing the anterior body to displace more water laterally. This resulted in a substantial loss of lateral hydrodynamic energy. In contrast, the phase-delay mechanism employed in $H_{E}H_{T}(60-41-220)$ reduced the overall lateral amplitude compared to $H_{E}H_{T}(1-1-250)$. In this mode, the body wave formed a flexible, mesh-like pattern, with the curvature wave gradually amplifying as it traveled from head to tail. This focused lateral oscillations in the caudal region, minimizing lateral hydrodynamic losses and improving overall propulsive efficiency.

\begin{figure}[htbp]
	\centering 
	\subfigbottomskip=2pt 
	\subfigcapskip=-5pt 
	\subfigure[$L_{E}L_{T}(139-1-90)$]{
		\includegraphics[width=0.48\linewidth]{f14a-eps-converted-to.pdf}}
	\subfigure[$H_{E}H_{T}(60-41-220)$]{
		\includegraphics[width=0.48\linewidth]{f14b-eps-converted-to.pdf}}
	\subfigure[$H_{E}H_{T}(1-1-250)$]{
		\includegraphics[width=0.48\linewidth]{f14c-eps-converted-to.pdf}}
	\subfigure[$L_{E}H_{T}(9-1-100)$]{
		\includegraphics[width=0.48\linewidth]{f14d-eps-converted-to.pdf}}
	\caption{Spatiotemporal distribution of hydrodynamic forces in the y-direction over two tail beat cycles (T denotes tail beat cycle).}
	\label{pic/f14}
\end{figure}

\subsection{Further discussion}

To further clarify the coupling characteristics of the muscle–skeleton–fluid system and the mechanisms underlying cross-scale energy transmission during steady swimming, based on the principle of single-variable control, we conducted a comparative analysis of several representative cases selected from the DRL optimization results.

\subsubsection{Effect of axial muscle activation range on undulatory swimming}

\begin{figure}[htbp]
	\centering
	\includegraphics[width=5in]{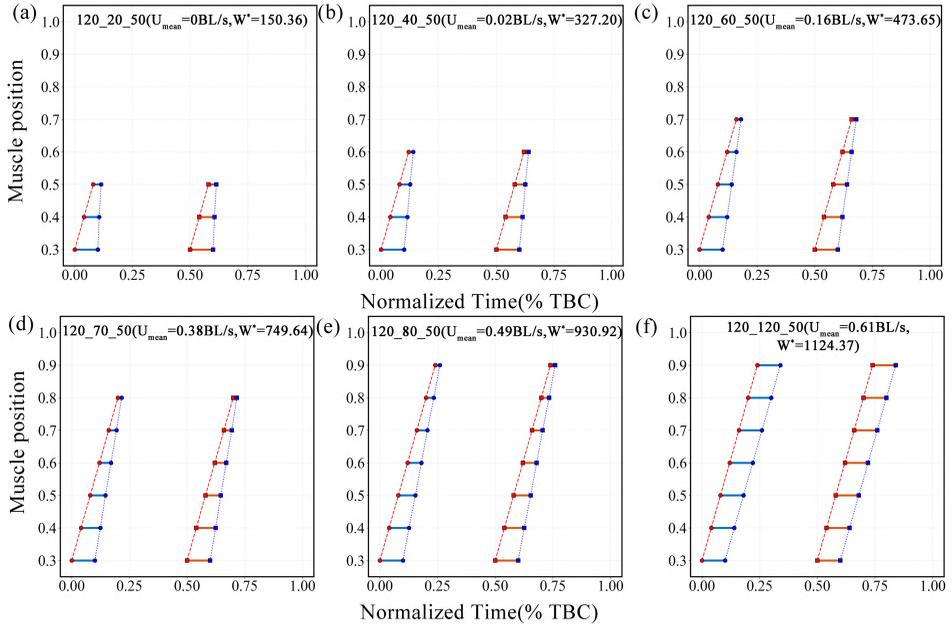}
	\caption{Spatiotemporal comparison of muscle activation patterns under different axial activation ranges.}
	\label{pic/f16}
\end{figure}

\begin{figure}[htbp]
	\centering
	\includegraphics[width=5in]{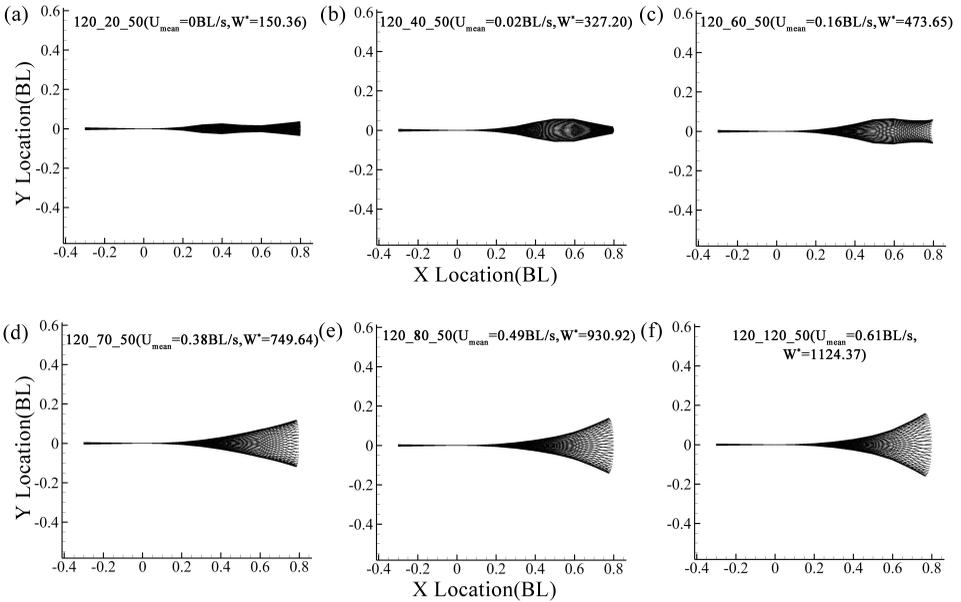}
	\caption{Body wave envelope diagrams corresponding to different axial muscle activation patterns.}
	\label{pic/f17}
\end{figure}

To isolate the effect of longitudinal muscle activation span, we selected cases with fixed ${t_{act}}$ and CDT, varying only ${t_{deact}}$ (single-variable control). These configurations span a spectrum of axial activation patterns, with spatiotemporal muscle activation maps and body wave envelopes shown in figures \ref{pic/f16} and \ref{pic/f17}. The results indicated that myomere distributed along the body axis function in a serially coupled manner, allowing for the dynamic integration of propulsive forces. Specifically, a modest curvature wave initiated by muscle contractions in the rostral (head) region was progressively amplified through the temporally coordinated, sequential activation of downstream muscle segments. This process culminated in the formation of a high-amplitude body wave at the caudal fin. When the axial span of muscle activation was less than 0.5 BL (figure \ref{pic/f16}-b), the curvature wave initiated near the rostrum experienced significant attenuation as it propagated forward, resulting in a substantial loss of amplitude before reaching the tail. To generate a quadratically increasing body wave envelope—where the wave amplitude grows from head to tail—the axial activation range must exceed 0.5 BL (figures \ref{pic/f16}-d and \ref{pic/f17}-d). If the axial activation range falls below this critical threshold, a stable travelling curvature wave cannot be formed, resulting in reduced propulsion efficiency and a significant decrease in swimming speed.

\subsubsection{Utilizing inertial forces, elastic forces, and fluid–structure interaction to reduce energy consumption in undulatory swimming}\label{subsubsec::energysave}

\begin{figure}[htbp]
	\centering
	\includegraphics[width=5in]{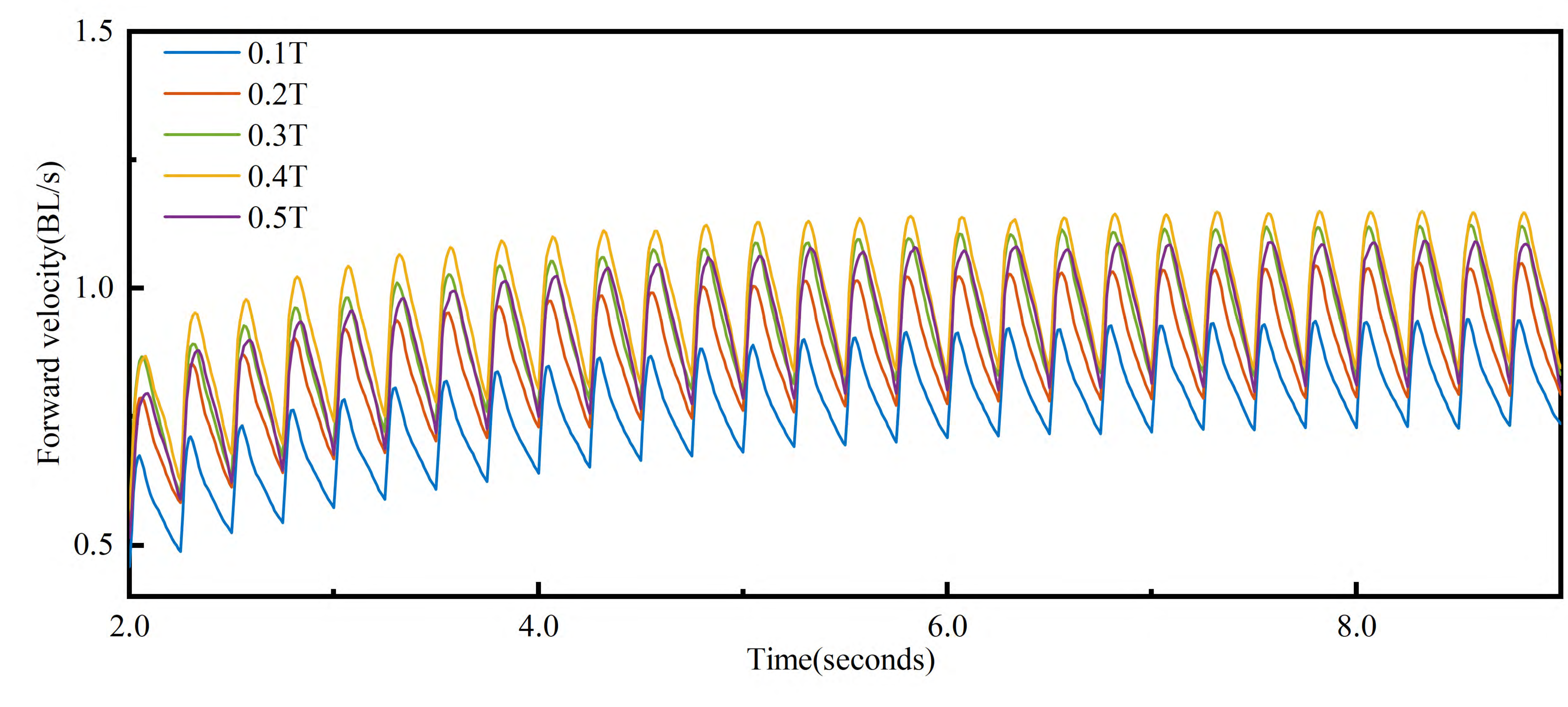}
	\caption{Forward swimming speed under varying muscle contraction durations.}
	\label{pic/f18}
\end{figure}

\begin{figure}[htbp]
	\centering
	\includegraphics[width=5in]{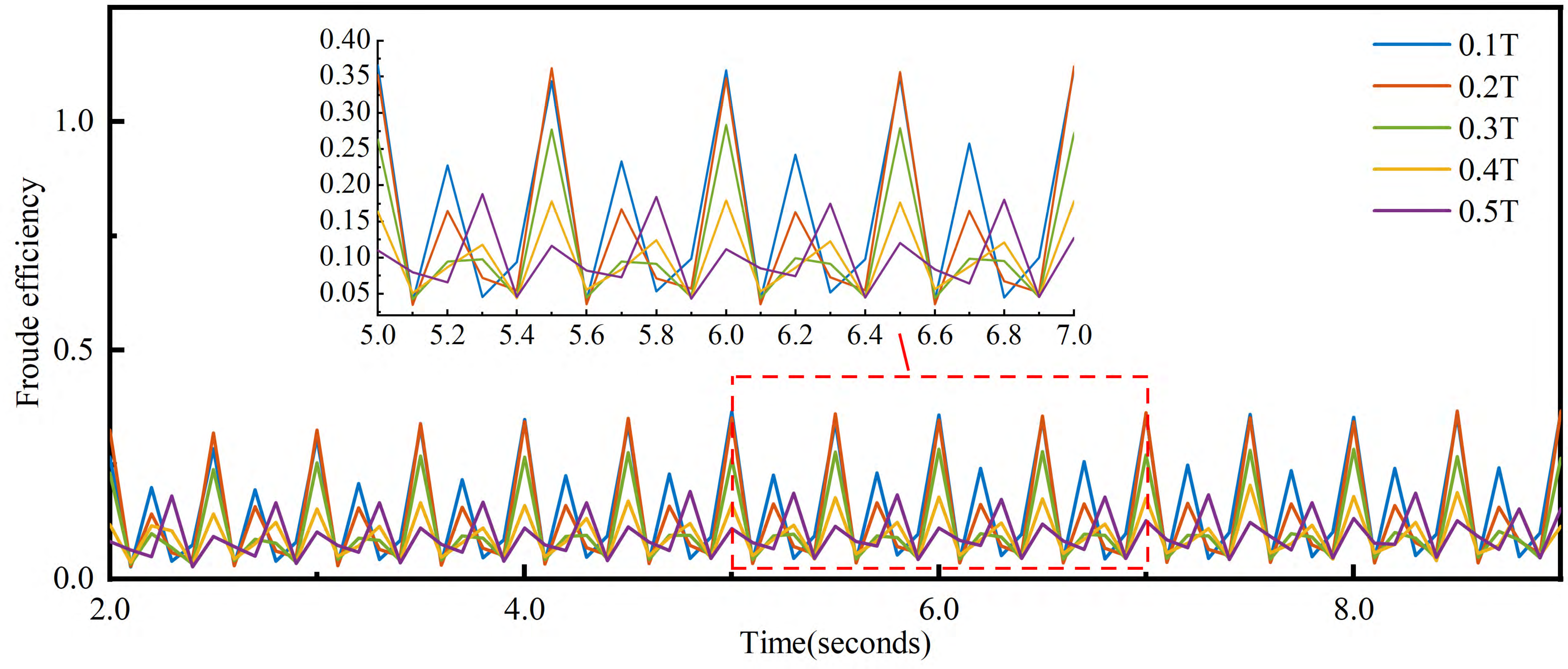}
	\caption{Froude efficiency of fish locomotion under varying muscle contraction durations.}
	\label{pic/f19}
\end{figure}

Optimization results from the DRL analysis show that, when ${t_{act}}$ and ${t_{deact}}$ are held constant, swimming speed is highly sensitive to the muscle CDT, which in this study ranges from 0.1 to 0.5 tail-beat cycles (TBC). This parameter range corresponds to a zero-phase-lag mode in which muscles on both sides of the body are activated synchronously (figure \ref{pic/f18}). Further analysis reveals that once CDT exceeds a critical threshold (0.3 TBC), the marginal benefit of extending CDT on swimming speed begins to diminish, while energy consumption increases sharply (figure \ref{pic/f20}). These findings suggest that in swimming modes characterized by alternating, simultaneous contraction of bilateral muscles, moderately reducing CDT is an effective strategy for optimizing energy efficiency. While extending CDT may offer slight gains in swimming speed, the resulting increase in energy consumption far outweighs these benefits, ultimately leading to a significant decrease in overall efficiency. The distribution of Froude efficiency across different CDT conditions (figure \ref{pic/f19}) further supports this conclusion. Cases with shorter CDT demonstrate higher average Froude efficiency, indicating superior hydrodynamic energy utilization.

\begin{figure}[htbp]
	\centering
	\includegraphics[width=5in]{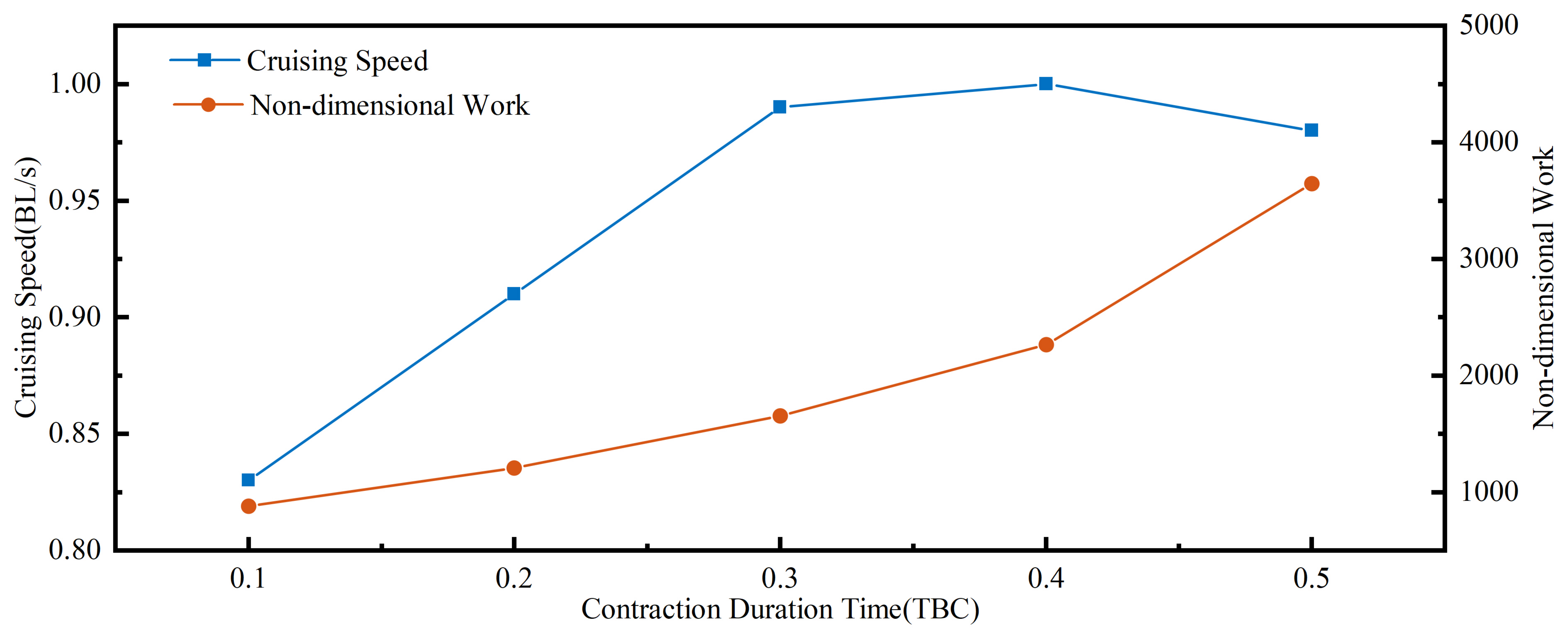}
	\caption{Cruising speed and energy consumption of the fish under different CDT conditions.}
	\label{pic/f20}
\end{figure}

The physical mechanisms underlying these phenomena can be better understood by examining instantaneous snapshots of the flow fields from representative cases, as shown in figure \ref{pic/f21}. In addition, we introduce deformation kinetic energy (DKE) to quantify the energy associated with body deformation, computed as $DKE = \int_l {{{\left| {{{\bf{u}}_d}} \right|}^2}/2ds} $, where $\mathbf{\dot{u}}$ denotes the body deformation velocity vector, and $l$ is ${\rm{x}} \in [0.0BL,1.0BL]$. At $t = 0.0\,T$, both the short contraction duration case (CDT = 0.2 TBC) and the long contraction duration case (CDT = 0.5 TBC) inherited residual kinetic energy from the previous cycle. However, in the short CDT case, the dominant propulsion mechanism shifted from active muscular force to inertia-driven passive oscillation, resulting in a DKE of 0.74. In contrast, the long CDT case remained within a muscle-driven regime, exhibiting a significantly higher DKE of 1.92. During the interval $t = 0.1\,T$ to $0.2\,T$, the left-side muscles were activated. In the short CDT case (0.2 TBC), the fish leveraged fluid–structure interaction and body elasticity to initiate pre-braking earlier. This led to a faster rate of reverse body deformation, with the DKE reaching 2.72 at $t = 0.1\,T$. In contrast, in the long CDT case (0.5 TBC), the prolonged contraction of the right-side muscles from the previous phase resulted in excessive momentum buildup. As a consequence, additional muscular effort was needed to reverse the motion, and the DKE at $t = 0.1\,T$ was measured at 1.97. Between $t = 0.3\,T$ and $0.4\,T$, the short CDT case entered a fully inertia-driven mode. The muscles were in a relaxed state while the body passively decelerates via fluid–structure interaction, with DKE values of 1.70 and 1.26 at $t = 0.3\,T$ and $0.4\,T$, respectively. In contrast, the long CDT case remained in an actively muscle-driven state, continuously generating high deformation speeds. At the same time points, its DKE values were 2.33 and 2.14, respectively. In order to reverse direction, this mode required additional muscular effort to counteract the accumulated body momentum. By $t = 0.5\,T$, the DKE in the short CDT case had dropped significantly to 0.71, while the long CDT case still maintained a high DKE of 1.77. These observations clearly indicate that in bilateral simultaneous contraction modes, a longer CDT does not necessarily yield better performance. In the short CDT case, after a brief active contraction, the fish can rely on inertial forces to complete the remainder of the body oscillation, while body elasticity and fluid–structure interaction facilitate passive braking. This strategy maximizes the contributions of hydrodynamic, elastic, and inertial forces, significantly reducing energy demands during direction reversal. Conversely, although a longer CDT can achieve greater deformation velocities and larger tail amplitudes, it also significantly increases the energy cost on the contralateral muscle when reversing body momentum. By utilizing this mechanism, the fish achieves 90\% of the cruising speed observed in the long CDT case while consuming significantly less energy. This represents a highly efficient swimming strategy that lies near the optimal region of the energy consumption–speed Pareto frontier (see figure \ref{pic/f8}).

\begin{figure}[htbp]
	\ContinuedFloat
	\centering 
	\subfigbottomskip=2pt 
	\subfigcapskip=-5pt 
	\subfigure[$t = 0.0T$($1-1-100$)]{
		\includegraphics[width=0.4\linewidth]{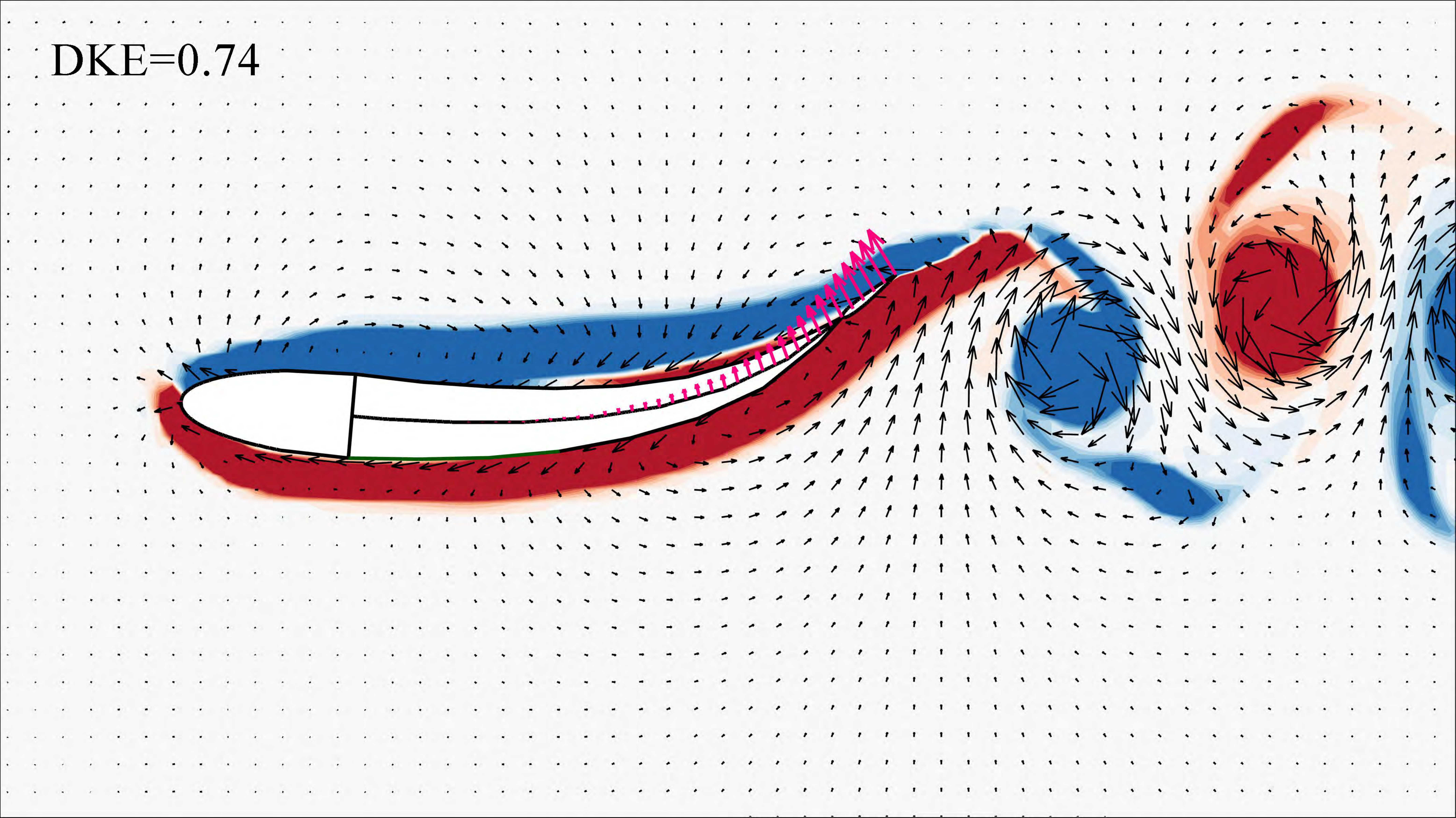}}
	\subfigure[$t = 0.0T$($1-1-250$)]{
		\includegraphics[width=0.4\linewidth]{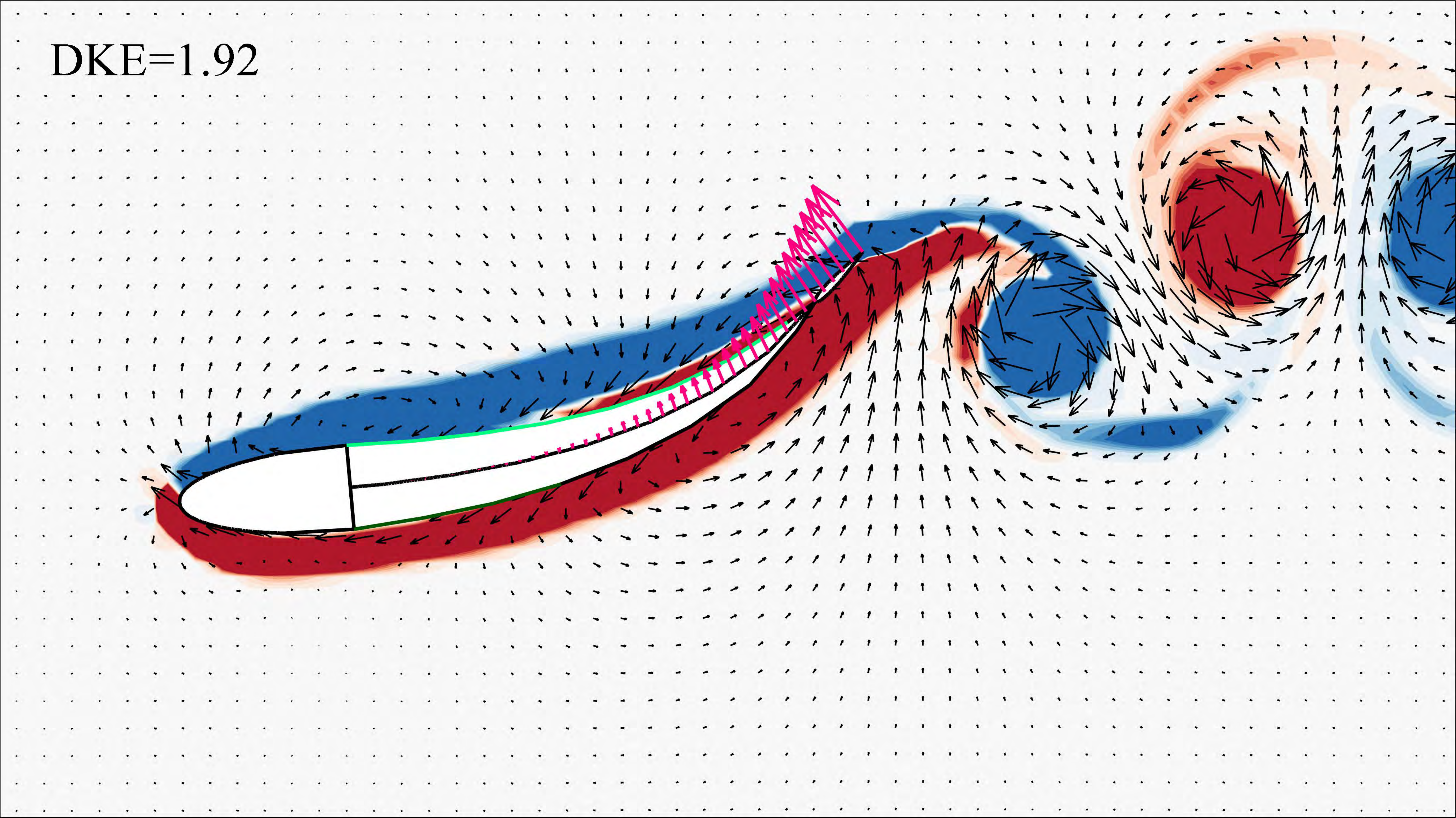}}
	\subfigure[$t = 0.1T$($1-1-100$)]{
		\includegraphics[width=0.4\linewidth]{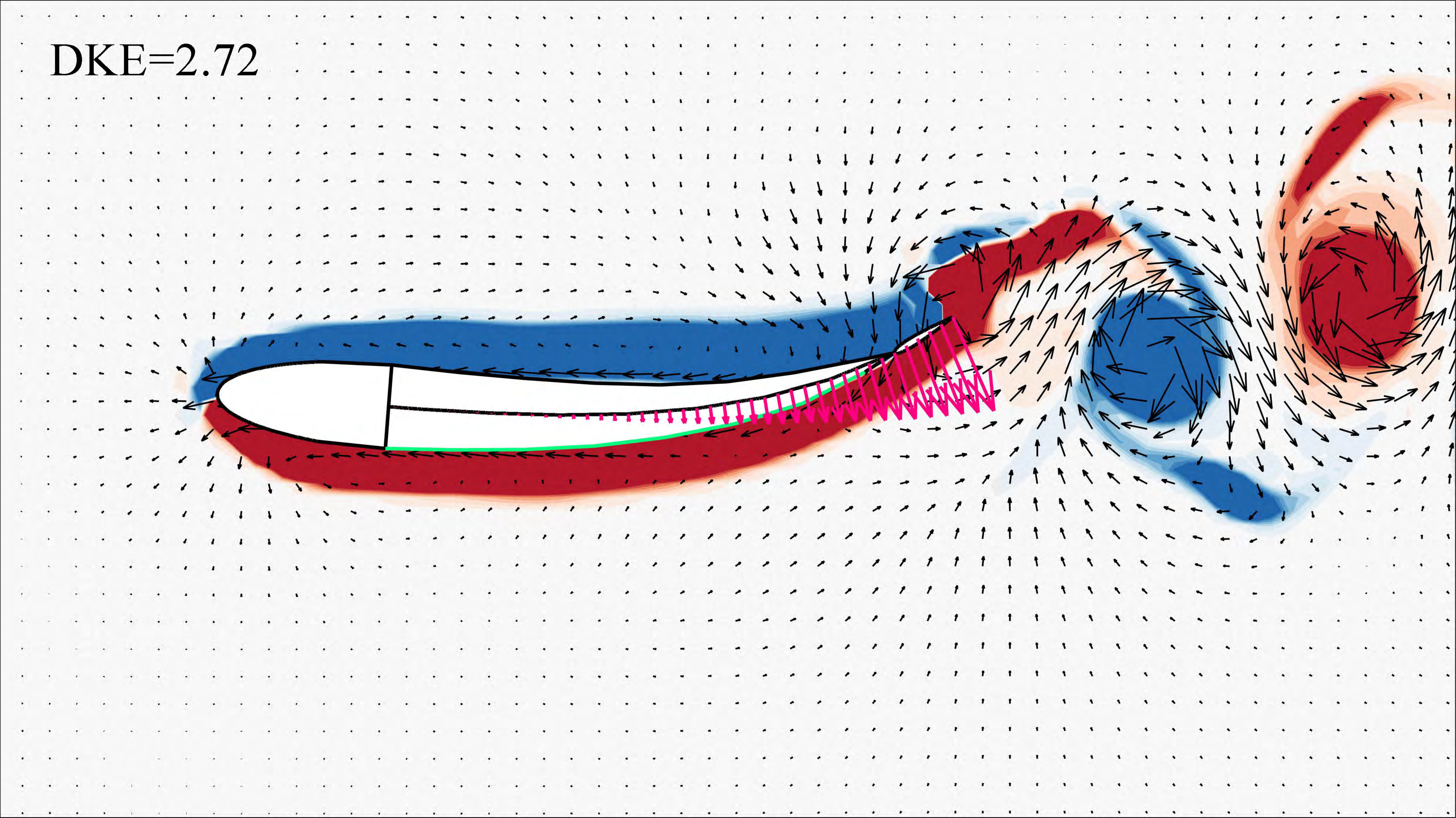}}
	\subfigure[$t = 0.1T$($1-1-250$)]{
		\includegraphics[width=0.4\linewidth]{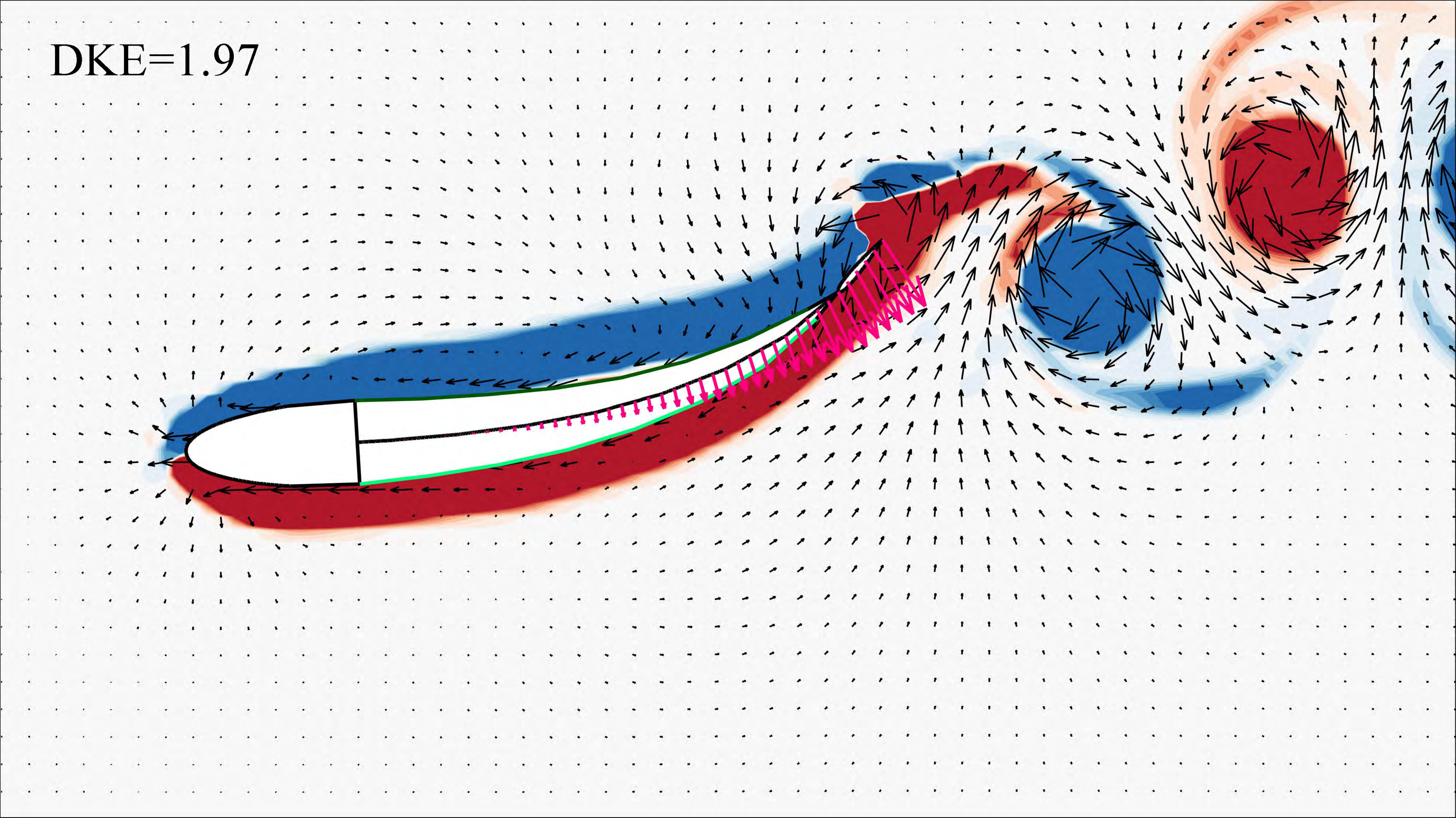}}
	\subfigure[$t = 0.2T$($1-1-100$)]{
		\includegraphics[width=0.4\linewidth]{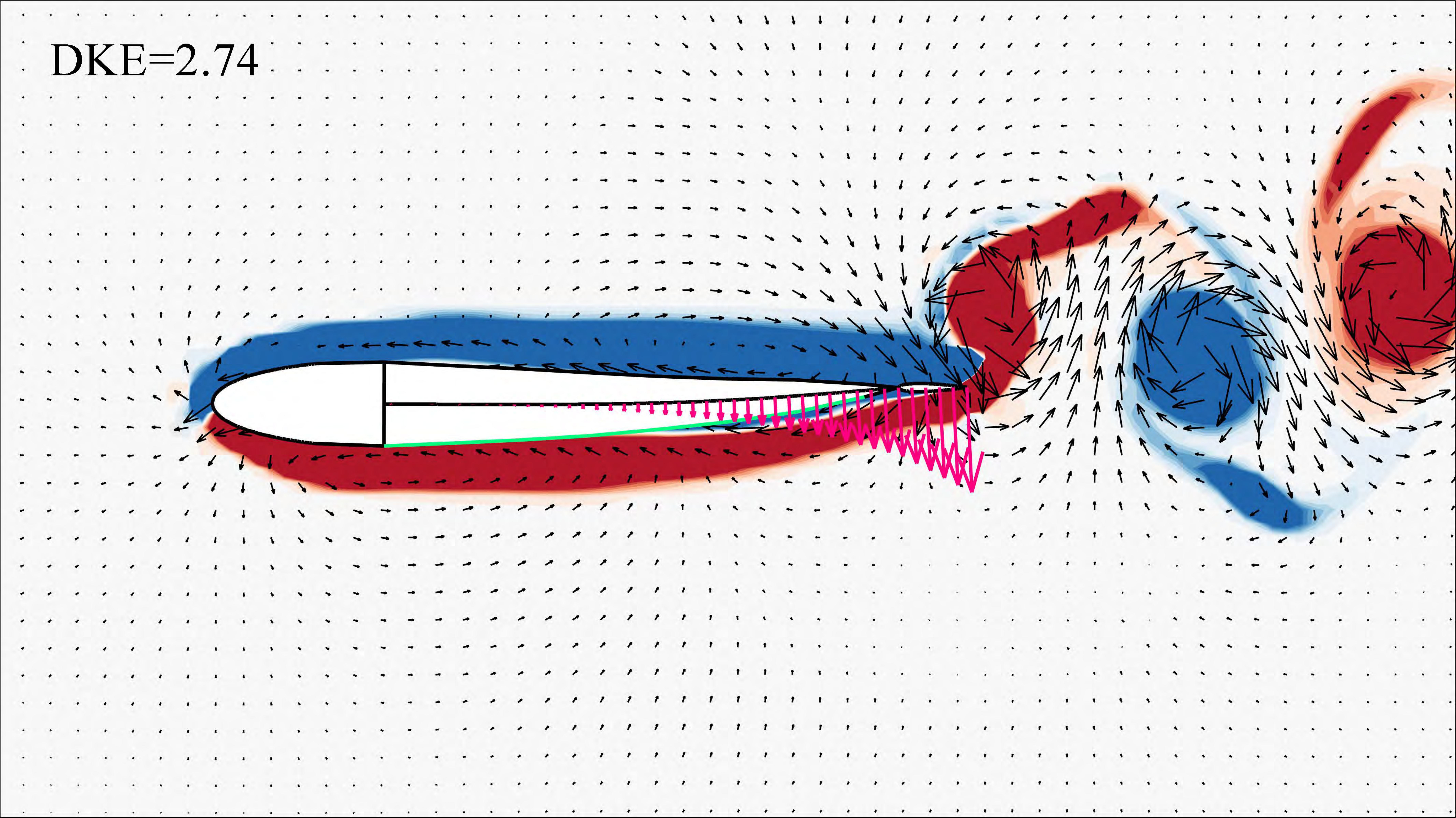}}
	\subfigure[$t = 0.2T$($1-1-250$)]{
		\includegraphics[width=0.4\linewidth]{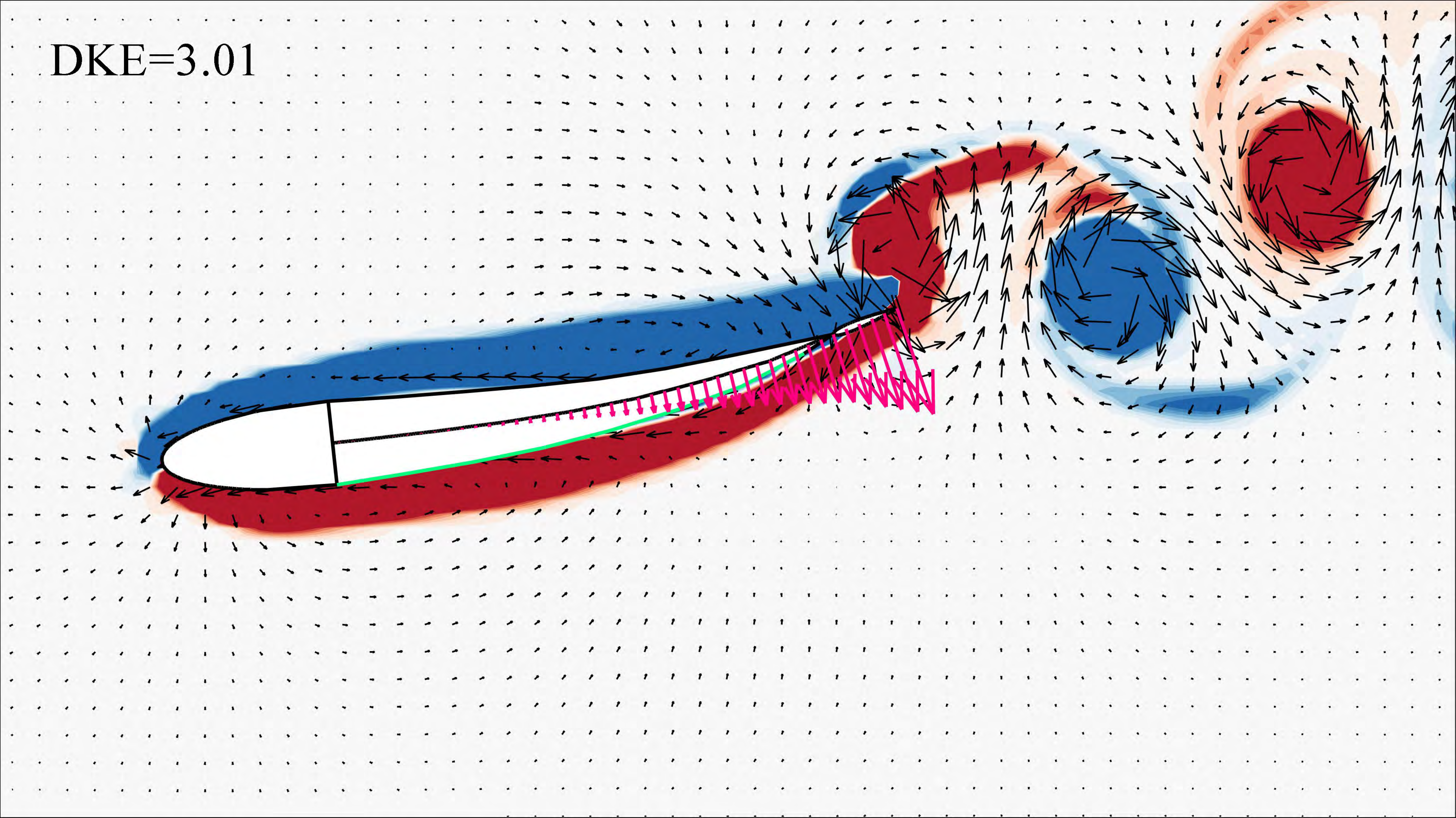}}
	\caption{Comparison of instantaneous flow vorticity fields (black indicates inactive muscles, bright green indicates strongly activated muscles, pink arrows represent body deformation velocity vectors, and black arrows represent flow velocity vectors; T denotes the tail beat cycle).}
\end{figure}

\begin{figure}[htbp]
	\centering 
	\subfigbottomskip=2pt 
	\subfigcapskip=-5pt 
\subfigure[$t = 0.3T$($1-1-100$)]{
	\includegraphics[width=0.4\linewidth]{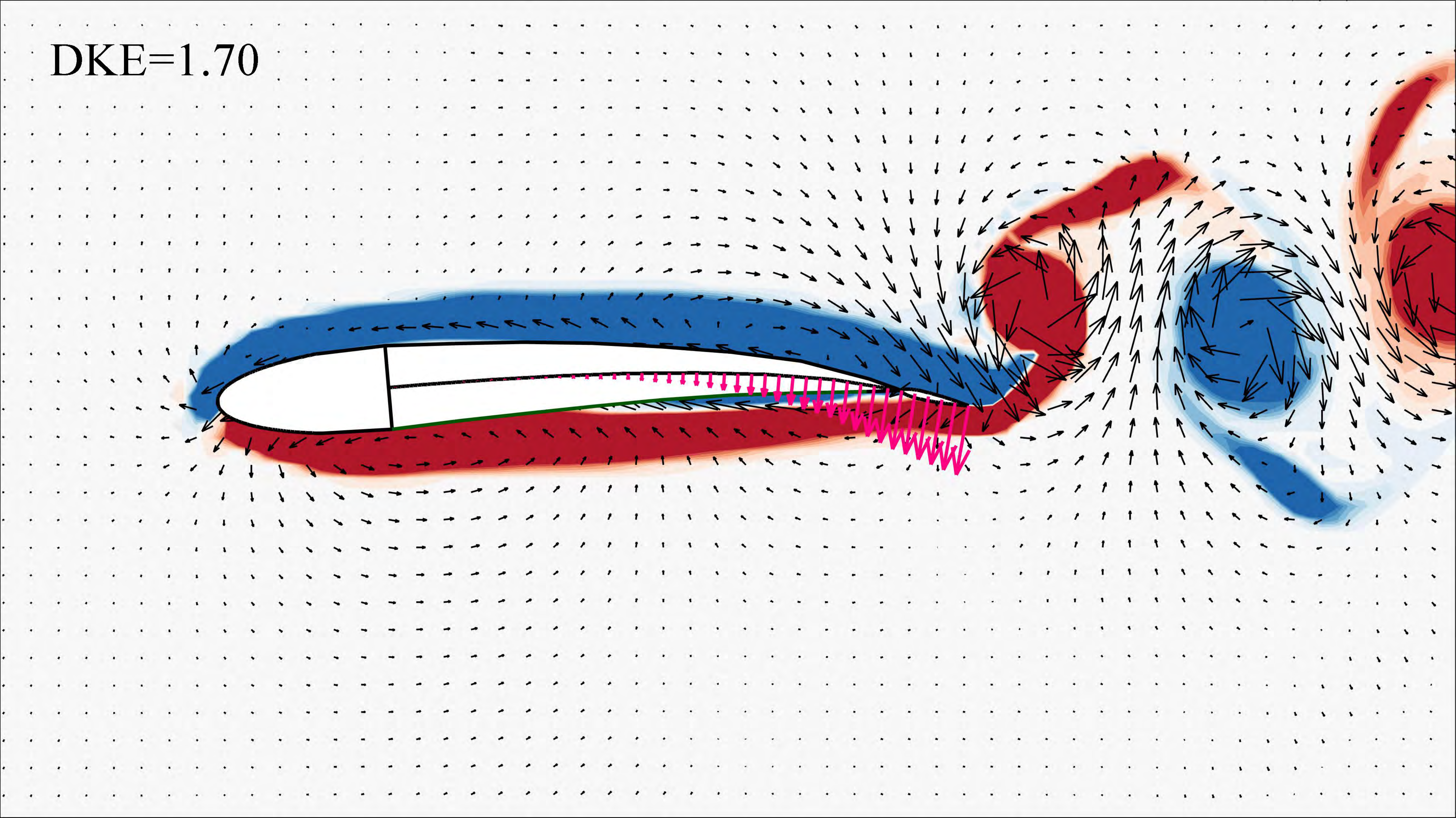}}
\subfigure[$t = 0.3T$($1-1-250$)]{
	\includegraphics[width=0.4\linewidth]{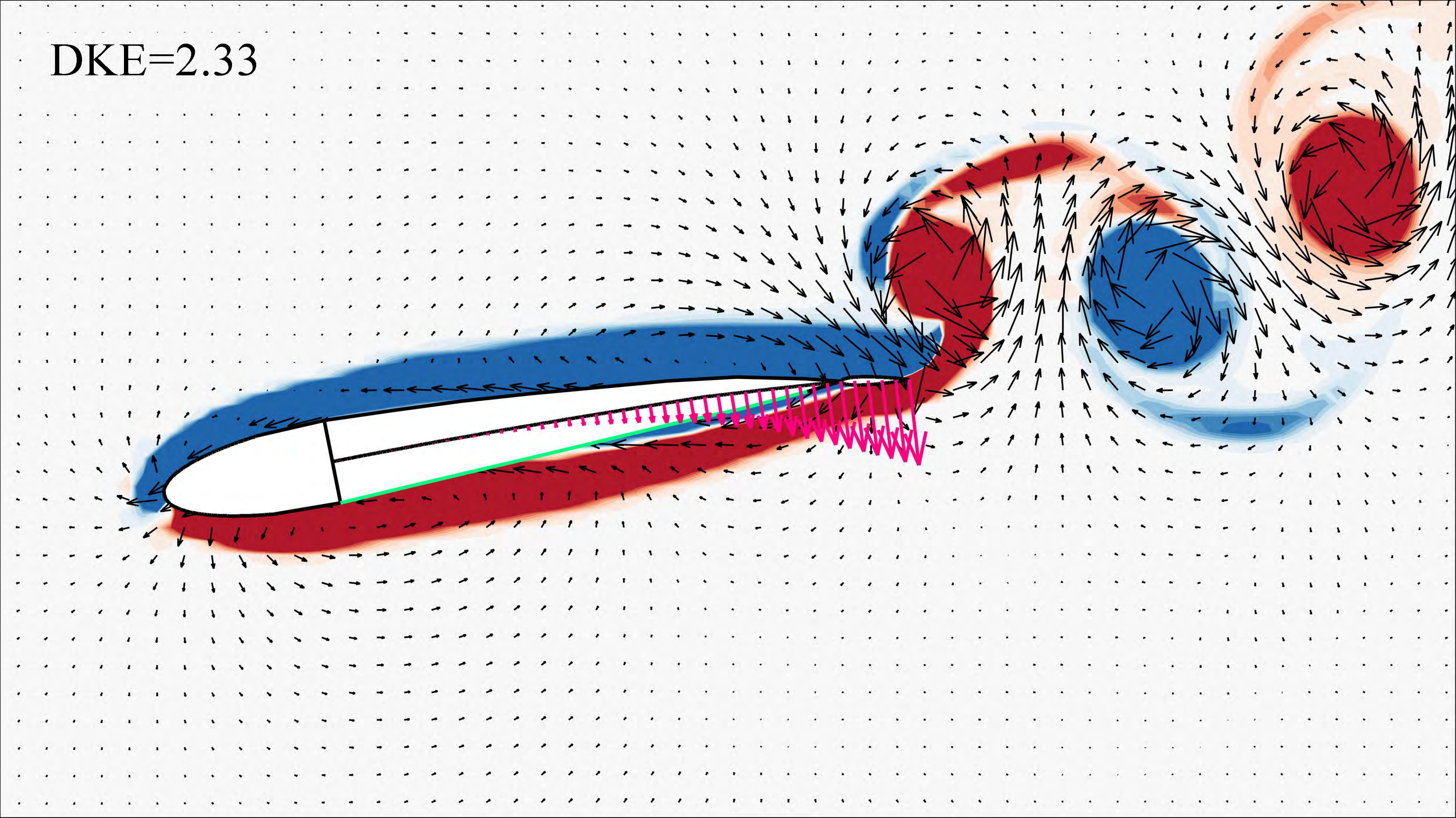}}
\subfigure[$t = 0.4T$($1-1-100$)]{
	\includegraphics[width=0.4\linewidth]{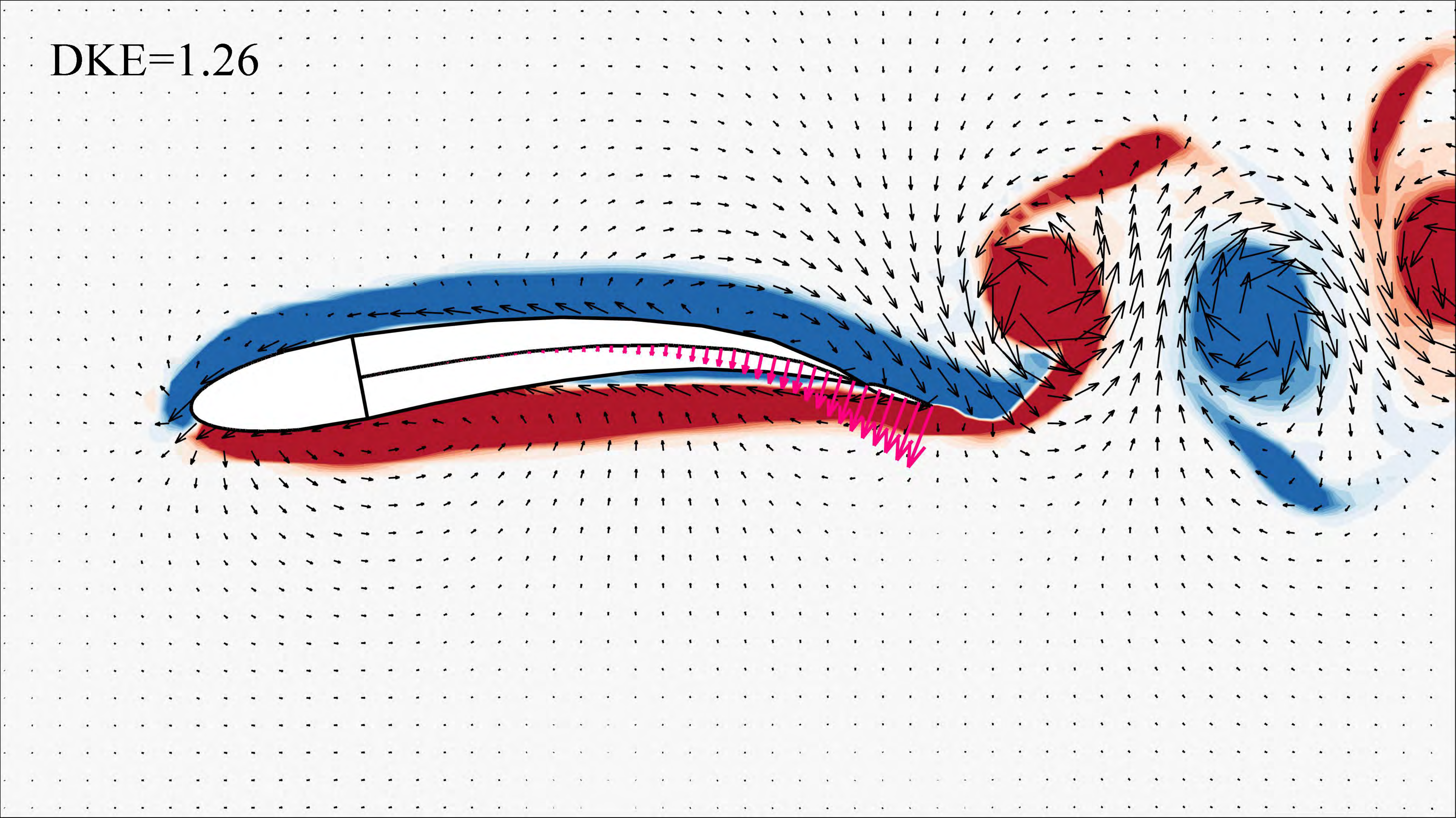}}
\subfigure[$t = 0.4T$($1-1-250$)]{
	\includegraphics[width=0.4\linewidth]{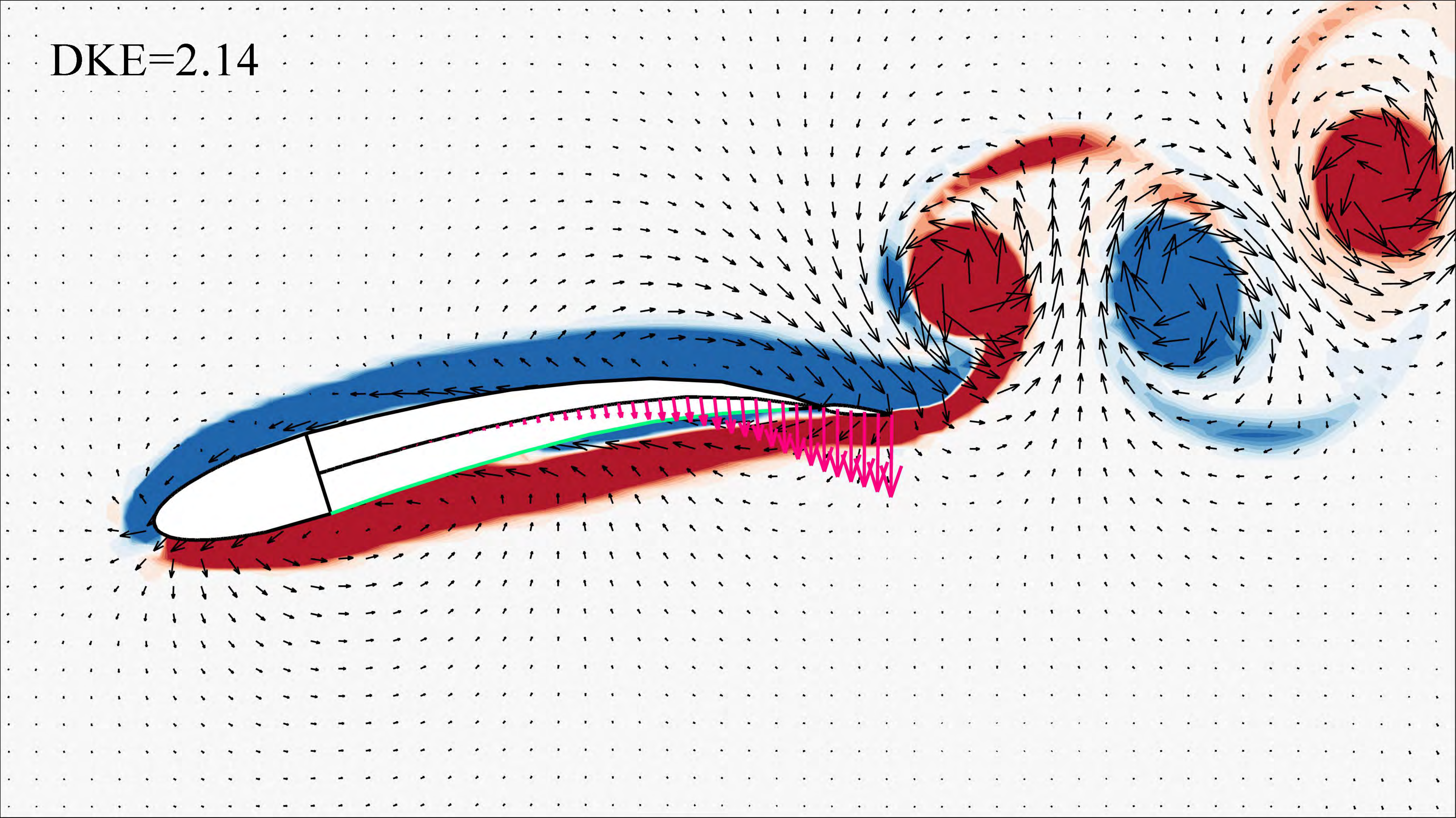}}
\subfigure[$t = 0.5T$($1-1-100$)]{
	\includegraphics[width=0.4\linewidth]{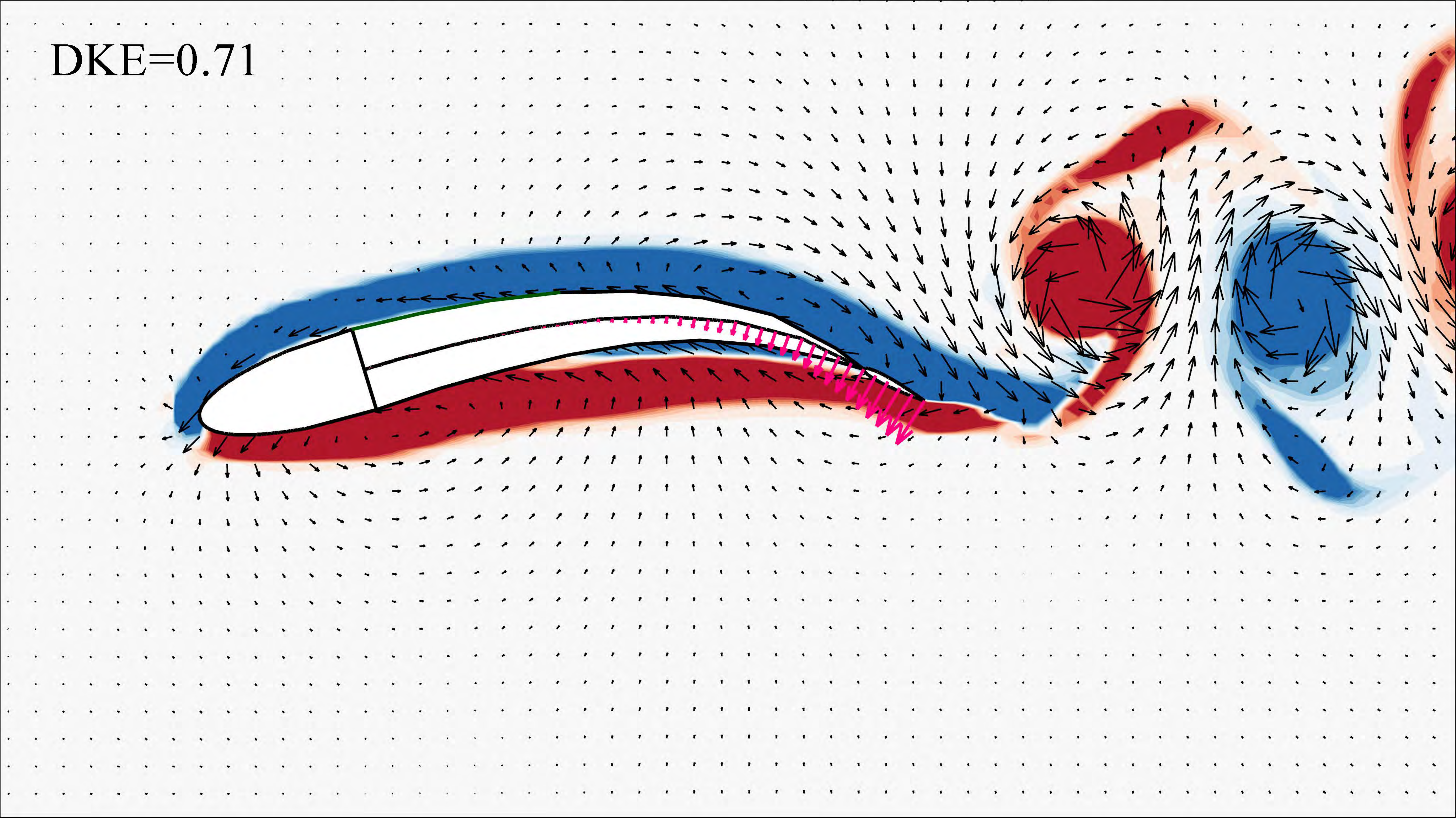}}
\subfigure[$t = 0.5T$($1-1-250$)]{
	\includegraphics[width=0.4\linewidth]{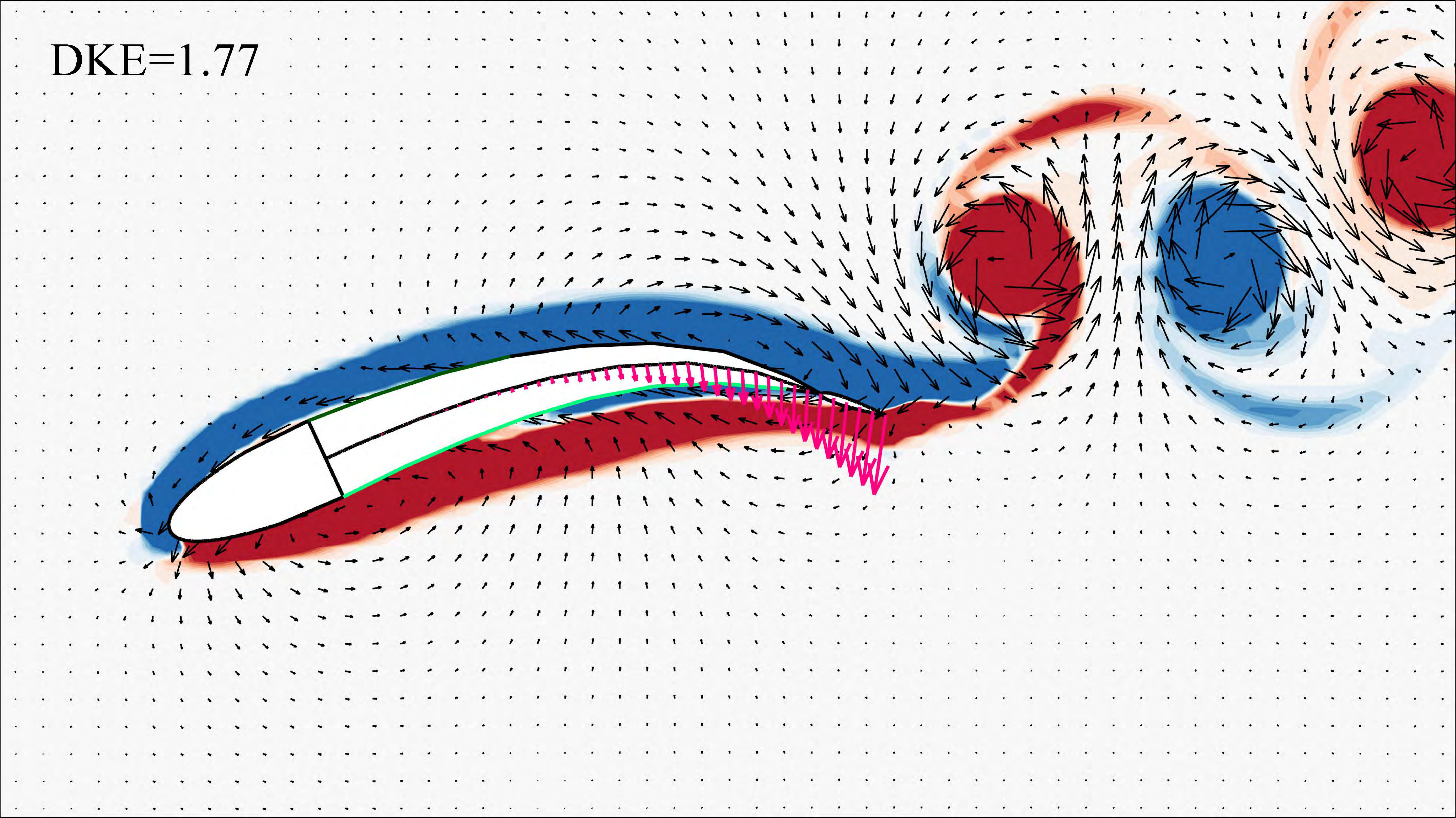}}
\subfigure[]{
	\includegraphics[width=5in]{color_bar_2-eps-converted-to.pdf}}
	\caption{(Continued.)}
\label{pic/f21}
\end{figure}

\subsubsection{The functional role of axial muscle phase lag in undulatory swimming}

\begin{figure}[htbp]
	\centering 
	\subfigbottomskip=2pt 
	\subfigcapskip=-5pt 
	\subfigure[Polar contour plots of swimming speed]{
		\includegraphics[width=0.48\linewidth]{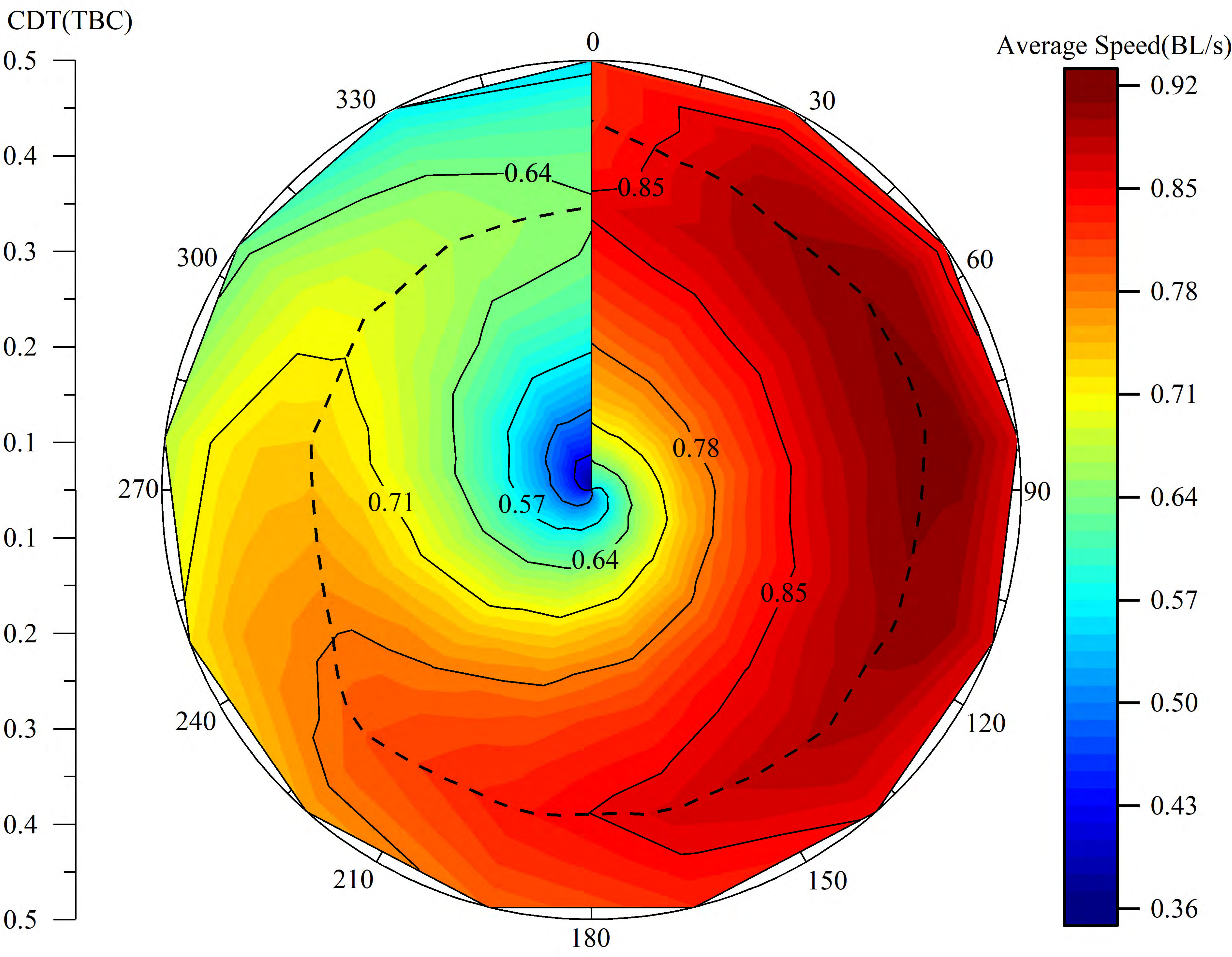}}
	\subfigure[Polar contour plots of muscular work]{
		\includegraphics[width=0.48\linewidth]{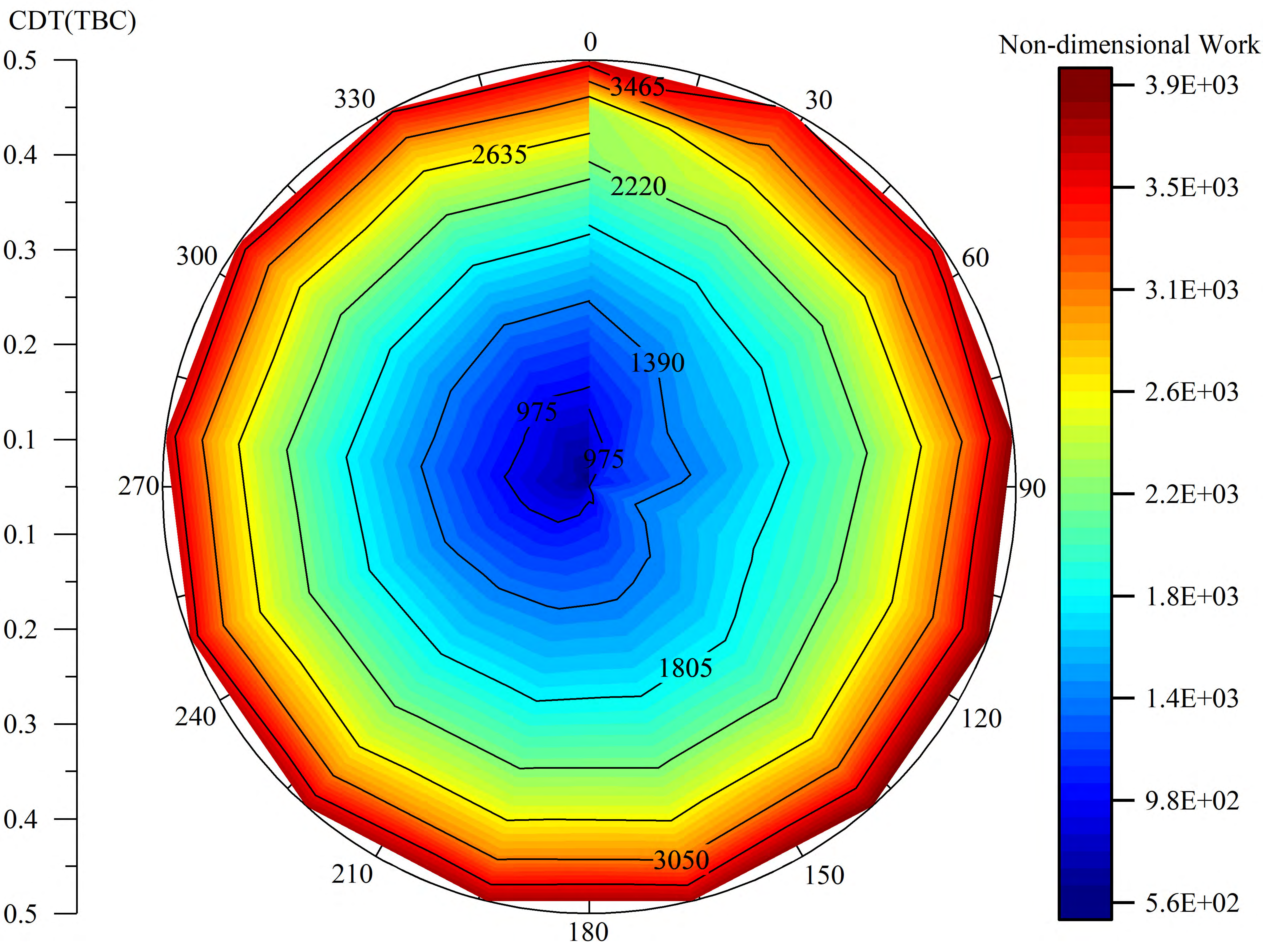}}
	\subfigure[Polar contour plots of RL reward]{
		\includegraphics[width=0.48\linewidth]{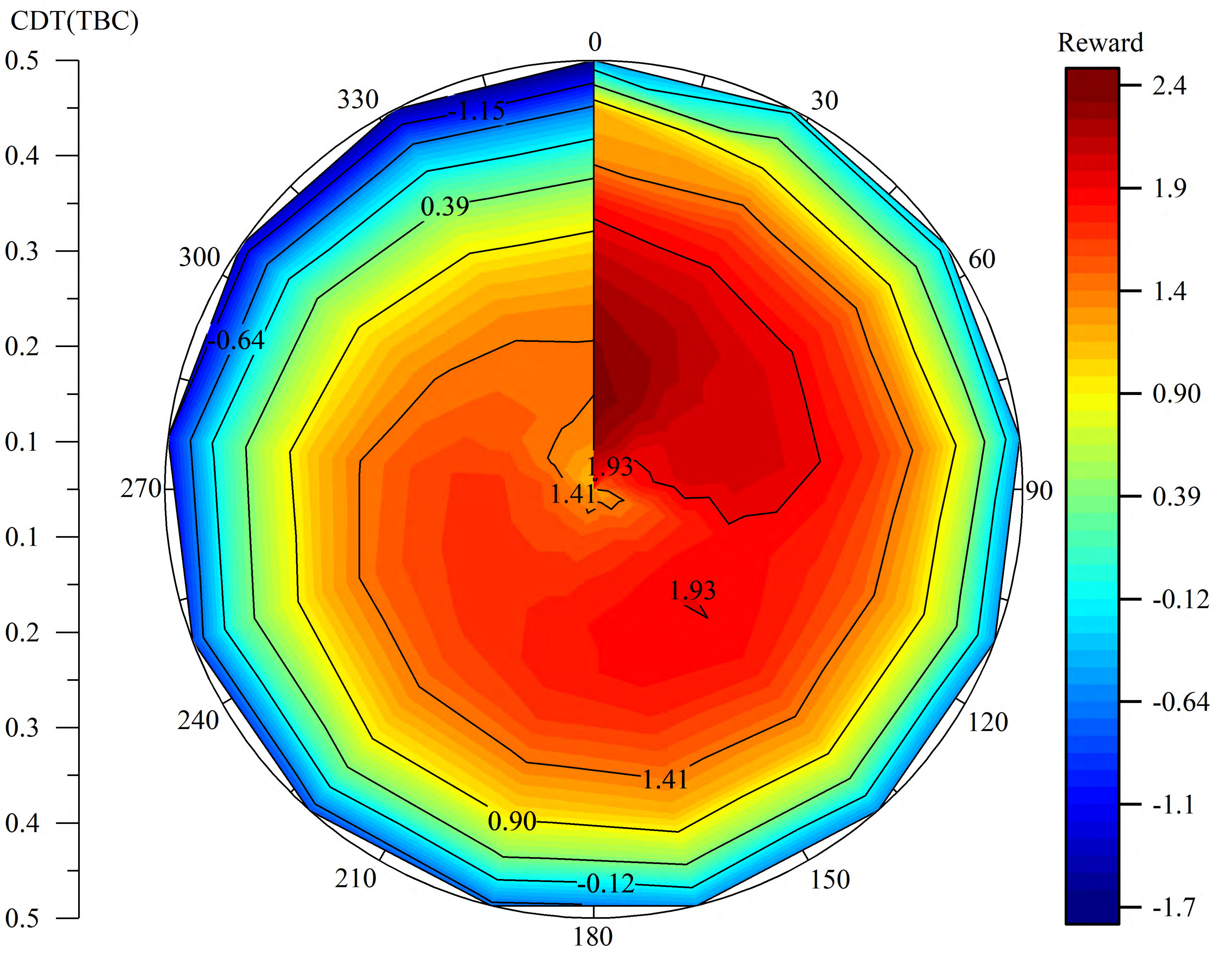}}
	\caption{Polar contour plots of swimming speed, muscular work, and reinforcement learning reward under different initial phases $\Phi$ and contraction durations (CDT).}
	\label{pic/f22}
\end{figure}

To investigate the role of muscle phase lag $\Phi$ in undulatory propulsion, we selected additional representative cases from the DRL training dataset. Figure \ref{pic/f22}-a presents a polar contour plot showing the distribution of average propulsion speed across the parametric space defined by contraction duration time (CDT) and muscle phase lag. Here, the phase lag is defined as: $\Phi = \frac{t_{\text{act}}}{0.5 \times \text{TBC}}$, where TBC denotes the characteristic tail-beat cycle of the fish. As shown in figure \ref{pic/f22}-a, for any fixed value of $\Phi$, propulsion speed initially rises with increasing CDT before eventually declining. The ridge line, which links the points of maximum propulsion speed for each radial direction, forms an inward spiral moving toward the center of the polar plot as $\Phi$ increases. When CDT is relatively short (i.e., less than 0.3 TBC), the mean propulsion speed decreases monotonically with increasing $\Phi$. However, when CDT exceeds 0.3 TBC, the relationship becomes non-monotonic: the mean propulsion speed initially increases with $\Phi$ before declining. The highest propulsion speed region (indicated by the dark red area in the plot) corresponds to a phase lag range of $\phi = 60^\circ\text{--}120^\circ$ and a moderately long contraction duration (CDT = 0.35 TBC--0.45 TBC). In contrast, the lowest propulsion speed region (deep blue "valley" area) appears when $\Phi = 270^\circ\text{--}360^\circ$ and CDT is extremely short ($CDT < 0.1 TBC$). Figure \ref{pic/f22}-b shows a polar contour plot of active muscular work (non-dimensional) across the same CDT--$\Phi$ parametric space. The results indicate a clear quasi-radial gradient increase, forming a near conical structure, which suggests a strong positive correlation between active muscular work and contraction duration time. Figure \ref{pic/f22}-c illustrates the distribution of reinforcement learning rewards across the same parametric space. In this study, the reward function $R$ was defined as a composite objective balancing propulsion speed and energy consumption. The highest reward region in figure \ref{pic/f22}-c corresponds to the $L_{E}H_{T}$ region identified earlier, and it significantly outperforms other regions in terms of average reward. This region is characterized by a small phase lag and short CDT, with its underlying energy-saving mechanism discussed in detail in section \ref{subsubsec::energysave}. Although the peak propulsion speed region identified in figure \ref{pic/f22}-a attains a moderate average reward ($\approx$ 1.2), it is notably lower than that of the $L_{E}H_{T}$ region, suggesting that while high-speed performance is attainable, it incurs a significantly higher energy cost. Additionally, holding CDT constant (i.e., along a fixed radial distance), the reward consistently declines as the muscle phase lag $\Phi$ increases. This trend suggests that excessive phase lag hinders the development of energy-efficient propulsion strategies.

\begin{figure}[htbp]
	\centering
	\includegraphics[width=5in]{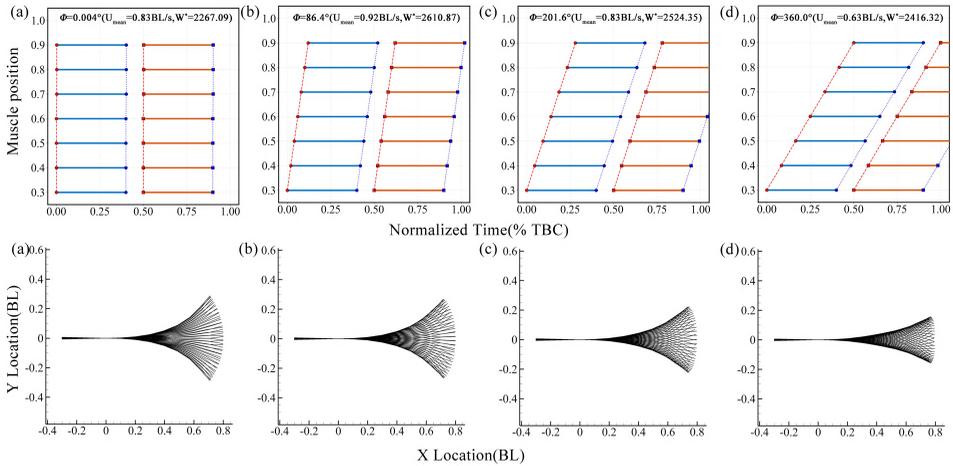}
	\caption{Comparison of muscle activation patterns and the envelop of body waves under different phase lag $\Phi$.}
	\label{pic/f23}
\end{figure}

Figure \ref{pic/f23} presents a comparison of muscle activation patterns and corresponding body wave envelope under different values of phase lag $\Phi$, with contraction duration time fixed at CDT=0.4 TBC. Given the similar CDT, the energy consumption among these modes did not differ significantly. In terms of body wave morphology, the synchronous activation mode produced a rigidly divergent waveform (figure \ref{pic/f23}-a), whereas modes with non-zero phase lag exhibited flexible mesh-like waveforms (figure \ref{pic/f23}-b). The synchronously activated waveform produced larger oscillation amplitudes but resulted in lower average swimming speed compared to the phase lag activation modes (figures \ref{pic/f23} a-b). This is because when bilateral muscles contracted simultaneously, both the anterior and posterior regions of the body swung almost in unison, leading to substantial lateral hydrodynamic losses (as shown in figure \ref{pic/f14}). As $\Phi$ increased, the amplitude of the flexible mesh-like body wave became progressively dampened, resulting in a corresponding decline in swimming speed (figures \ref{pic/f23} b-d).

\begin{figure}[htbp]
	\centering 
	\subfigbottomskip=2pt 
	\subfigcapskip=-5pt 
	\subfigure[$t = 0.2T$($\Phi=0.004^\circ $)]{
		\includegraphics[width=0.4\linewidth]{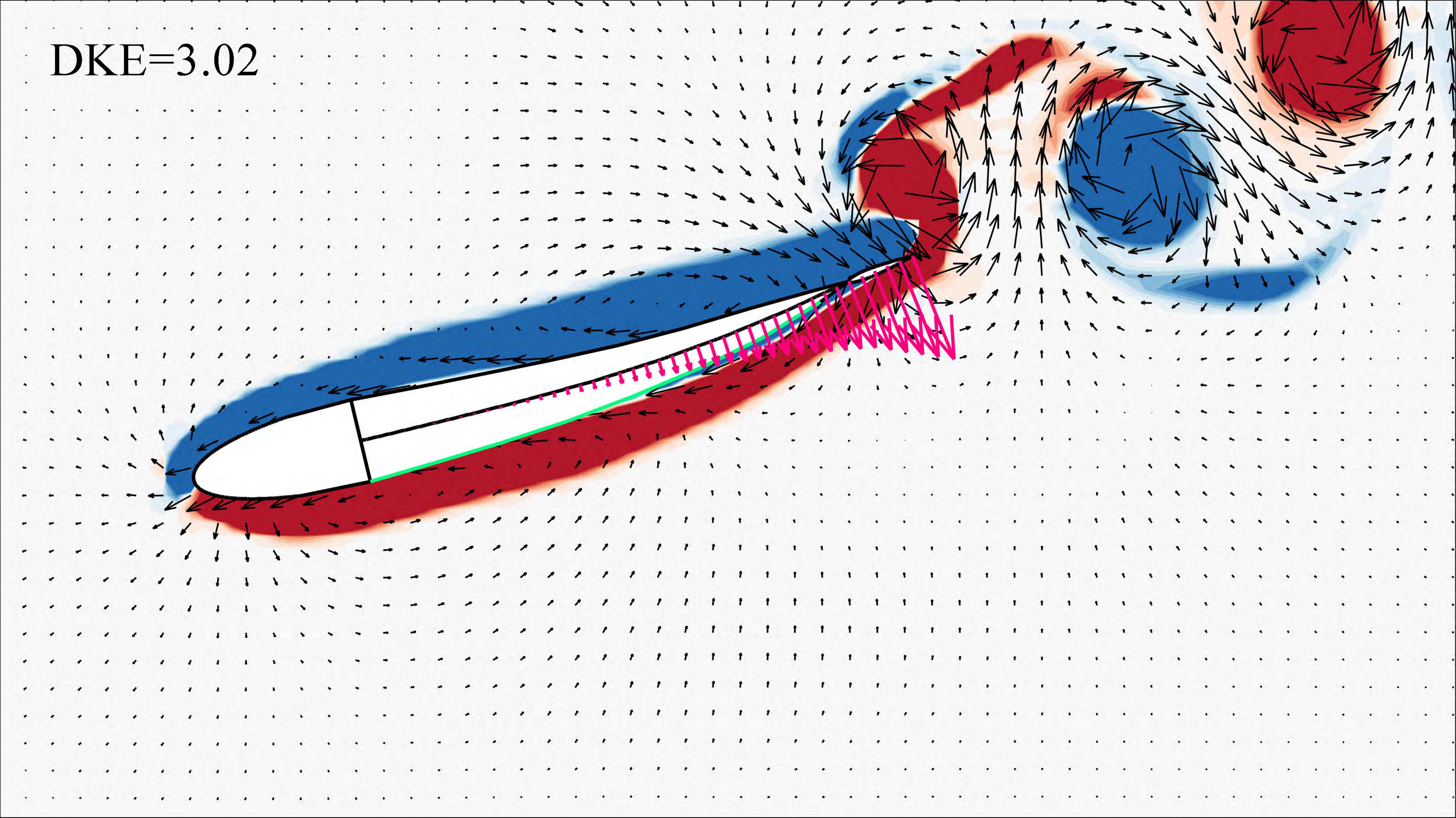}}
	\subfigure[$t = 0.2T$($\Phi=86.4^\circ $)]{
		\includegraphics[width=0.4\linewidth]{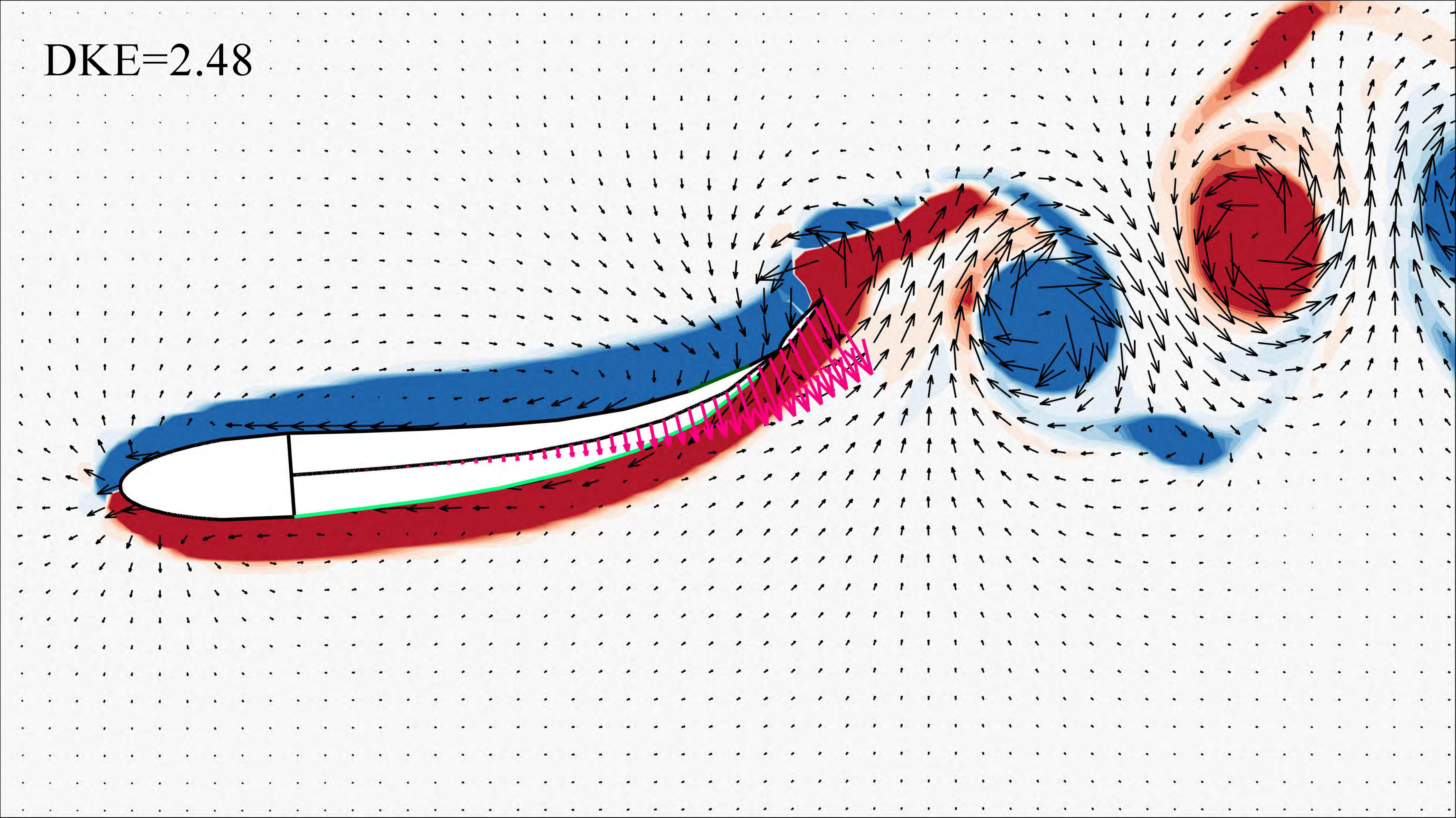}}
	\subfigure[$t = 0.4T$($\Phi=0.004^\circ $)]{
		\includegraphics[width=0.4\linewidth]{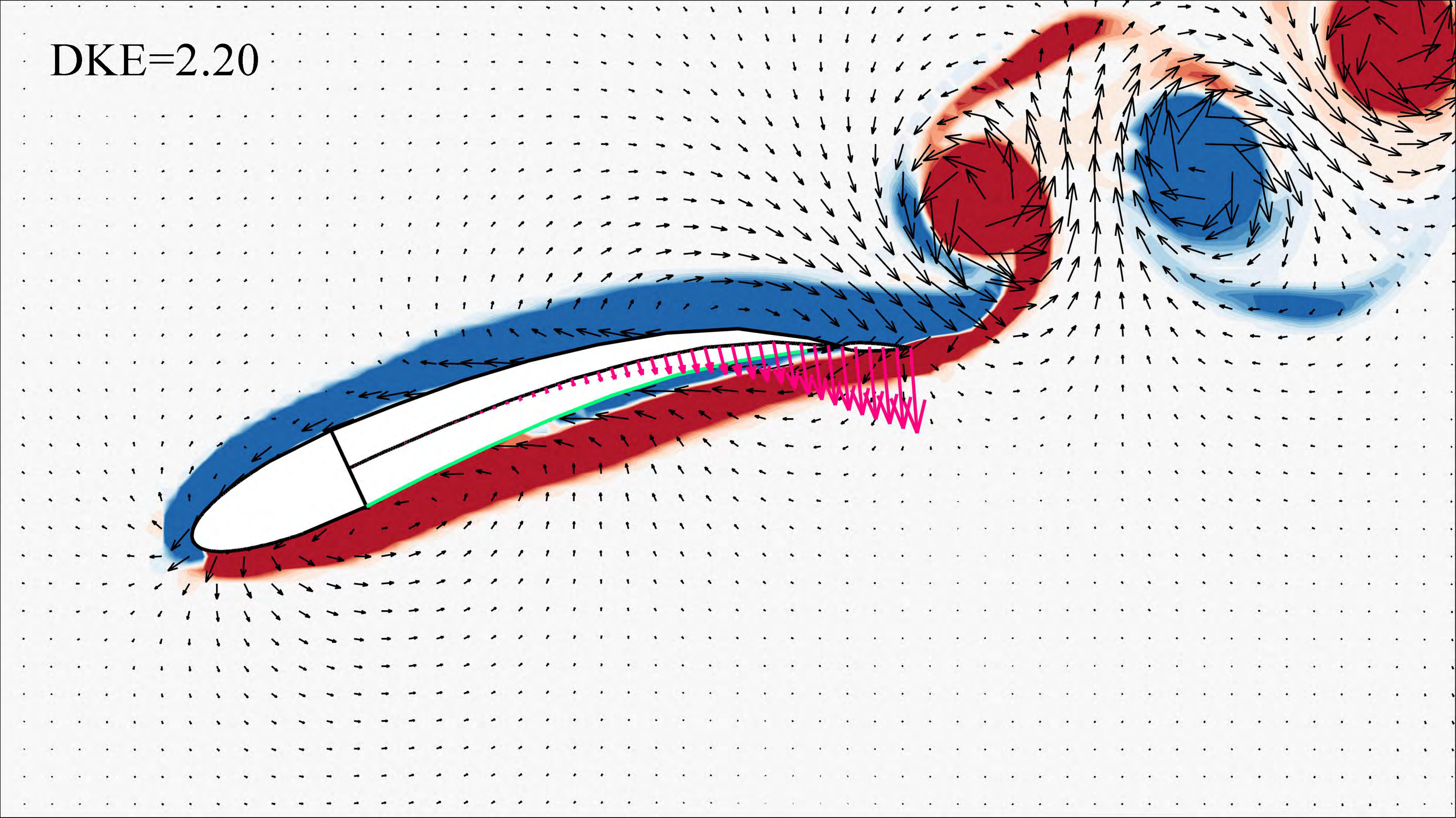}}
	\subfigure[$t = 0.4T$($\Phi=86.4^\circ $)]{
		\includegraphics[width=0.4\linewidth]{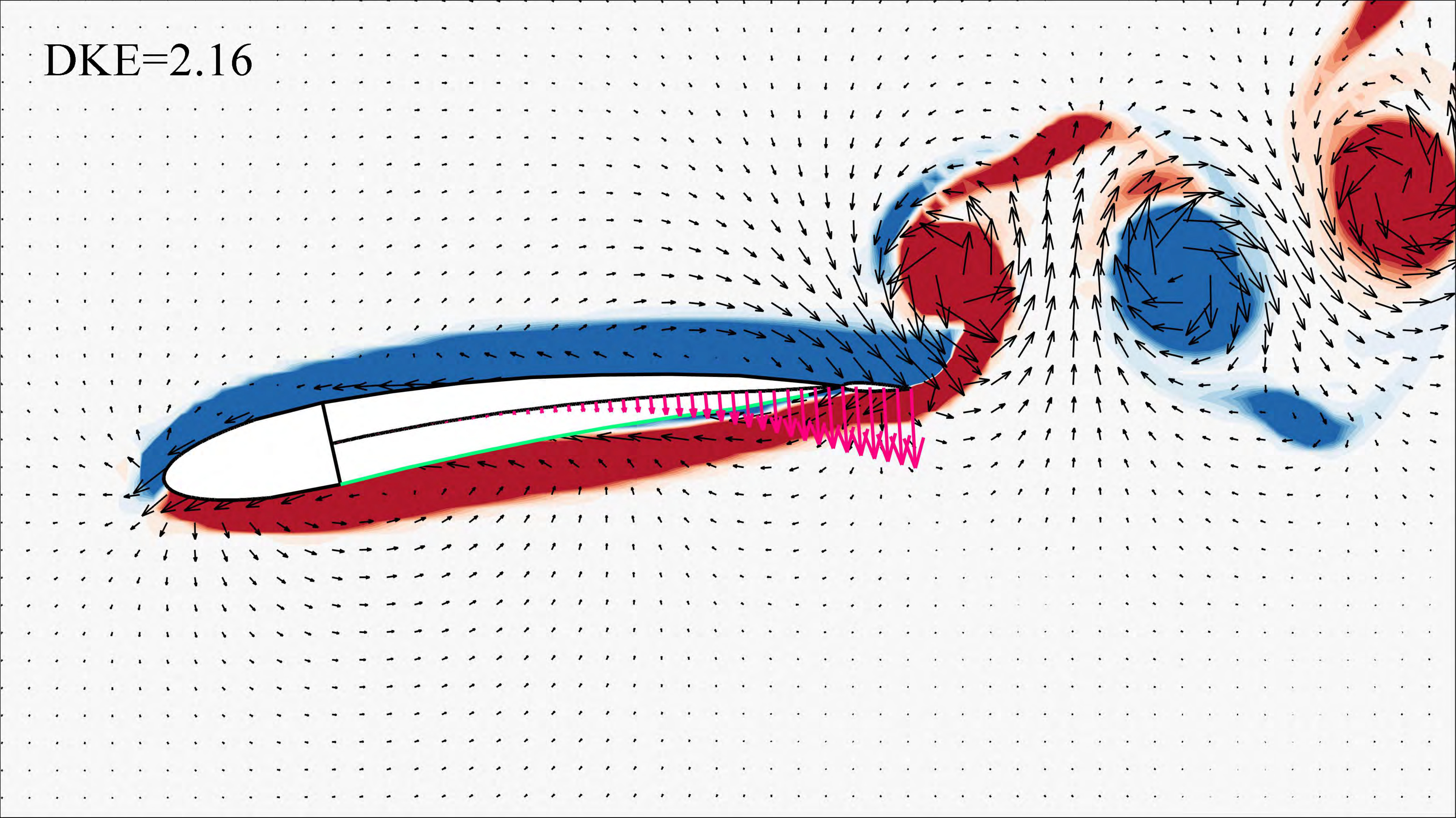}}
	\subfigure[$t = 0.6T$($\Phi=0.004^\circ $)]{
		\includegraphics[width=0.4\linewidth]{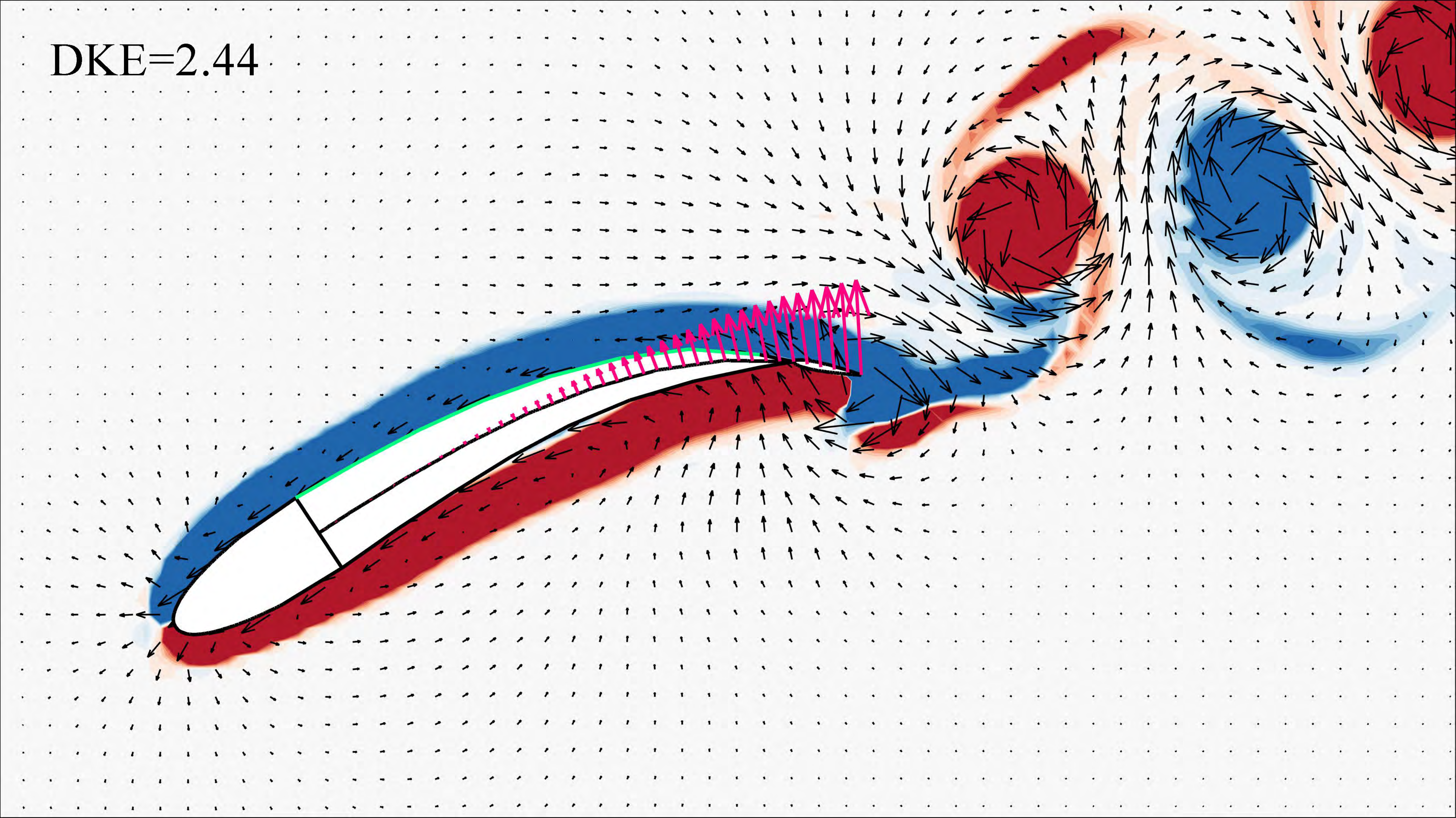}}
	\subfigure[$t = 0.6T$($\Phi=86.4^\circ $)]{
		\includegraphics[width=0.4\linewidth]{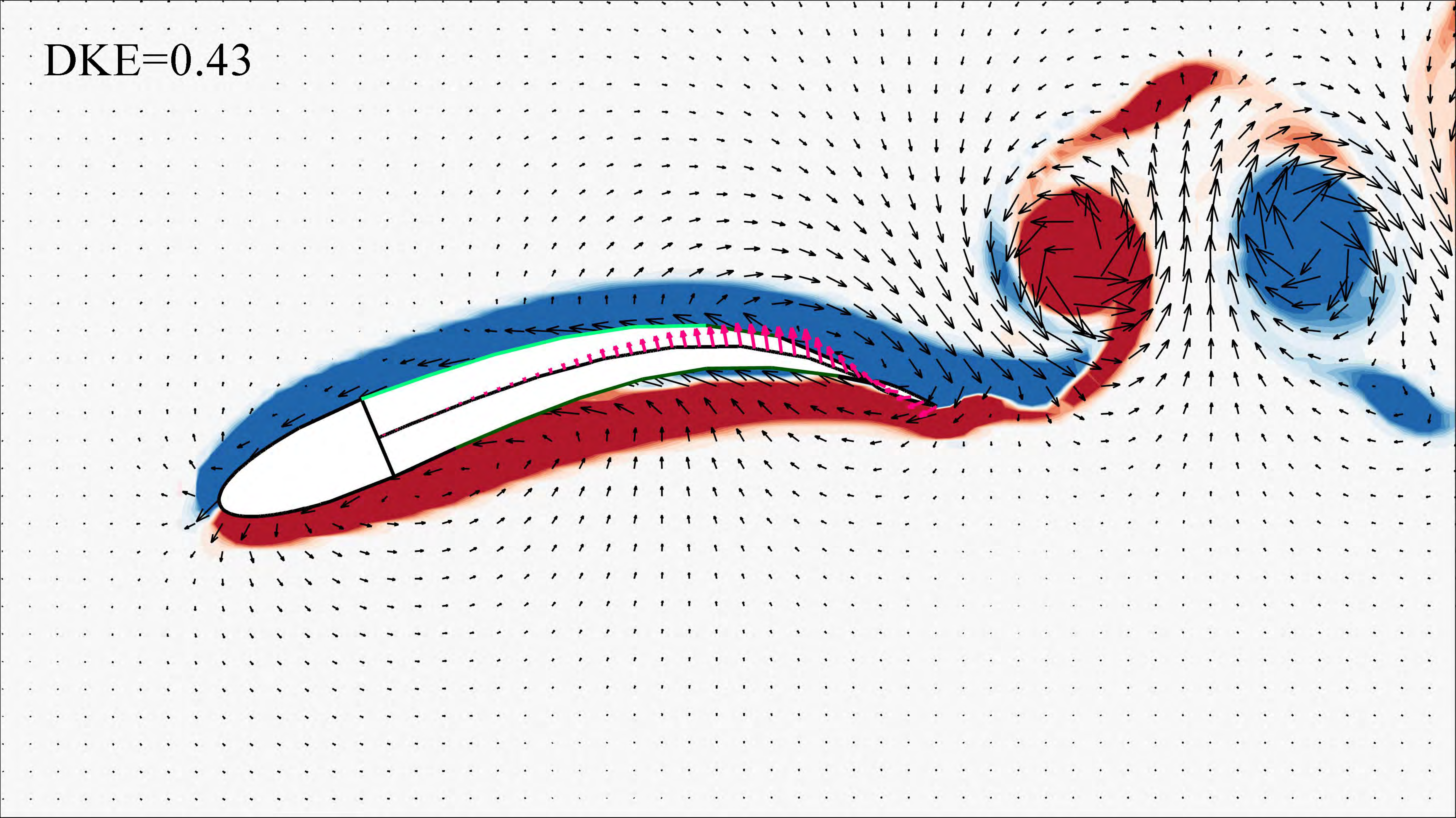}}
	\subfigure[$t = 0.8T$($\Phi=0.004^\circ $)]{
		\includegraphics[width=0.4\linewidth]{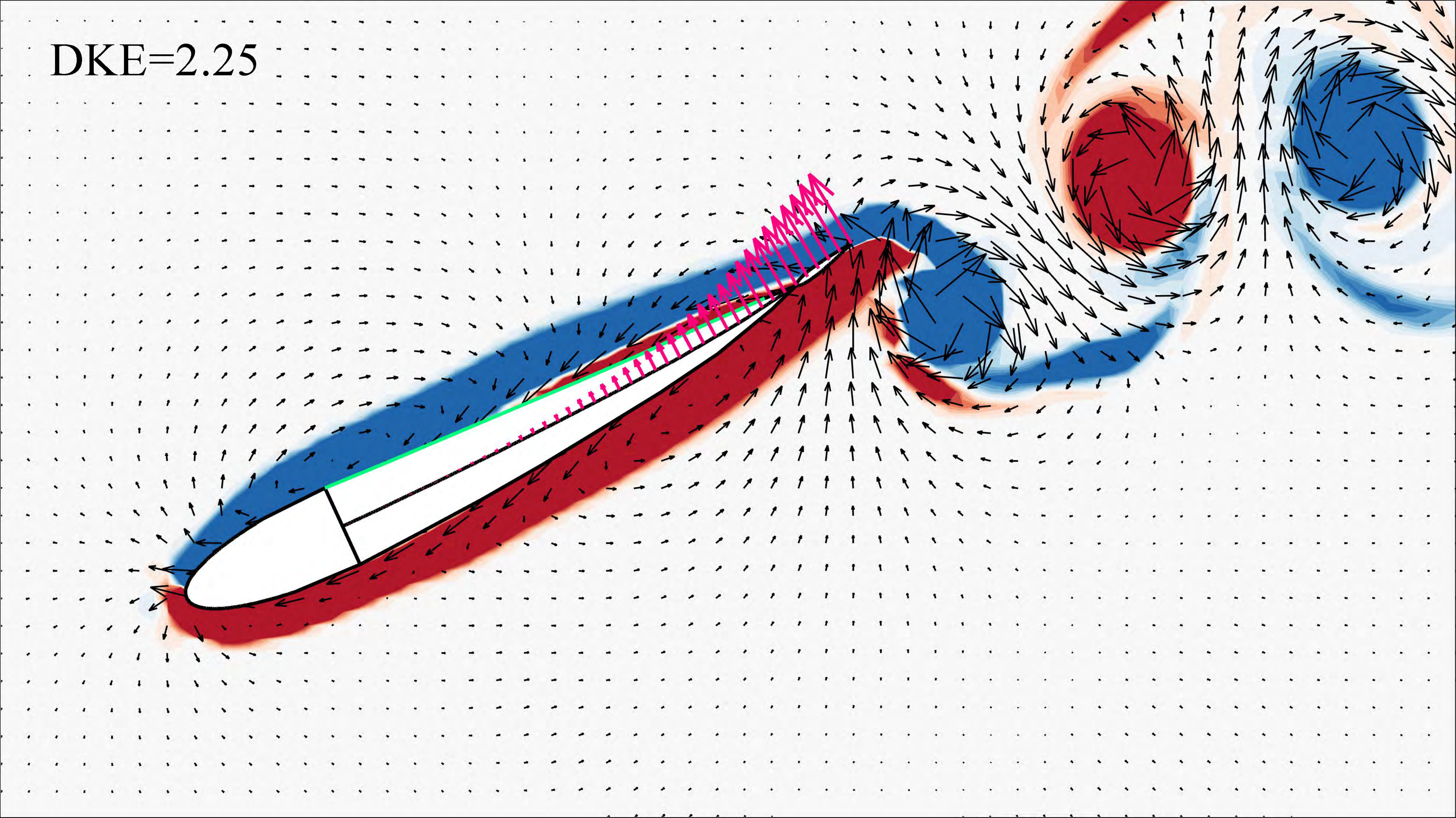}}
	\subfigure[$t = 0.8T$($\Phi=86.4^\circ $)]{
		\includegraphics[width=0.4\linewidth]{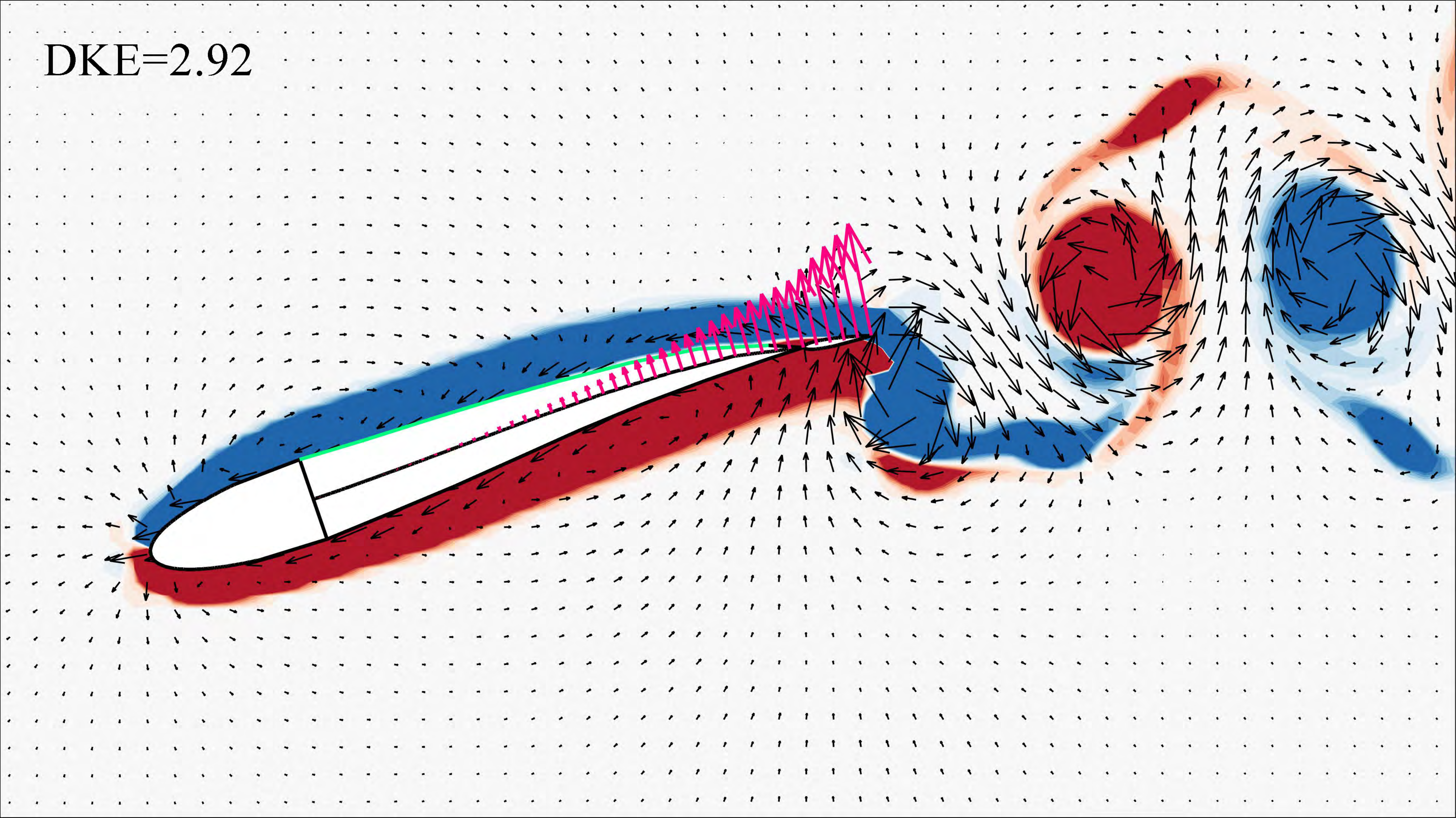}}
	\subfigure[$t = 1.0T$($\Phi=0.004^\circ $)]{
		\includegraphics[width=0.4\linewidth]{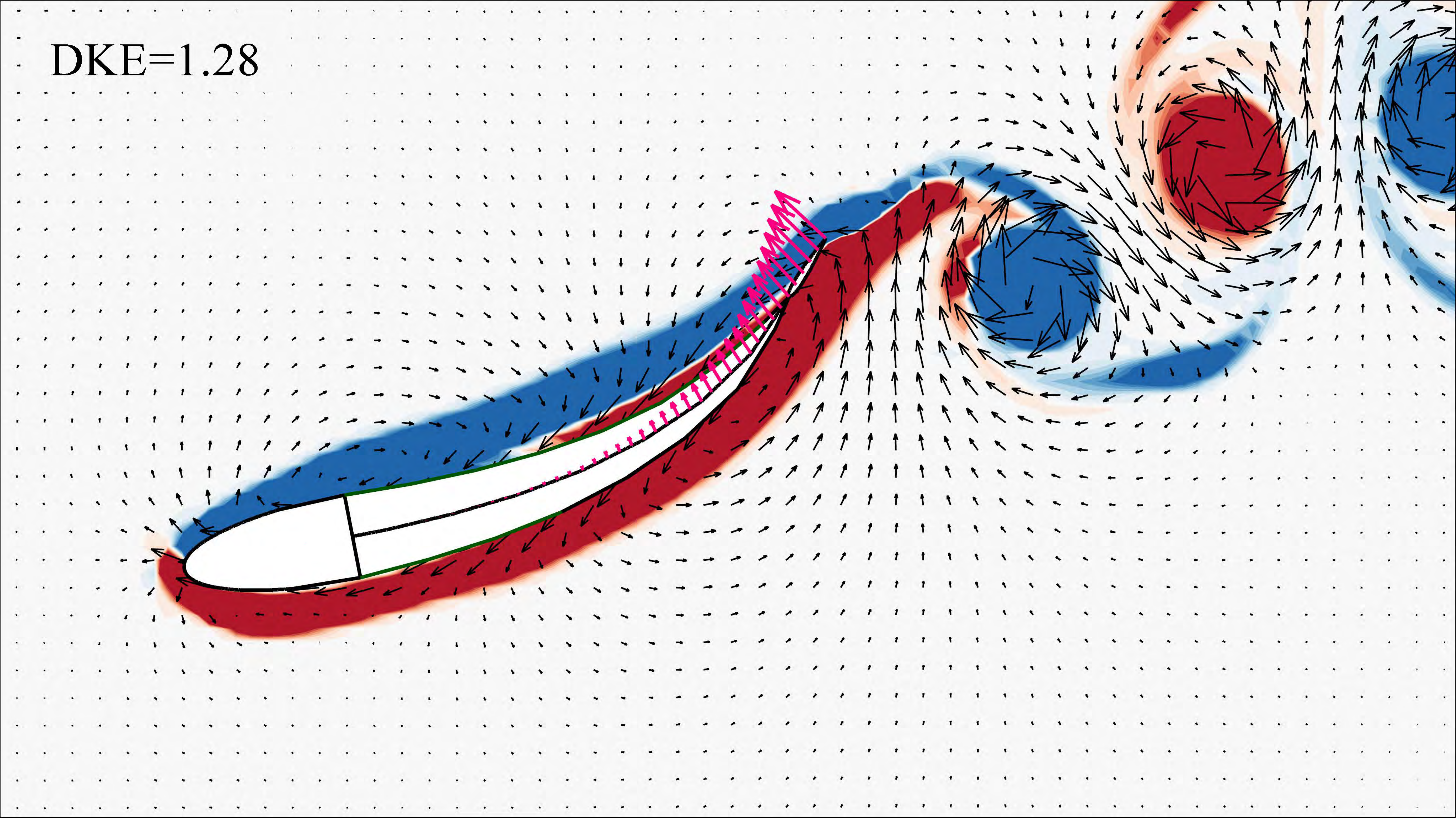}}
	\subfigure[$t = 1.0T$($\Phi=86.4^\circ $)]{
		\includegraphics[width=0.4\linewidth]{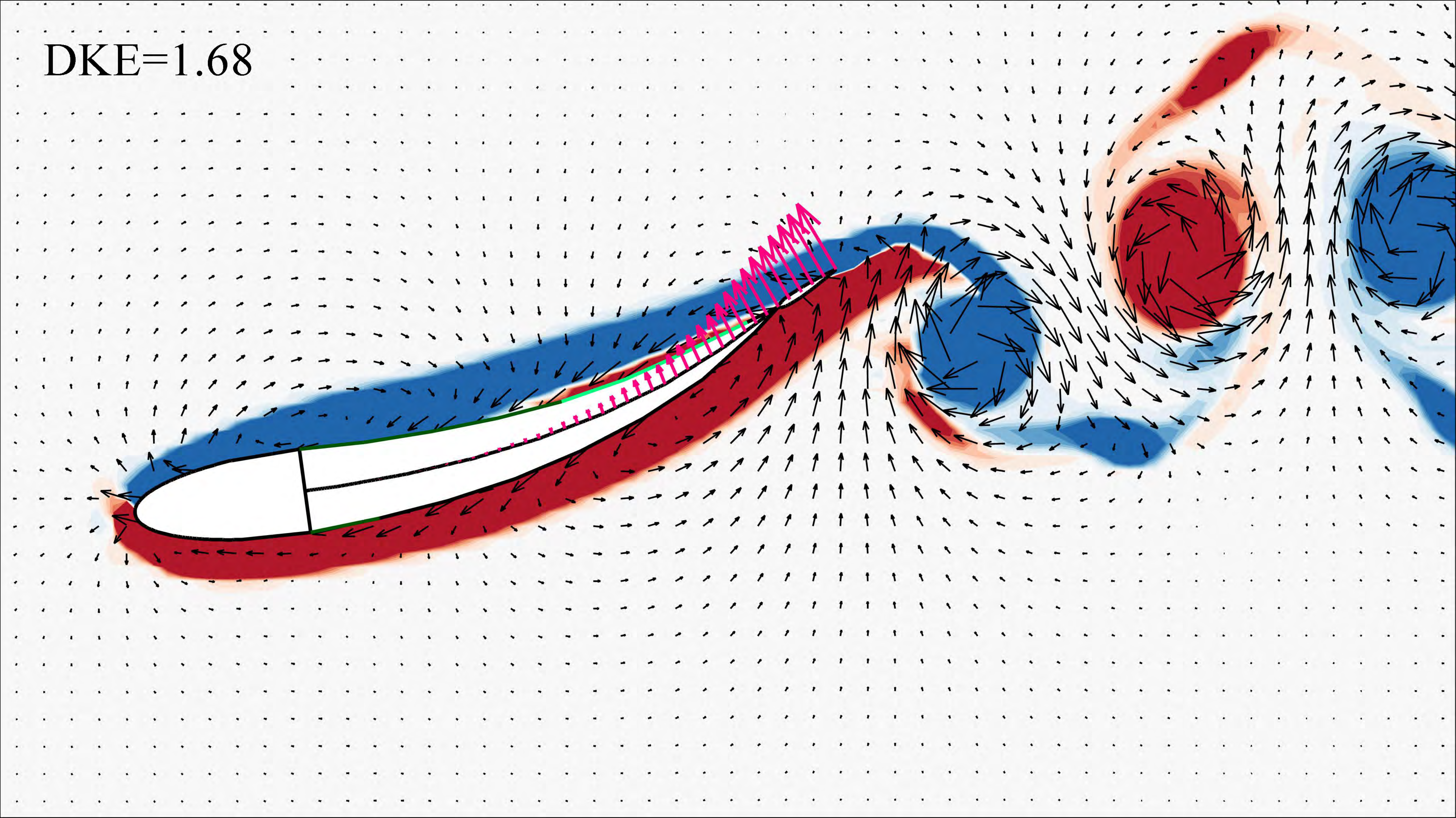}}
	\subfigure[]{
		\includegraphics[width=5in]{color_bar_2-eps-converted-to.pdf}}
	\caption{Comparison of instantaneous flow vorticity fields with varying $\Phi$ (black indicates inactive muscles, bright green indicates strongly activated muscles, pink arrows represent body deformation velocity vectors, and black arrows represent flow velocity vectors; T denotes the tail beat cycle).}
	\label{pic/f24}
\end{figure}

\begin{figure}[htbp]
	\centering 
	\subfigbottomskip=2pt 
	\subfigcapskip=-5pt 
	\subfigure[$t = 0.2T$($\Phi=201.6^\circ $)]{
		\includegraphics[width=0.4\linewidth]{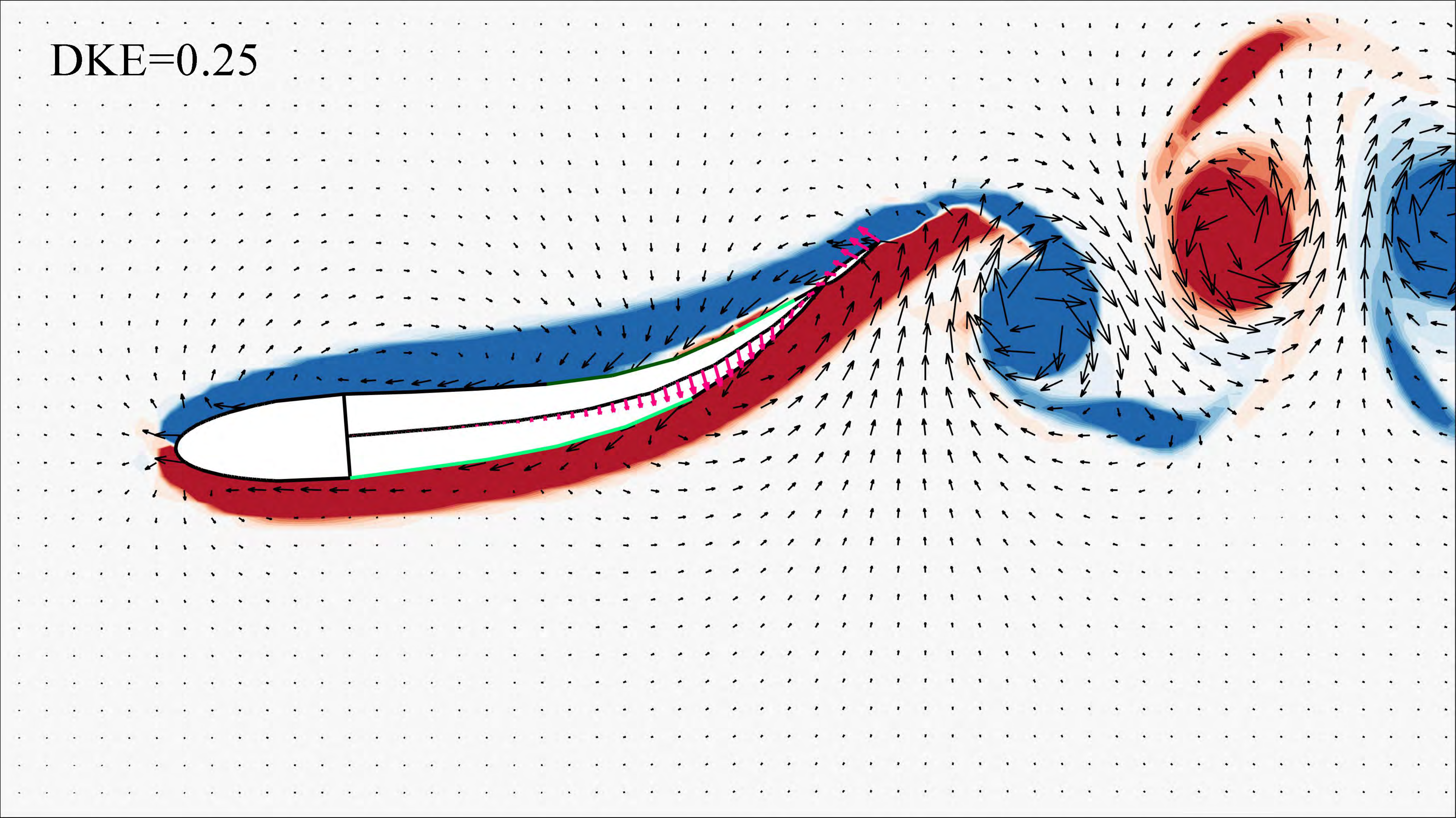}}
	\subfigure[$t = 0.2T$($\Phi=360.0^\circ $)]{
		\includegraphics[width=0.4\linewidth]{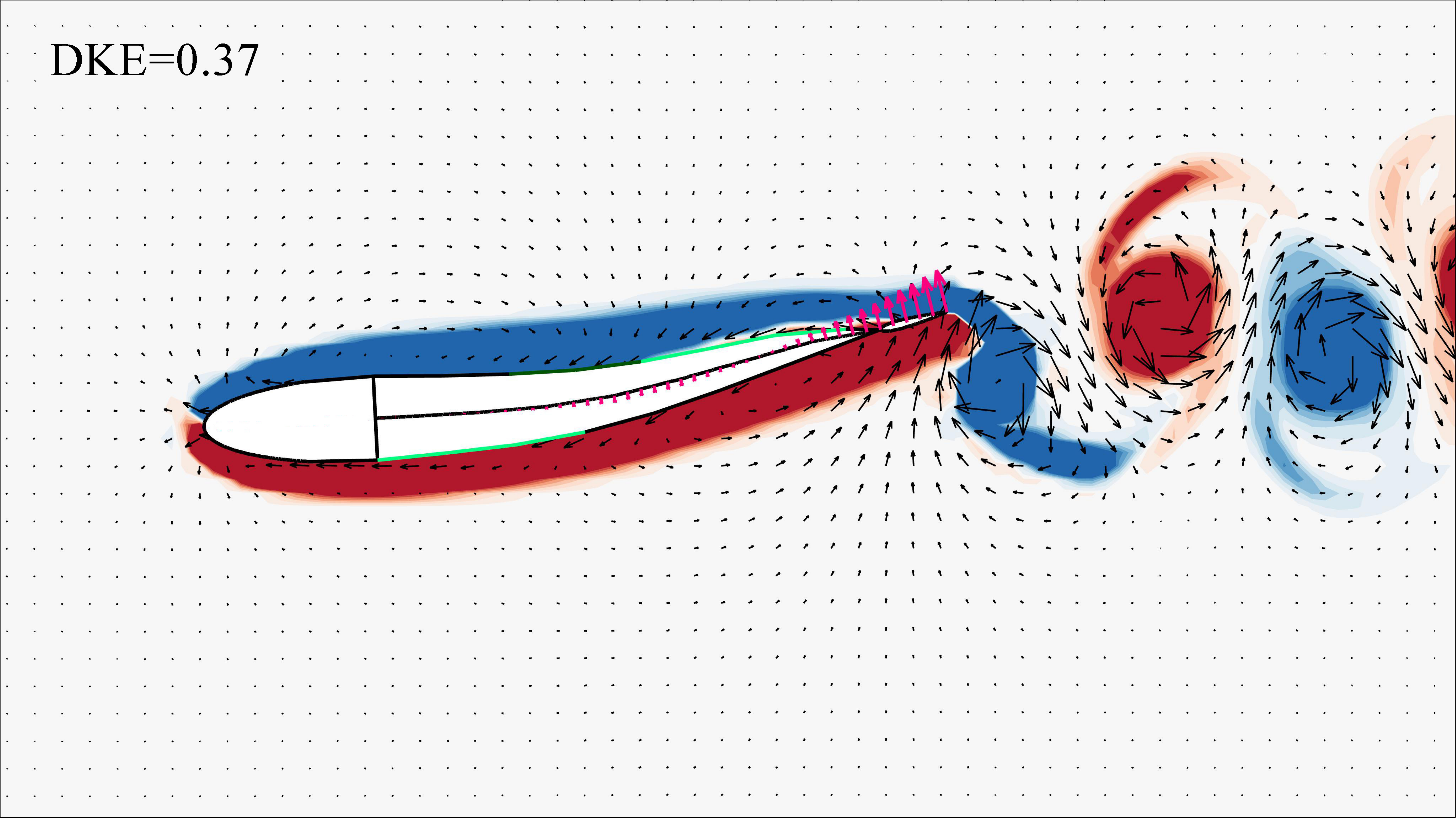}}
	\subfigure[$t = 0.4T$($\Phi=201.6^\circ $)]{
		\includegraphics[width=0.4\linewidth]{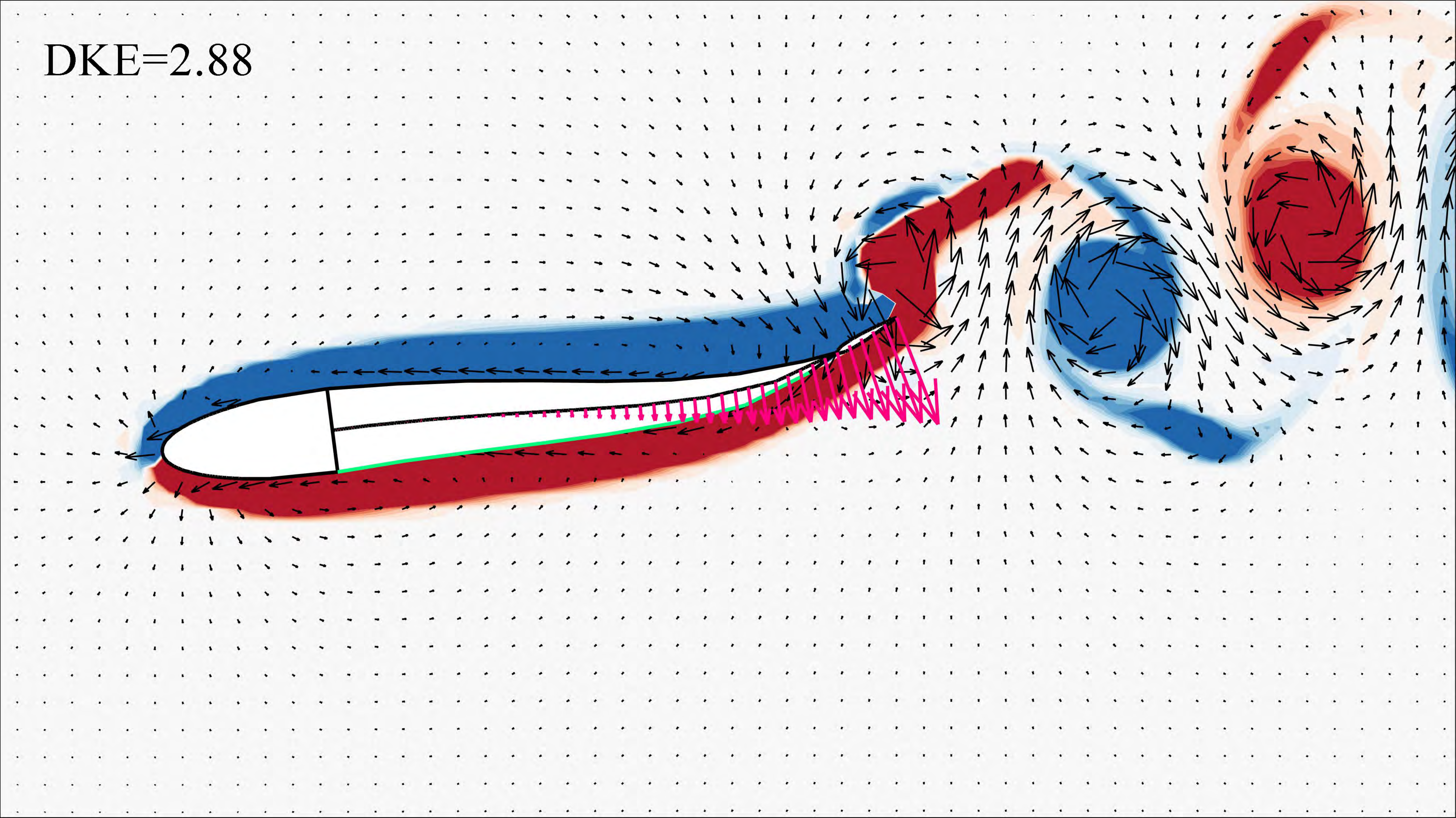}}
	\subfigure[$t = 0.4T$($\Phi=360.0^\circ $)]{
		\includegraphics[width=0.4\linewidth]{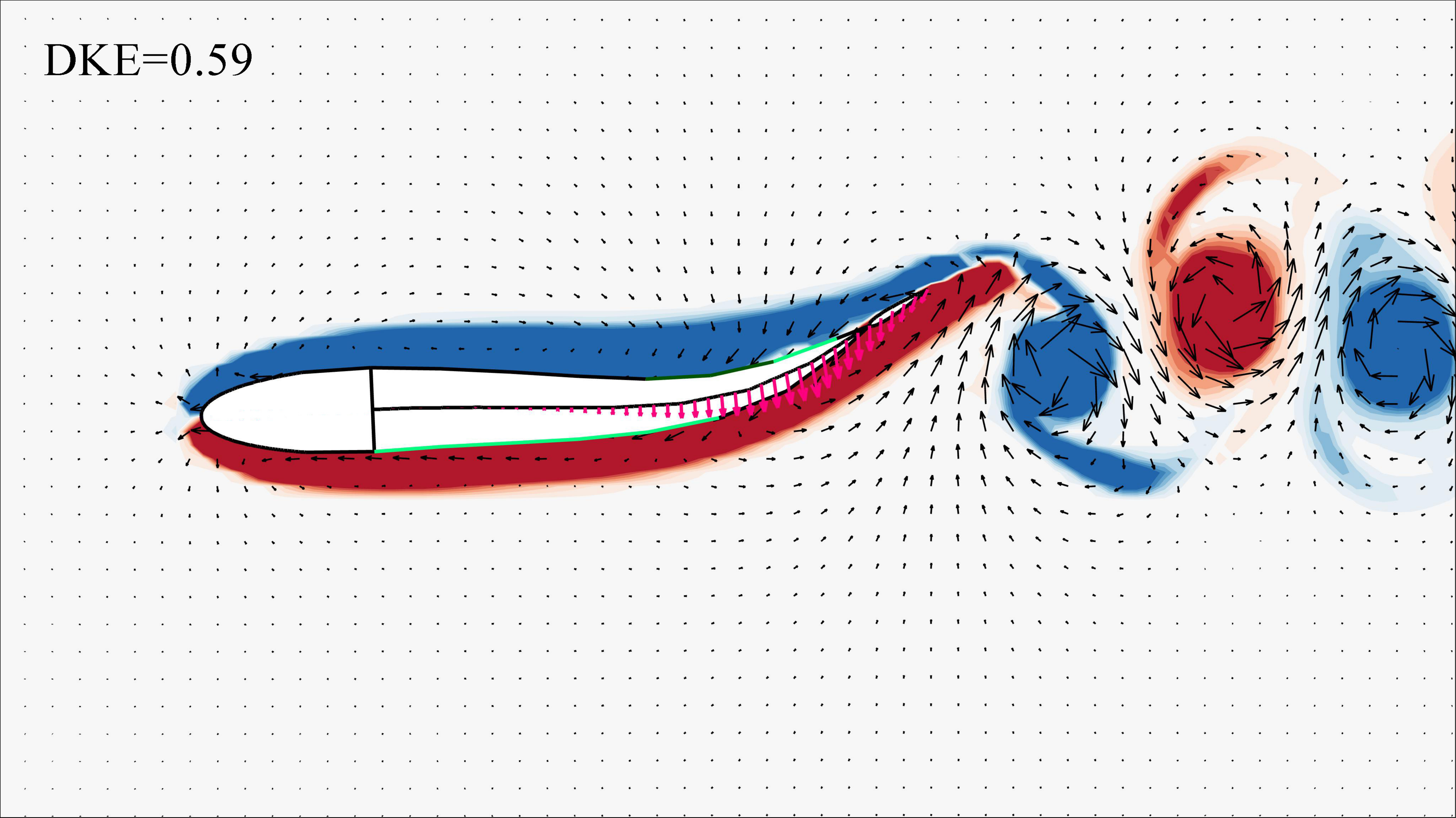}}
	\subfigure[$t = 0.6T$($\Phi=201.6^\circ $)]{
		\includegraphics[width=0.4\linewidth]{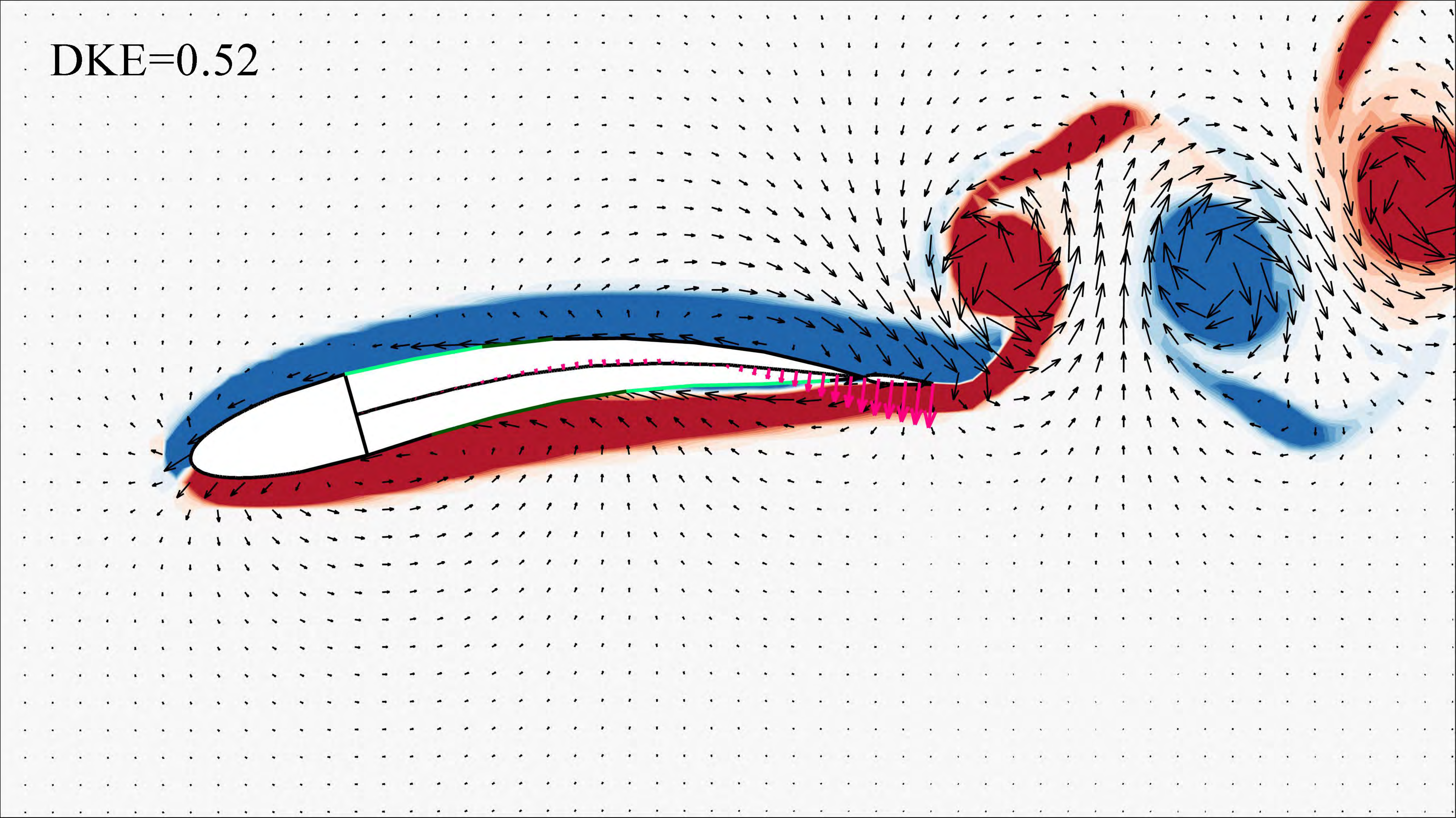}}
	\subfigure[$t = 0.6T$($\Phi=360.0^\circ $)]{
		\includegraphics[width=0.4\linewidth]{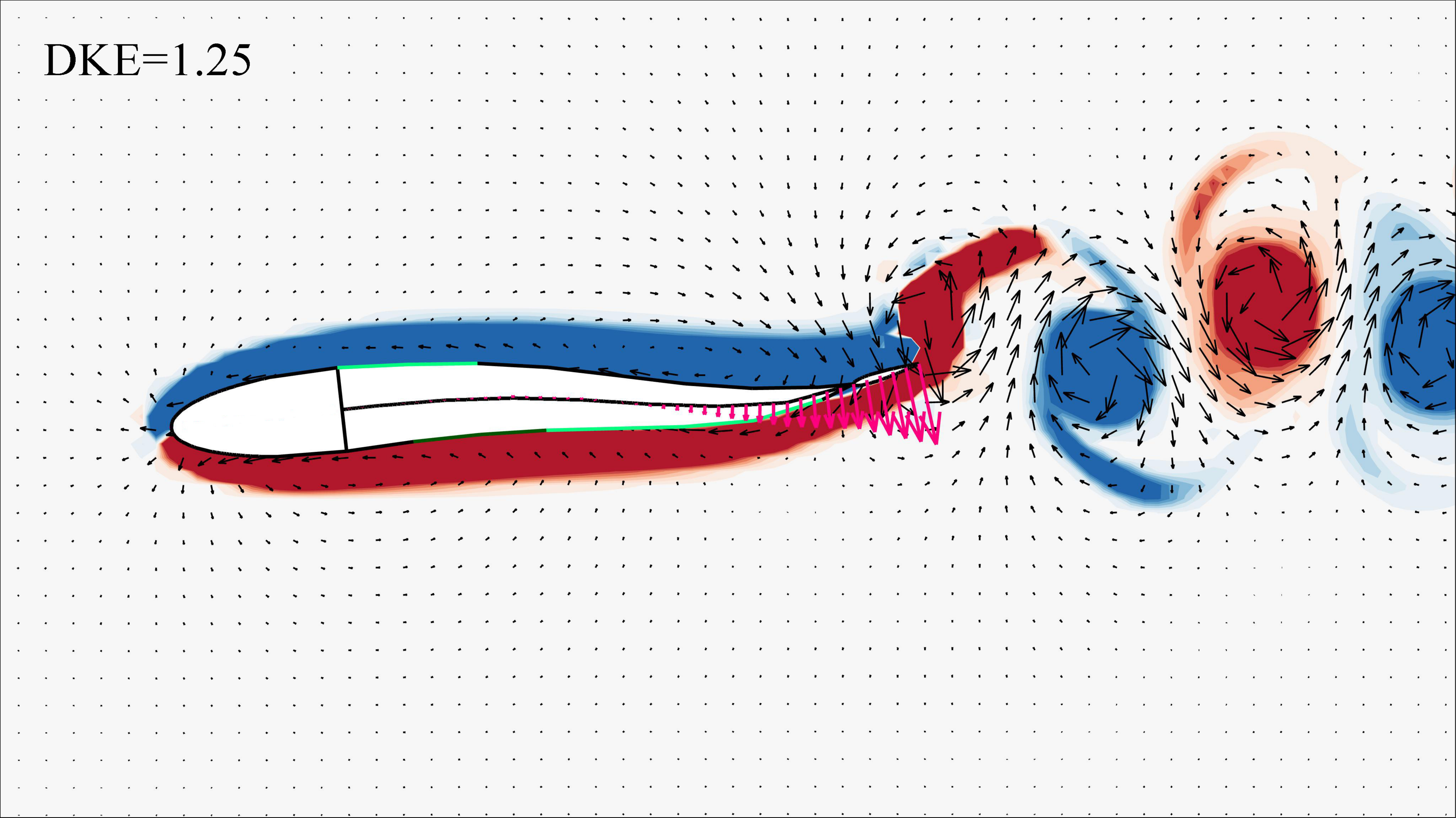}}
	\subfigure[$t = 0.8T$($\Phi=201.6^\circ $)]{
		\includegraphics[width=0.4\linewidth]{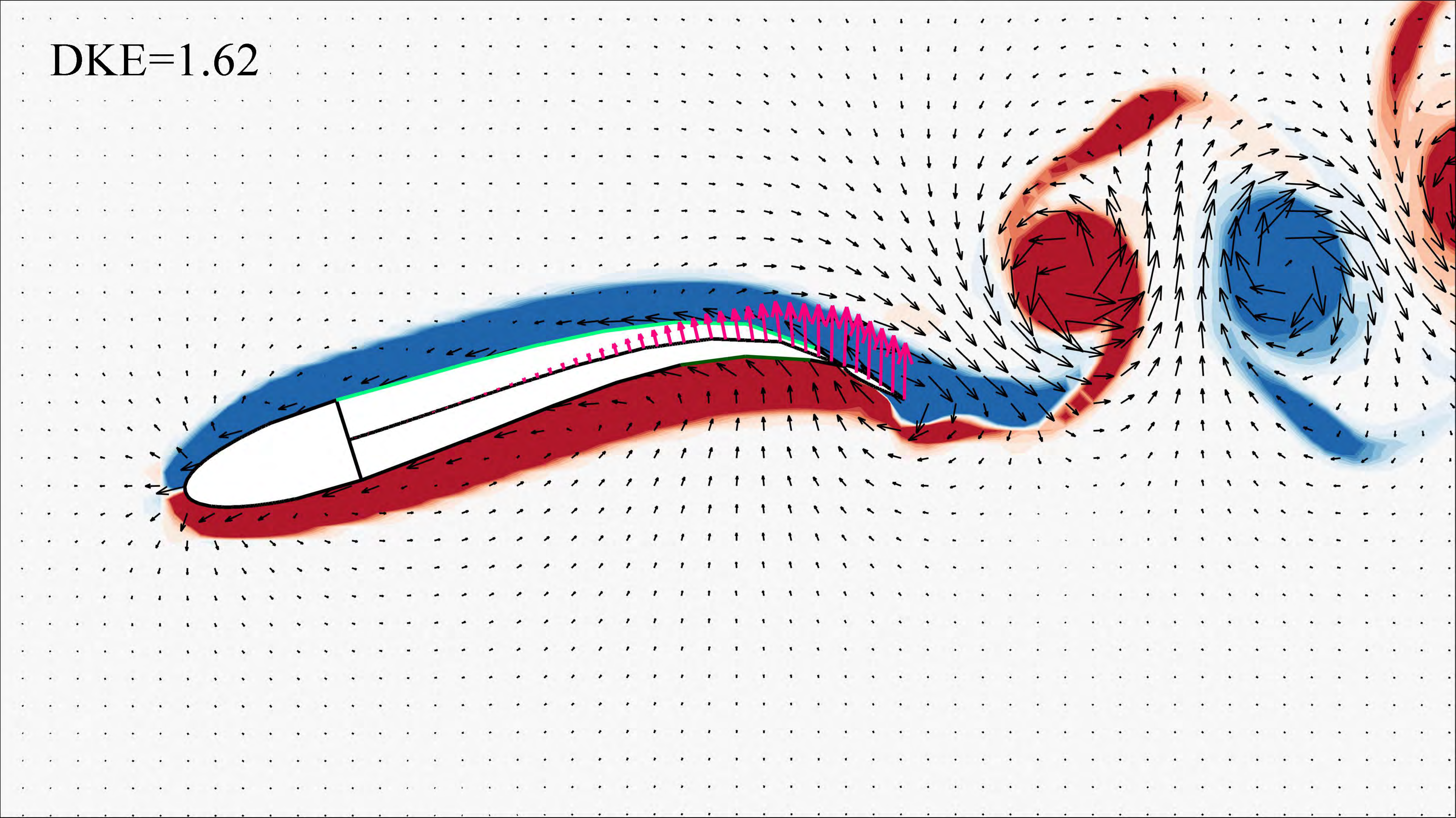}}
	\subfigure[$t = 0.8T$($\Phi=360.0^\circ $)]{
		\includegraphics[width=0.4\linewidth]{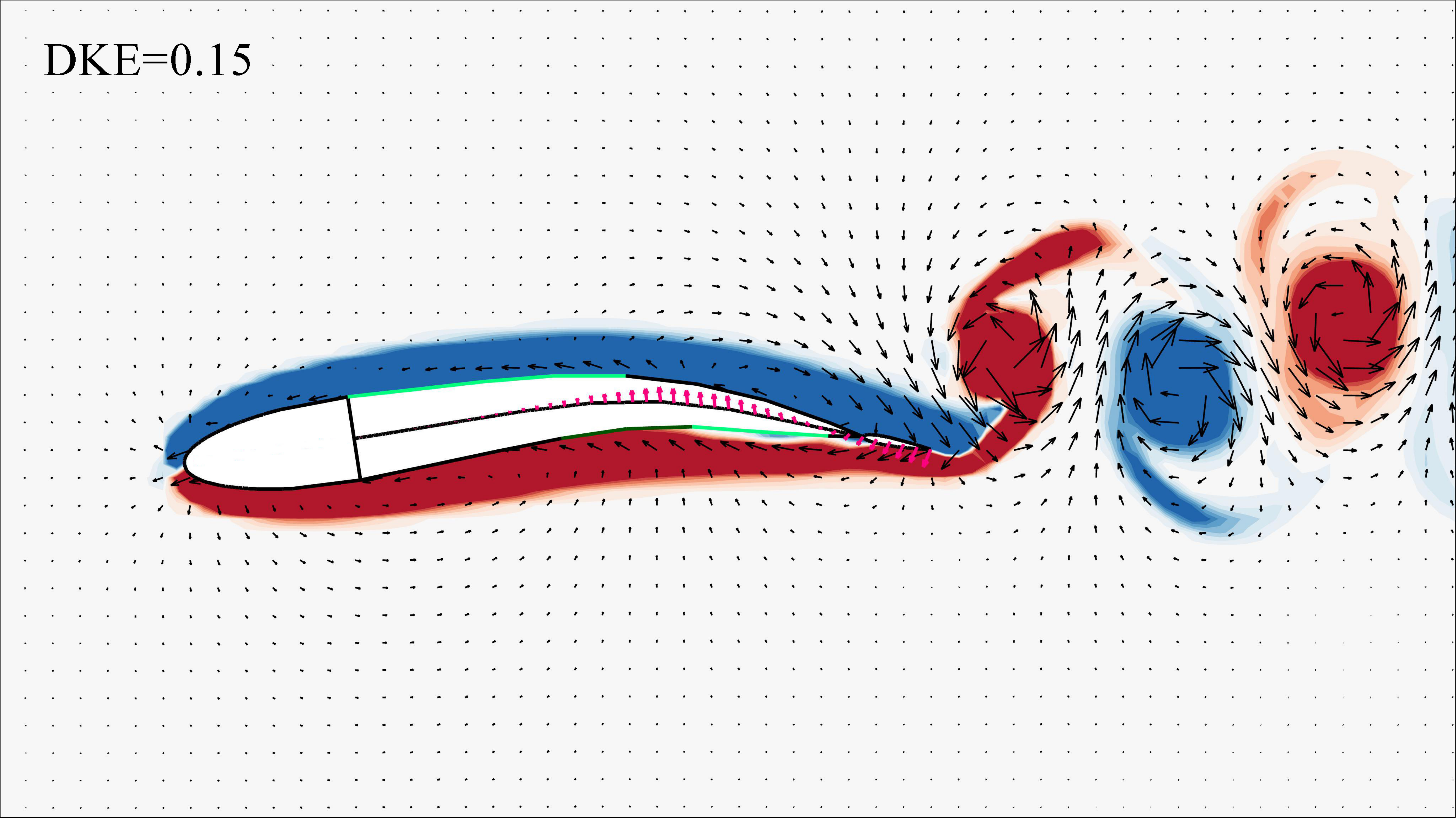}}
	\subfigure[$t = 1.0T$($\Phi=201.6^\circ $)]{
		\includegraphics[width=0.4\linewidth]{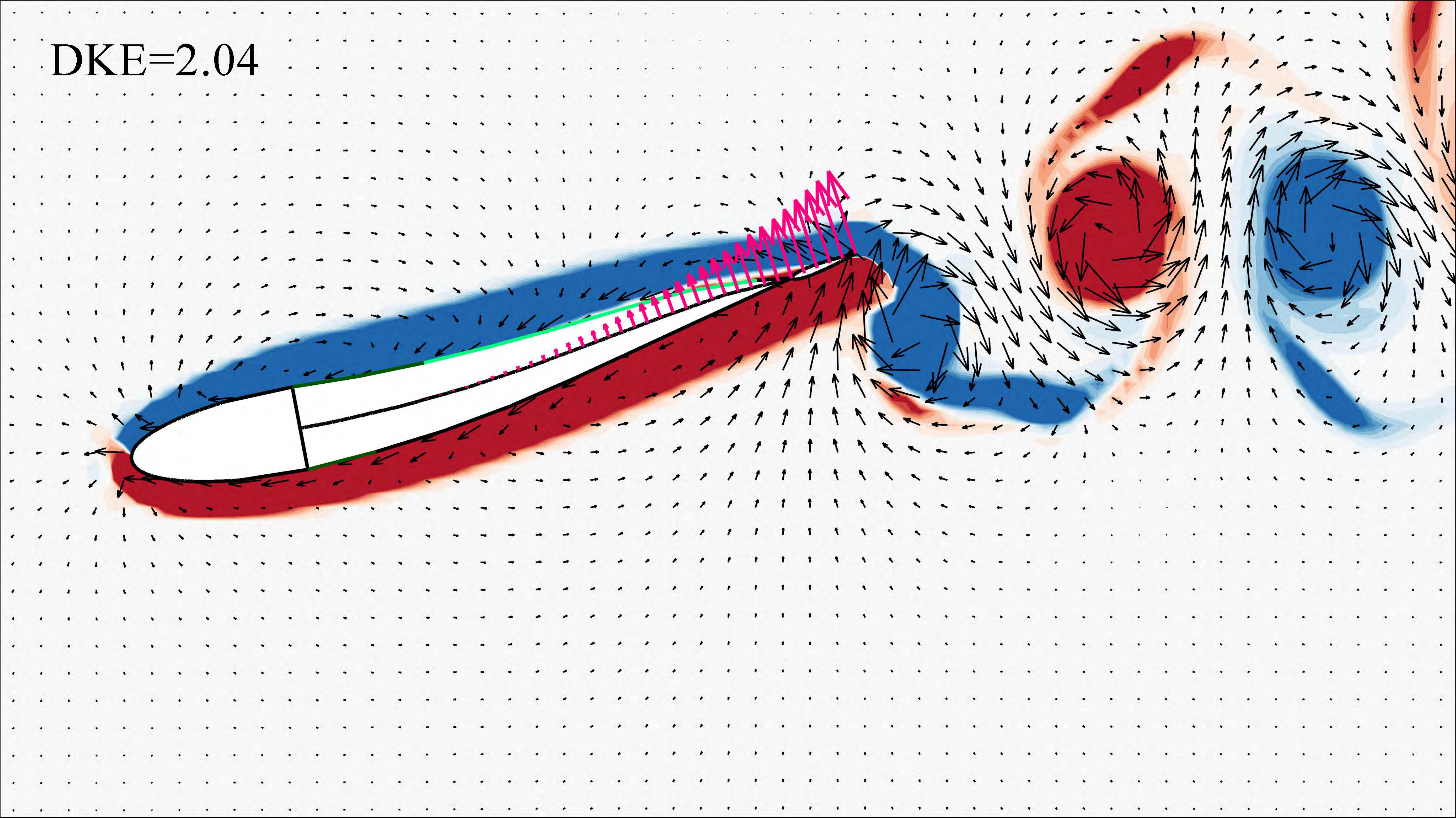}}
	\subfigure[$t = 1.0T$($\Phi=360.0^\circ $)]{
		\includegraphics[width=0.4\linewidth]{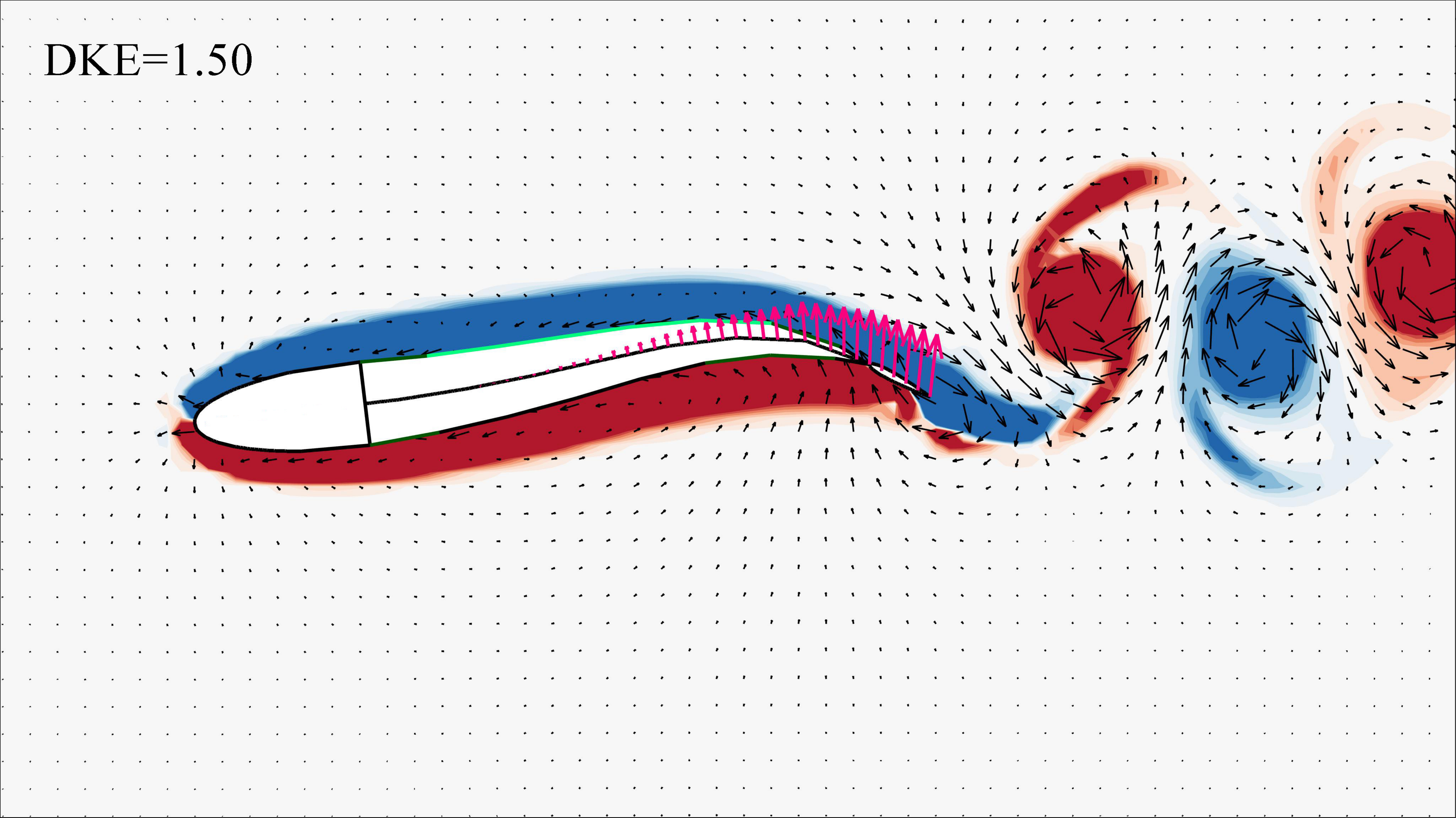}}
	\subfigure[]{
		\includegraphics[width=5in]{color_bar_2-eps-converted-to.pdf}}
	\caption{Comparison of instantaneous flow vorticity fields with varying $\Phi$ (black indicates inactive muscles, bright green indicates strongly activated muscles, pink arrows represent body deformation velocity vectors, and black arrows represent flow velocity vectors; T denotes the tail beat cycle).}
	\label{pic/f25}
\end{figure}

Figures \ref{pic/f24} and \ref{pic/f25} present the body deformation velocity vectors and instantaneous flow fields snapshot under varying muscle phase lag $\Phi$, with contraction duration time fixed at $\text{CDT} = 0.4\,\text{TBC}$. The results indicated that excessively high phase lag can lead to competition between contralateral myomere activation and body wave propagation. This competition suppressed both the velocity and amplitude of tail-edge motion. For instance, at $t = 0.2\,T$, leftward reverse body motion had already begun under several conditions: $\Phi = 0.004^\circ$ (figure \ref{pic/f24}-a, DKE = 3.02), $\Phi = 86.4^\circ$ (figure \ref{pic/f24}-b, DKE = 2.48), $\Phi = 201.6^\circ$ (figure \ref{pic/f25}-a, DKE = 0.25). However, under the condition $\Phi = 360^\circ$, due to the excessive phase lag, the left-side muscle contraction began before the right-side tail muscles had started relaxation, impeding the initiation of tail reversal. As a result, DKE was only 0.37, and the body remained in a residual motion phase from rightward tail beating (figure \ref{pic/f25}-b). At $t = 0.4\,T$, the reverse motion in the high phase lag case ($\Phi = 360^\circ$) finally began. However, the activation wave had not yet propagated to the left-side caudal peduncle, and the right-side caudal muscles were still actively contracting. This prevents the tail from undergoing secondary acceleration, resulting in low deformation kinetic energy (figure \ref{pic/f25}-d, DKE = 0.59). A similar suppression effect was observed at $t = 0.8\,T$ during left-to-right motion reversal (figure \ref{pic/f25}-h), where DKE was only 0.15. From a wave dynamics perspective, motion modes with appropriately tuned muscle phase lag produce sequential axial curvature waves that create flexible, mesh-like body undulations. Compared to modes with synchronous activation of ipsilateral muscles (i.e., low $\Phi$), which generate stiff, divergent body waves, sequential activation allows the organism to achieve comparable swimming speeds while significantly reducing wave amplitude. Moreover, these flexible mesh-like waves induce weaker flow disturbances (figure \ref{pic/f15}), which may benefit fish in predator evasion and stealth locomotion (\cite{stewart2014prey}).

\section{Conclusion}\label{sec:conclusion}

In this study, we aimed to investigate the integrated locomotor chain of fish swimming from multiple perspectives—including internal biomechanics, external hydrodynamics, and multiscale internal–external coupling—with the goal of uncovering the multiscale locomotion patterns and cross-scale energy transfer mechanisms, from the muscular level to body undulations level and ultimately to macroscopic efficient propulsion. Using trout (Oncorhynchus mykiss) as the model organism, we developed a high-fidelity digital trout model in a virtual environment that closely replicates the morphology and biomechanics of a live trout. This model integrated the Hill-type muscle model, multibody dynamics, and a fluid–structure interaction algorithm to achieve multi-scale modeling of the fish's locomotor chain. Muscle activation patterns at the neural system level were controlled using deep reinforcement learning, enabling us to study, from an embodied intelligence perspective, how different muscle activation modes influence swimming speed and energy consumption. The main findings are summarized as follows:

(1) During steady swimming, the myomeres distributed along the body axis function in a serially coupled manner, enabling effective integration of propulsion throughout the swimming process. Small-amplitude curvature waves generated by relatively minor muscle contractions in the rostral region can be progressively amplified through the temporally sequenced activation and coordinated contraction of downstream myomeres along the body axis. This amplification culminates in the formation of high-amplitude body waves at the caudal fin. To maintain the stability of the curvature wave and support the characteristic quadratic growth of the wave envelope from head to tail, the longitudinal span of muscle activation must exceed 0.5 body lengths (BL). Activation ranges below this critical threshold lead to significant amplitude attenuation during posterior wave propagation, destabilization of the wave structure, and a notable decline in swimming speed. Therefore, when designing multi-stage artificial muscle-driven biomimetic robotic fish using materials such as Electroactive Polymers (EAPs), Shape Memory Alloys (SMAs), and Pneumatic Artificial Muscles (PAMs), the longitudinal deployment span of multiple actuators along the body axis constitutes a critical design consideration.

(2) When the phase lag parameter $\Phi$ is relatively small, inertial forces, elastic forces, and fluid–structure interactions play a dominant role in determining both the fish’s energy consumption and swimming speed. Notably, high-reward locomotion modes are concentrated in this region of the parametric space. In such modes, when the contraction duration time is relatively short, the fish body quickly transitions into an inertia-driven regime after inheriting residual kinetic energy from the previous cycle. During this phase, passive braking and directional reversal are achieved through the combined effects of body elasticity and fluid–structure coupling. In this process, the interaction between the fish body and the surrounding fluid acts as a passive damping system, enabling deceleration without active muscular input. This mechanism substantially reduces the energetic cost of body reversal, placing this locomotion mode in an advantageous position on the energy consumption–speed Pareto frontier. Specifically, it allows the fish to achieve 90\% of the cruising speed of a long-CDT mode with only 33\% of the energy consumption. Conversely, when CDT is long, the body accumulates excessive momentum due to prolonged active muscle contraction. Consequently, additional energy must be expended to reverse the direction of body motion.

(3) Phase coordination plays a crucial role in shaping body wave morphology and determining propulsion characteristics. When the phase lag $\Phi$ is small, the muscles on the same side of the body contract nearly simultaneously, resulting in a rigidly divergent body wave characterized by large lateral amplitudes. At moderate values of $\Phi$, the induced sequential muscle activation generates flexible mesh-like body waves that propagate from head to tail. Compared to rigid divergent waves, these flexible mesh-like waves achieve higher propulsion speeds at lower amplitudes, indicating greater hydrodynamic efficiency. However, when $\Phi$ becomes excessively large, it induces spatiotemporal competition between contralateral muscle contractions. This leads to a suppression of caudal acceleration and a delay in body directional reversal. Although the body wave may still exhibit a flexible mesh-like structure, its amplitude is substantially reduced. As a result, this mode exhibits weakened thrust vectors and a noticeable decline in swimming speed.

At present, most studies focus on the kinematic perspective, examining how fish utilize external flows and vortex structures to achieve energy-efficient and high-performance swimming. However, few studies have investigated the cross-scale energy transmission processes from a dynamic perspective, as explored in this study. Energy-efficient swimming in fish is the result of a complex interplay of multiple factors. Prior research has demonstrated that fish can exploit external flow conditions to reduce locomotor energy consumption (\cite{liao2003fish, liao2003karman}). The findings of this study, however, reveal that fish may also proactively leverage their own bodily structural advantages—from the standpoint of emergent embodied intelligence—to conserve energy during swimming. This insight could offer valuable implications for the design of bio-inspired underwater robotic systems.

\section*{Acknowledgement.}

The authors acknowledge the support provided by the National Key Research and Development Program of China (Grant No. 2023YFC3205900), Chongqing
Water Conservancy Science and Technology Project (Grant No. CQSLK-2024021), the Project of Xinjiang Ecological Water Conservancy Research Center (Grant No. 2024B002), the Guizhou Province Science and Technology Project (Grant No. [2024]116). 

%
%
%
%
%
%

\clearpage



\bibliographystyle{jfm}
\bibliography{reference}


\end{document}